\documentclass[aps,amsmath,amsfonts,10pt]{revtex4}
\usepackage{epsfig,graphicx}
\usepackage{setspace} 
\usepackage[english]{babel}
\usepackage{amsfonts}
\usepackage{amsmath}
\usepackage{latexsym}
\usepackage{graphics,bm}
\usepackage{natbib}
\usepackage{dcolumn}
\usepackage{bm}
\usepackage{rotating}
\usepackage{amsmath}
\usepackage{braket}
\usepackage{epstopdf}

\begin{document}

\title{Cavity quantum electrodynamics of ultracold atoms in optical and optomechanical cavities}

\author{Kamanasish Debnath$^1$}
\author{Aranya B Bhattacherjee$^2$}
\address{$^1$Institute of Applied Sciences, Amity University, Noida - 201303 (U.P.), India \\
 $^2$School of Physical Sciences, Jawaharlal Nehru University, New Delhi- 110067, India}

\maketitle

\tableofcontents

\section{Introduction}
	The phase transition in an ideal gas of identical bosons near absolute zero still remains the most amazing prediction of quantum statistical mechanics made in 1926 by S. N. Bose and A. Einstein. The Bose Einstein statistics allows the integral spin particles to share the same lowest energy state, thereby exceeding the thermal de Broglie wavelength from the mean spacing between the particles. It was predicted that if it could be possible to cool a collection of N non interacting bosonic particles near absolute zero, all the particles would come to the lowest energy state and would coalesce into a single quantum mechanical entity, described by a single wave function. Such a phase transition leads to the formation of the fifth state of matter called the Bose Einstein Condensate. Such a state is believed to exist nowhere in the universe other than the highly sophisticated quantum labs.\\

	With few decades of experimental advancements backed by theoretical proposals, the condensate was prepared in 1995 by Eric Cornell and Carl Wieman in JILA \citep{1} and Wolfgang Ketterle in MIT \citep{2}, thereby jointly sharing the 2001 Nobel Prize for Physics. In the breakthrough experiment of JILA, approximately 2000 spin polarised $^{87}$Rb atoms were produced in a magnetic trap for 15- 20 seconds at a temperature of around 170 nK, with a density of 2.6 $\times$ 10$^{12}$ cm$^{-3}$. The advent of advanced cooling techniques like laser and evaporative cooling and realization of efficient magneto optical traps (MOT) by three pairs of lasers and anti Helmholtz coils produced the condensate of 20 micron diameter. Laser cooling applies the technique of Doppler Effect to hit the atoms from opposite direction with frequency apt for the atomic transition, thereby exciting the atom to higher level and decreasing its velocity due to the recoil momentum after collision. Repeating the process many times ultimately leads to decrease in average velocity of the particles, which is actually the measure of the temperature of the system. Atoms cooled so far are too hot for the required phase transition. Evaporative cooling by altering the trap potential through rf field allows the high energy particles to leave the trap which leads to further cooling of the system to nanokelvins and forming the condensate. Similar techniques were used later to demonstrate BEC of trapped vapours of $^{23}$Na \citep{3}, $^{7}$Li \citep{4}, $^{84}$Sr \citep{5}, $^{133}$Cs \citep{6} and many other Alkali metals.\\

	Soon after the experimental realisation of BECs, there was a sudden outburst of theoretical and experimental interest in ultra cold ensembles for probing into its properties and study its applications in diverse fields like quantum computation, communications and even detection of weak forces. Laser became the versatile tool to prepare and manipulate such atoms. Ultra cold atoms combined with the tools of cavity quantum electrodynamics have made it possible to study them extensively in optical cavities \citep{9, 11} and even optomechanical cavities \citep{10}. The lasers form an effective optical trap for microscopic atoms by interference pattern of the external pump and cavity beams. Although the probability of a photon to be scattered by a particle in free space is negligible, the case changes when the light is confined between high finesse optical resonators. Multiple round trips of beam inside the resonator with ultra cold atoms enhance the backaction of the atoms on the cavity light, creating an interesting interaction scenario between light and atoms. Such systems treats cavity dynamics and atomic motion on equal footing and the phase shift induced due to the interaction of the cavity light with atoms, makes it a perfect tool to study quantum dynamics of trapped ultracold atoms. Such ensembles in optical resonators are one of the most intriguing systems where quantum and solid state phenomena can be explored using the power of atomic and laser physics. Historically, cavity quantum electrodynamics was designed for studying the radiation properties of atoms in the presence of boundaries. For cold atoms in a laser driven optical cavity, the dynamics of the system becomes complex when the atom field coupling strength increases. The effective coupling strength between the cavity field and the particles often varies with the square root of the particle number. Since the local intensity of the cavity field experienced by single atom is dependent on the position of all other atoms, complexity arises when there is relative motion. This gives rise to a global atom- atom interaction, described by overall dispersive shifts. The long range interaction induced by the cavity gets a different character when the cavity mode and the driving field mode are not identical. In such cases the interference between the intracavity fields becomes crucial.\\

	Further developments in microengineering techniques brought fore the coupling of optical and mechanical mode through optomechanical cavities. Once the fixed mirror of the cavity is replaced by a cantilever, ground state of the mechanical mirror can be achieved through Doppler cooling. Correspondingly, the mirror behaves as a harmonic oscillator generating discrete phonons due to radiation pressure. Successful integration of BECs with optomechanical cavities, led to realization of effecting coupling between the optical mode, mechanical mode and atomic motion. In general, the ground state cooling of the mechanical mirror plays a crutial role in connecting the regimes. The exchange of momentum between light and matter leads to the alteration of motional degrees of freedom of the mechanical system. Changes induced by the mechanical motion of the mirror on the resonant frequency and its back action connects the quantum world with the macroscopic objects \citep{12, 13, 14}, leading to realisation of phenomenon like quantum entanglement \citep{7, 8, 15}, controllable phase transitions etc. \\

	In case of transversally driven ultracold atoms within a cavity, if the laser driven Bose Einstein Condensate is coupled to the vacuum field of the cavity, a quantum phase transition is observed between a superfluid and a self organized state above a critical threshold frequency. This corresponds to the open system realization of the Dicke Hamiltonian and its quantum phase transition \citep{18, 19}. This self organization refers to the onset of the superradiance in an effective non equilibrium Dicke model \citep{16} and can also be considered as a supersolid resulting from a broken Ising type symmetry. Experimental realisation of such Dicke models has made it possible to observe the superradiant phases and attractors of the system predicted theoretically. On replacing the fixed mirror of the cavity with a mechanical one, the mirror frequency shifts the usual critical transition point. In the present cold atom settings, the splitting of the two distinct momentum states of the BEC is controlled by the atomic recoil energy, and this enables the phase transition to be observed with optical frequencies with light. This is quite similar to the theoretical approach proposed by Dimer $\textit{et al.}$ \citep{17}, for attaining Dicke phase transitions using Raman pumping schemes between the hyperfine levels. More complexity arises in highly degenerate multimode cavities \citep{20} which are presumed to exhibit interesting physics and applications in the field of quantum information \citep{24} and simulation \citep{22}. In a two mode Dicke model an electromagnetic superradiant phase exists when the two atom photon coupling strengths were made equal leading to continuous U (1) symmetry breaking and the existence of Nambu Goldstone mode \citep{21}. This can also be realised experimentally by combining Bragg spectroscopy and cavity enhanced Bragg scattering.

	In this review, we shall illuminate the recent theoretical developments and some breakthrough experiments in cavity QED systems \citep{25, 26, 27} and also focus on the experimental realisations \citep{28, 29} of the theoretical proposals. Briefly summarizing, in the first section we shall discuss the integration of ultracold atoms in optical cavities and the concept of cavity cooling. We shall study the optomechanical system and bring out the dynamics of the movable mirror and the dependency of the mirror spectrum on the position of the condensate and also the mixing of the mechanical motion with the fluctuation of the mirror and the condensate with finite two body interaction. We shall finally discuss the non equilibrium Dicke models and Jaynes Cumming model and relate the theoretical analysis and phase portraits with the experimentally observed data.

\section{Optical Cavities }

An optical cavity is an arrangement of high finesse mirrors that forms a standing wave by superposition of two waves of same frequency and amplitude. Light confined within the cavity experiences multiple reflections from the mirrors to form a standing wave to trap atoms or molecules within them. Early development of such sytems were devoted for studying radiation pressure effects of light. Successful integration of such optical cavities with atoms and ultracold ensembles made it possibe to study the light matter interaction to a large extend in a microscopic level. Successful cavity QED experiments revealed the forces of cavity light to deflect the slowly moving atoms within the cavity. With the advancements in microscopic engineering, it is now possible to study such systems with higher precision and control. The optical quantum electrodynamics plays an important role in exploring quantum dynamics of such systems where evolution rate dominates the dissipating process. A complete theoretical description of the atom field dynamics is given by Domokos and Ritsch \citep{23}. In this section, we initially highlight a simple model of single two level atom in an optical cavity and discuss the concept of cavity cooling and friction coefficients. Next, we integrate the ultracold atoms within the optical cavity and probe into the properties and behaviour of such ensembles. Finally, we produce the results of some important breakthrough experiments by Tilman Esslinger and his group \citep{25, 29} showing the experimental realization of energy spectrum and bistable behaviour of such coupled BEC- cavity systems.

\subsection{Two level atom in an optical cavity- Particle dynamics, Friction and cavity cooling}

We consider a single two level atom coupled to the optical mode inside an optical resonator with transition frequency $\omega_{A}$ and resonance frequency $\omega_{C}$. Cavity detuning $\Delta_{C}$ and atomic detuning $\Delta_{A}$ is defined with respect to the frequency of the external laser $\omega$ as $\Delta_{C}$ = $\omega$ - $\omega_{C}$ and $\Delta_{A}$= $\omega$- $\omega_{A}$. The two levels of the atomic system refers to the ground state  $\Ket{g}$ and excited state  $\Ket{e}$. The system is excited with the coherent laser field and the dissipative cooling of the atomic modes requires a suitable closed atomic transition, since the kinetic energy carried away by the atom after collision by exciting to higher levels is very small and has to be repeated several number of times. Basic properties of such systems are governed by the atom coupling strength (g), cavity photon loss $\kappa$, atomic decay rate $\gamma$ and the atom field interaction time $\tau$. Strong coupling of such systems is defined as g$\ge$ $\kappa$, $\gamma$, $\tau^{-1}$, so that the atom field interaction predominates the dissipative process. Two dissipative processes, namely the photon decay rate ($\kappa$) and emission rate ($\gamma$) has been considered in the present scenario. Considering a single two level atom moving along the axis of a single mode, the mode function and the inhomogeneous coupling has been defined as $f(x)= cos (kx)$ and $g(x)= gf(x)$ respectively. Neglecting the noise terms, the semi classical equations of motion for field amplitude and force, read as

\begin{equation}
\dot{\alpha}= [\textit{i}\Delta_{C}- \kappa- (\textit{i}U_{0} + \Gamma_{0}) cos^{2} (\textit{kx})]\alpha + \eta,
\end{equation}

and 

\begin{equation} 
\dot{p}= \hbar U_{0}\mid\alpha\mid^{2}sin (\textit{2kx}),
\end{equation}

where $\eta$ represents the frequency of the external pumping source. U$_{0}$ defines the light- shift coefficient of the cavity resonance and hence determines the dipole force on the CM motion, thereby yielding a potential with half of the wavelength as its period. $\Gamma_{0}$ refers to the extra position dependent cavity loss due to the incoherent scattering of the cavity photons to the side. The plot (fig.1) shows the dynamical behaviour of the coupled system where the dotted, solid and dashed lines corresponds to the field intensity, particle position and momentum respectively. Initially, as seen from the figure, the atom is moving fast along the cavity axis, modifying the intracavity field. Shortly after the atom passes the antinode of the mode (horizontal lines), the field intensity (dotted) reaches a maximum value. On average, the atom has to climb steeper hills than it rolls down thereby eventually coming to rest and finally oscillates in a single well where the kinetic energy slowly damps further. Of course the discussion here is completely classical and in the real quantum world the cooling or damping effect will be accompanied by the noise terms in the above dynamical equations.

\begin{figure}[h!]
\includegraphics[width=0.39\textwidth]{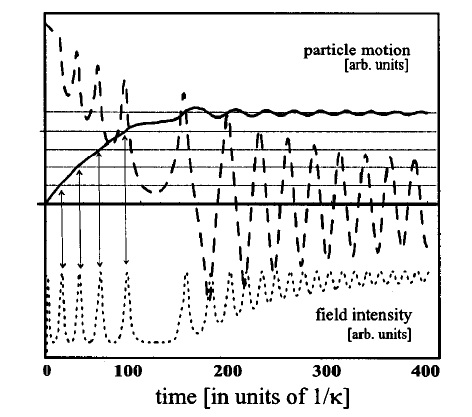}
\caption{The time evolution of particle position (solid), field intensity (dotted) and particle momentum (dashed) for a single atom. Parameters used were $U_{0}$= -3$\kappa$, $\gamma$= 0.1$\kappa$, $\Delta_c$= -4$\kappa$ and $\eta$= 1.5$\kappa$. The position of the field antinodes are marked with horizontal mark lines. Reprinted with permission from Peter Domokos $\textit{et al.}$, J. Opt. Soc. Am. B/ Vol. 20/ No. 5 (May, 2003) \citep{23}.}
\centering
\end{figure}

Realization of controlled cavity cooling through dissipation of kinetic energy by cavity photon loss has been a subject of interest since the advent of optical cavity ensembles. Early ideas relied on modification of the spectral mode density of the radiation field in the presence of spatial boundary conditions. To describe the cavity cooling mechanism by simple picture of friction coefficient, we explore two regimes, namely the good cavity and bad cavity limit specified by $\kappa <<$ g and $\kappa >>$ g respectively. When the coupling parameter is much much smaller than the cavity photon loss rate, the photon leak out of the cavity fast, such that no appreciable dynamics can be found for time scale smaller than $\kappa^{-1}$. The optical cavity works for reshaping the radiative environment of the atom, thereby increasing spontaneous emission rate around the cavity frequency. Although the atomic dipole is linear in electric field for bad cavity limit, yet inelastic scattering can occur because of CM motion and compensate for the energy difference of the outgoing and incoming photons. For direct field mode pumping, one encounters a situation analogous to free space Doppler cooling in a stationary wave. The frictional force upto first order takes the form: -

\begin{equation} 
F^{(1)}= \nu2\hbar[f'(x)]^{2}g^{2}\mid\alpha\mid^{2}\frac{2\gamma\Delta_{A}}{(\Delta_{A}^{2} + \gamma^{2})^{2}},
\end{equation}

It is evident from the above equation that the frictional parameter is dependent crucially on the sign of the detuning $\Delta_A$. For the other case of pumping the atom solely (atom pumping), the cavity photon number remains too low to observe Doppler cooling. However as the scattering can be inelastic due to CM motion, spontaneous emission is favoured at the cavity frequency, which is higher than the incoming photon frequency, thereby creating cooling through loss of kinetic energy. The same mechanism acts in reverse, leading to heating (blue contours in fig.2) for $\Delta_C>$0. Hence for atom pumping the sign of $\Delta_C$ remains crucial and the equation for frictional force upto first order takes the form: -

\begin{equation}
F^{(1)}= \nu2\hbar[f'(x)]^{2}g^{2}\mid\ s\mid^{2}\frac{2\kappa\Delta_{C}}{(\Delta_{C}^{2} + \kappa^{2})^{2}},
\end{equation}

\begin{figure}[h!]
\includegraphics[width=0.8\textwidth]{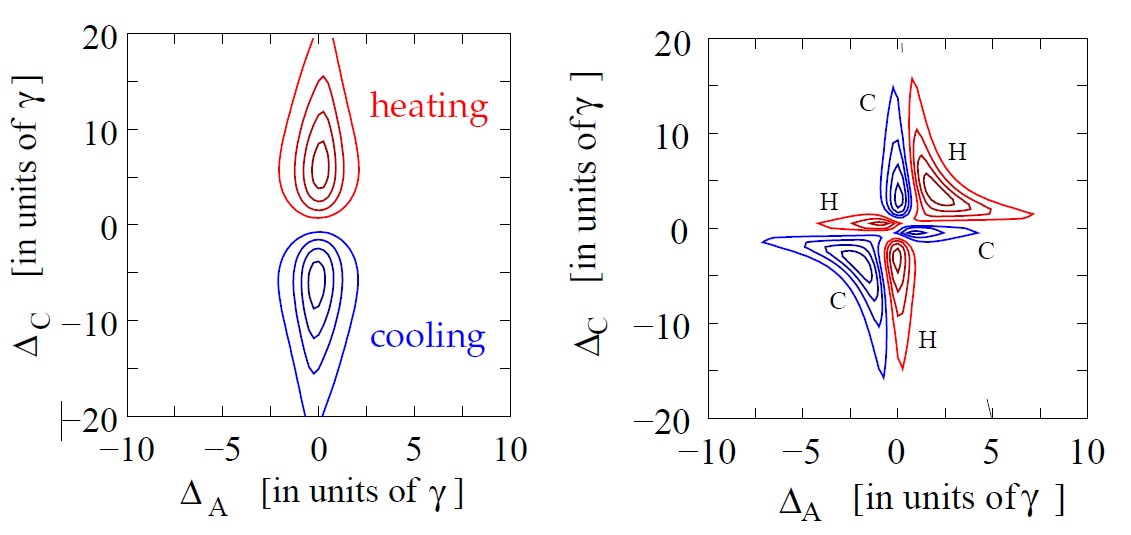}
\caption{Cooling and heating regions as a function of $\Delta_C$ and $\Delta_A$. Left (Right) contours corresponds to bad (good) cavity limit. Reprinted with permission from Helmut Ritsch $\textit{et al.}$, Rev. Mod. Phys, 85, 553- 601 (2013) \citep{30}.}
\centering
\end{figure}	

The frictional force represented by the above equations for good and bad cavity limit has been plotted as a function of cavity detuning ($\Delta_C$) and atomic detuning ($\Delta_A$) in fig. 2. However, for the good cavity limit, the scattering approach does not apply. The system spends a considerable time in the dressed state given as: -

\begin{equation} 
\Ket{+}= cos \theta \Ket{e, 0}+ i sin \theta \Ket{g, 1}
\end{equation}

and 

\begin{equation}
\Ket{-}= i sin \theta \Ket{e, 0}+ cos \theta \Ket{g, 1},
\end{equation}

where $\Ket{g}$ and $\Ket{e}$ denotes the bare atomic ground and excited states respectively with $\Ket{n}$ ($n$= 0, 1) representing the photon number states of the field mode. Accordingly, Sisyphus- type argument can be used to describe the cooling mechanism. As seen from fig. 3, when the excited state is $\Ket{+}$, a slowly moving atom has to climb up the potential hill, thereby increasing potential energy before it decays into the ground state. Hence kinetic energy is converted to potential energy before any spontaneous emission. Conversely for the case when $\Delta_A \approx $ 0, the lower state $\Ket{-}$ is pumped resonantly at the node, and as the schematic representation suggests, the atom is likely to descend potential wells, leading to heating. The upper dressed state is excited at the node, if pumped and this state has no $\Ket{g, 1}$ component, thereby removing any possibility of excitation from the ground state under pumping. The cooling process for this frequency setting is thus much powerful, which explains the appearance of narrow peaks at $\Delta_A$= 0, which is otherwise absent for cavity pumping plots (fig. 2). The left contour of fig. 2, corresponds to bad cavity limit ($g= \gamma/2$, $\kappa$= 10$\gamma$) and the right presenting the cooling and heating regions for good cavity limit ($g$= 3$\gamma$, $\kappa$= $\gamma$).

\begin{figure}[h!]
\includegraphics[width=0.8\textwidth]{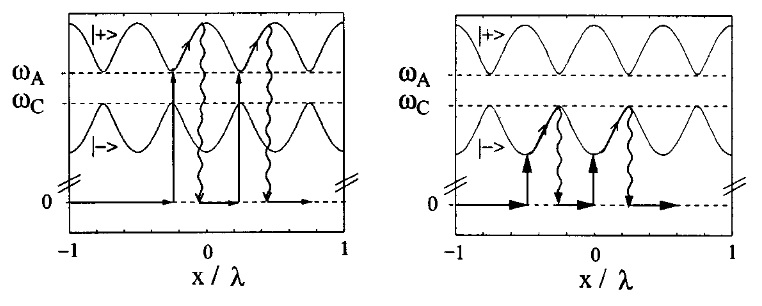}
\caption{Schematic representation of the Sisyphus cooling mechanism for $\Delta_ A \approx$ 0, $\Delta_C>$ 0 (left) and $\Delta_A$, $\Delta_C >$, $\Delta_A \Delta_C\approx g^2$ (right). Reprinted with permission from Peter Domokos $\textit{et al.}$, J. Opt. Soc. Am. B/ Vol. 20/ No. 5 (May, 2003) \citep{23}.}
\centering
\end{figure}

Experimentally cavity cooling has been observed with both blue \citep{32} and red \citep{33} detuned probe light. The experimental scheme developed by Maunz $\textit{et al.}$ \citep{32} consists of an optical cavity with a high finesse resonator. $^{85}$Rb atoms were trapped within the longitudinal cavity with a trap field which was red detuned with respect to the atom. The cooling, induced with a weak, blue detuned probe light brought a cooling rate which exceeded the rate achieved in free space cooling methods atleast by a factor of 5. The intracavity intensity was maximum for an atom placed at the node of the standing wave since it does not couple to the cavity field mode. On the contrary, an atom shifts the cavity resonance towards higher frequency when placed at the antinode, resulting in a reduced intracavity intensity. However, the intensity cannot drop instantly when the atom moves towards an antinode. Before the photons are able to leak out of the cavity, the induced blueshift of the cavity frequency leads to an increase of the energy stored in the field, which occurs at the expense of kinetic energy of the atoms thereby inducing cavity cooling. The reverse effect for an atom moving towards a node is much weaker since the cavity remains initially out of resonance with the probe light and a very small number of photon undergoes corresponding redshift. This argument by Maunz $\textit{et al.}$ \citep{32} brings out a delicate correlation between the atomic motion and the photon number variation underlying the cavity cooling technique. In the realistic experiment, atoms injected into the cavity were trapped at the field antinodes and on inducing cavity cooling, the resulting confinement was determined from the transmitted signal. Time- resolved detection of the transmitted signal allowed one to determine the cooling rate for the experiment to be 21 KHz, which was very much large as compared to the Doppler cooling (1.5 KHz) or blue detuned Sisyphus cooling (4 KHz). A detailed discussion of the experimental results and plots for the average cavity transmission of probe light can be found in Maunz $\textit{et al.}$ \citep{32}.

\section{Cold atomic ensembles in optical cavities}
Successful integration of ultracold atoms with optical cavities opened new directions in cavity QED for probing into the properties of trapped atoms. Coupling of the atoms to the cavity field in many body configurations created the possibility to implement atom- atom interaction over large distances mediated through cavity radiation field. The size of the cavity mode determines the interaction range, which in some cases can be macroscopic also. Such systems exhibit phenomenon of solid state physics like the formation of energy bands, Bloch oscillations and Josephson effects. In high Q- cavity, the quantum effects of the system become important and the atoms move in quantized potentials created by the interference of the pump and cavity beams. The strong coupling of the cold atoms to the cavity mode changes the resonant frequency of the cavity and the driving field in the cavity can significantly enhance the localization and cooling properties of the system. Due to the strong coupling of the condensate to the cavity mode, a band structure of the condensate leads to a band structure of the intracavity light fields. This influences the Bloch energies, effective mass and Bogoliubov excitations of the BEC.\\

	To study a generalized cold atom Bose Hubbard model where the periodic optical potential in formed by the cavity field with quantum properties, we consider an optical cavity with N two- level atoms with mass $m$ and transition frequency $\omega_a$. The atoms strongly interact with a single standing wave cavity mode of frequency $\omega_c << \omega_a$. A laser with frequency $\omega_p$ and amplitude $\eta$ coherently drives the system which is also illuminated transversally to the cavity axes with transverse pump of amplitude $\zeta$ as shown in the schematic representation below (fig.4). Here $\kappa$ represents the cavity photon loss rate.

\begin{figure}[h!]
\includegraphics[width=0.75\textwidth]{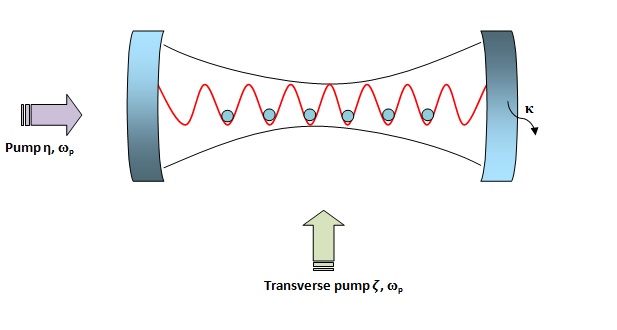}
\caption{The schematic representation of the setup with N two level atoms trapped between the optical cavity.}
\centering
\end{figure}

The Hamiltonian of such system under rotating wave approximation and adiabatic elimination of the excited states, takes the form \citep{34}: -

\begin{equation}
H= \frac{p^2}{2m}+ cos^2(kx)(\hbar U_0 a^{\dagger} a+ V_{cl})- \hbar\Delta_c a^{\dagger}a- i\hbar\eta(a- a^{\dagger})+ \hbar\eta_{eff}cos(kx)(a+ a^{\dagger}),
\end{equation}

where $a^{\dagger} (a)$ is the creation (annihilation) operator for cavity photons, $\Delta_c= \omega_p- \omega_c$ is the cavity pump detuning. $U_0= g_0^2/ \Delta_a$ denotes the optical lattice depth per photon and the effective pump through atomic scattering is represented by $\eta_{eff}= g_0 h_0 \zeta/\Delta_a$. The cavity, along x forms an optical lattice potential with period $\lambda/2$ and depth $\hbar U_0a^{\dagger}a$. $V_{cl}$ is the classical potential added for the sake of generality and in the later sections, we shall approximate it to unity. Along the x- axis, the atom- field coupling is set to $g(x)= g_0$ $ cos(kx)$, while the amplitude of the standing wave formed by the transverse pumping is denoted as $h(x)= h_0$ $ cos(k_p y)$.\\

The Bloch states of a single atom can be expanded inside the lattice using the localized Wannier functions with $b_k$ and $b_k^{\dagger}$ corresponding to the annihilation and creation operator of an atom at site $k$ and $\psi (x)= \sum\limits_{i} b_{i} w(x- x_i)$. The modified Hamiltonian with $U= \frac{4\pi a_s \hbar^2}{m} \int d^3x \mid (x) \mid^4$, representing the on- site interaction of two atoms can be written as: -

\begin{eqnarray} 
H&=& \sum\limits_{k, l} E_{k, l}b_k^{\dagger}b_l+ (\hbar U_0 a^{\dagger} a+ V_{cl}) \sum\limits_{k, l} J_{k, l}b_k^{\dagger}b_l + \hbar\eta_{eff}(a+ a^{\dagger}) \sum\limits_{k, l} \tilde{J}_{k, l}b_k^{\dagger}b_l- i\hbar\eta(a- a^{\dagger})\nonumber \\
&+& \frac{U}{2}\sum\limits_{k}b_k^{\dagger}b_k (b_k^{\dagger}b_k- 1)- \hbar\Delta_c a^{\dagger}a,
\end{eqnarray}

where the coupling matrix elements can be referred from Appendix A. In the case of transverse pumping, the two adjacent potential wells acquire different depths since the cosine term in the previous Hamiltonian changes sign periodically. Introducing the number operator $\hat{N}= \sum\limits_{k}= \hat{n}_k= \sum\limits_{k} b_k^{\dagger} b_k$ and  jump operator $\hat{B}= \sum\limits_{k} \Big(b_{k+1}^{\dagger}b_k+ h.c.\Big)$, the Hamiltonian reads: -

\begin{eqnarray}
H&=& E_0\hat{N}+ E\hat{B}+ (\hbar U_0a^{\dagger}a+ V_{cl})\Big(J_0\hat{N}+ J\hat{B}\Big)+ \hbar\eta_{eff}(a+ a^{\dagger})\tilde{J_0}\sum\limits_{k}(-1)^{k+1}\hat{n}_k- \hbar\Delta_c a^{\dagger}a \nonumber \\
&-& i\hbar\eta(a- a^{\dagger})+ \frac{U}{2}\sum\limits_{k}\hat{n}_k (\hat{n}_k- 1),
\end{eqnarray}

with $E_0$, $J_0$, $\tilde{J}_0$ denoting the on- site matrix element and $E$ and $J$ representing the site to site hopping element.

\subsection{Potential depth and energy difference for two atoms in two wells}

To investigate the dynamics of the system, we calculate the Heisenberg equations of motion for the light field as: -

\begin{equation} 
\dot{a}= \Big[i \Big(\Delta_c- U_0\Big(J_0\hat{N}+ J\hat{B}\Big)\Big)- \kappa\Big]a+ \eta- i\eta_{eff}\tilde{J}_0\sum\limits_{k}(-1)^{k+ 1}\hat{n_k}.
\end{equation}

\begin{figure}[h!]
\includegraphics[width=0.8\textwidth]{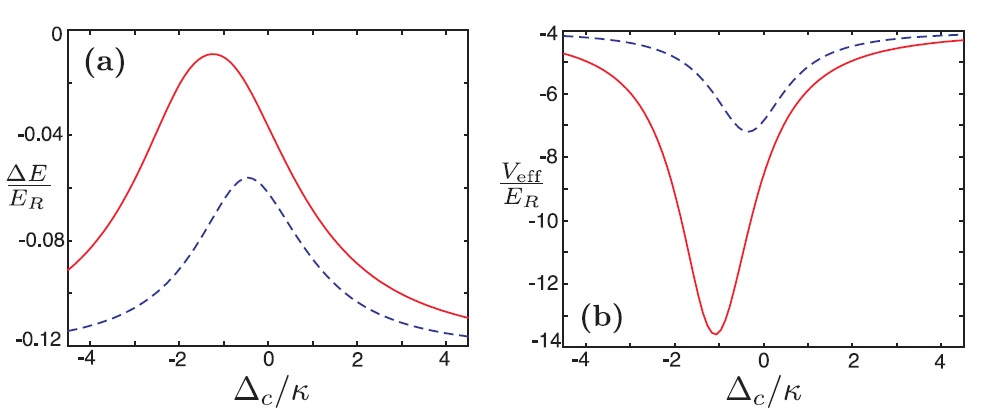}
\caption{Left- Energy difference $\Delta$E as a function of $\Delta_c$ for a single atom in two wells $U_0$= $-1.2\kappa$ ($-0.4\kappa$) for solid (dashed) line.\\
Right- The associated lattice depth for the same parameters. Reprinted with permission from Christoph Maschler $\textit{et al.}$, Phys. Rev. Lett., 95, 260401 (2005). \citep{34}.}
\centering
\end{figure}

It is evident that both number of atoms $\hat{N}$ and coherences via $\hat{B}$ determines the quantum state of the field. We consider here a simple setup by setting $\eta_{eff}$= 0 in the absence of transverse pumping. In the bad cavity limit, for a fixed atom number $N$= $<\hat{N}>$, expanding $a$ to second order in the small tunnelling matrix element $J$ gives: -

\begin{equation} 
a \approx \frac{\eta}{\kappa- i \Delta_c'} \Big [ 1- \frac{i U_0 J}{\kappa- i \Delta_c'}\hat{B}- \frac{U_0^2J^2}{(^2\kappa- i\Delta_c')}\hat{B}^2\Big],
\end{equation}

where $\Delta_c'= \Delta_c- U_0J_0N$ is the rescaled detuning. Considering the simplest case of a single particle in two wells, the energy difference between the eigenstates of the symmetric and anti symmetric superposition is given as: -

\begin{equation} 
\Delta E= 2\Big[ E+ J\Big( V_{cl}- \hbar U_0 \eta^2 \frac{\kappa^2- \Delta_c'^2}{(\kappa^2+ \Delta_c'^2)^2} \Big)\Big],
\end{equation}

which is strongly dependent on the cavity parameters as shown by Christoph Maschler $\textit{et al.}$ \citep{34} from the plots above (fig. 5). In simple words, the cavity- pump detuning can be used as a tool to alter the tunnel coupling and atom confinement. The symmetric and antisymmetric eigenstates are associated with different lattice depths. Adding more atoms to the system, the interaction term comes into play and the ground state remains as the superposition of different atomic configurations.

\subsection{Band structure of cavity field, effective mass and superfluid fraction}

To discuss the properties in detail, we simplify the previous model by approximating the classical potential to be unity and $\zeta$= 0. In the absence of the transverse pump, the corresponding Hamiltonian in rotating wave approximation and under adiabatic elimination of the excited state takes the form \citep{35}: -

\begin{eqnarray}
H&=& E_0\sum\limits_{j} b_j^{\dagger}b_j+ E\sum\limits_{j} \Big( b_{j+1}^{\dagger} b_j+ b_{j+1}b_j^{\dagger}\Big)+ \hbar U_0(a^{\dagger}a+ 1)\Bigg( J_0\sum\limits_{j} b_j^{\dagger}b_j+ J\sum\limits_{j}\Big(b_{j+1}^{\dagger}b_j+ b_{j+1}b_j^{\dagger}\Big)\Bigg)\nonumber \\
&-& \hbar\Delta_c a^{\dagger}a- i\hbar\eta(a- a^{\dagger})+ \frac{U}{2} \sum\limits_{j} b_j^{\dagger}b_j^{\dagger}b_j b_j,
\end{eqnarray}

where we have retained only the lowest band with nearest neighbour interaction. The constants $U, E_0, E, J_0$ and $J$ can be referred from Appendix A and all other variables have the same usual meaning as in previous section. The nearest neighbour interaction term are generally neglected due to their small magnitude as compared to the onsite interaction. The Heisenberg equation of motion for cavity photons $a$ and bosonic field operator $b$ can be written as: -

\begin{equation}
\dot{a}= -iU_0\Bigg( J_0 \sum\limits_{j} b_j^{\dagger} b_j+ J\Big(b_{j+1}^{\dagger}b_j+ b_{j+1}b_j^{\dagger}\Big) \Bigg)a+ \eta+ i\Delta_c a- \kappa a
\end{equation}

and

\begin{equation}
\dot{b}_j= -iU_0(1+ a^{\dagger}a) J\Big(b_{j+1}+ b_{j-1}\Big)- \frac{iE}{\hbar}\Big(b_{j+1}+ b_{j-1}\Big)- \frac{iUn_0I}{\hbar}b_j.
\end{equation}

where $I$ represents the occupied lattice sites and $\kappa$ represents the field damping rate. $n_0$= $N/I$ denotes the number of atoms per lattice site.

\subsubsection{Optical lattice potential}

Working in the bad cavity limit, where $\kappa$ is the fastest time scale, the intracavity field adiabatically follows the field. Assuming tight binding approximation in the steady state condition ($\dot{a}= 0$), $b_j$ can be replaced by $\phi_j= u_k exp(ikjd) exp(-i\mu t/\hbar)$ and we look for solutions in the form of Bloch waves, where $\mu$ and $d$ are the chemical potential and periodicity of the lattice respectively. In the tight binding approximation, keeping the pump- cavity detuning fixed at $\Delta_c \approx 2Jn_0 IU_0 cos(kd)$, the Heisenberg equation of motion for $a$ yields: -

\begin{equation}
a= \frac{\eta}{\kappa- i(\Delta_c- 2Jn_0 IU_0 cos(kd))}.
\end{equation}

We find that the quantum state of the cavity field varies with the Brillioun zone due to the atomic back action. Analogous to the photonic band gap materials, the cavity photons develop a band structure due to the strong coupling with the condensate. The atomic back action also modifies the effective optical lattice potential as $V_{op}= \hbar U_0(1+ a^{\dagger}a)$. As the condensate moves along the Brillioun zone, the optical lattice gets continiously modified as a result of variation in atom- field interaction. The final expression for optical lattice potential takes the form: -

\begin{equation}
V_{op}= \hbar U_0 \Big(1+ \frac{\eta^2}{\kappa^2+ [\Delta_c- 2Jn_0 I U_0 cos(kd)]^2}\Big).
\end{equation}

The plot for the effective optical lattice potential as a function of $kd$ has been plotted below (fig. 6) for $\frac{\Delta_C}{\kappa}=$ 5.5 (bold, red) and 3.5 (dashed, black). Evident from the plot (fig. 6), that a decrease in the pump- cavity detuning $\Delta_c$, increases the atomic back action and the potential increases. A larger pump- cavity detuning also reduces the effective optical potential height and in the absence of the pump, the $k$ dependency vanishes. The quantum state of the cavity field varies along the Brillioun zone due to the backaction of the cold atoms. This strong coupling between the cavity field and condensate generates the band like structure as seen from the plot below (fig. 6). A larger pump cavity reduces the effective optical potential height (red, dashed curve). The barrier height touches maximum when $\Delta_C$= $2Jn_0I U_0 cos(kd)$.

\begin{figure}[h!]
\includegraphics[width=0.5\textwidth]{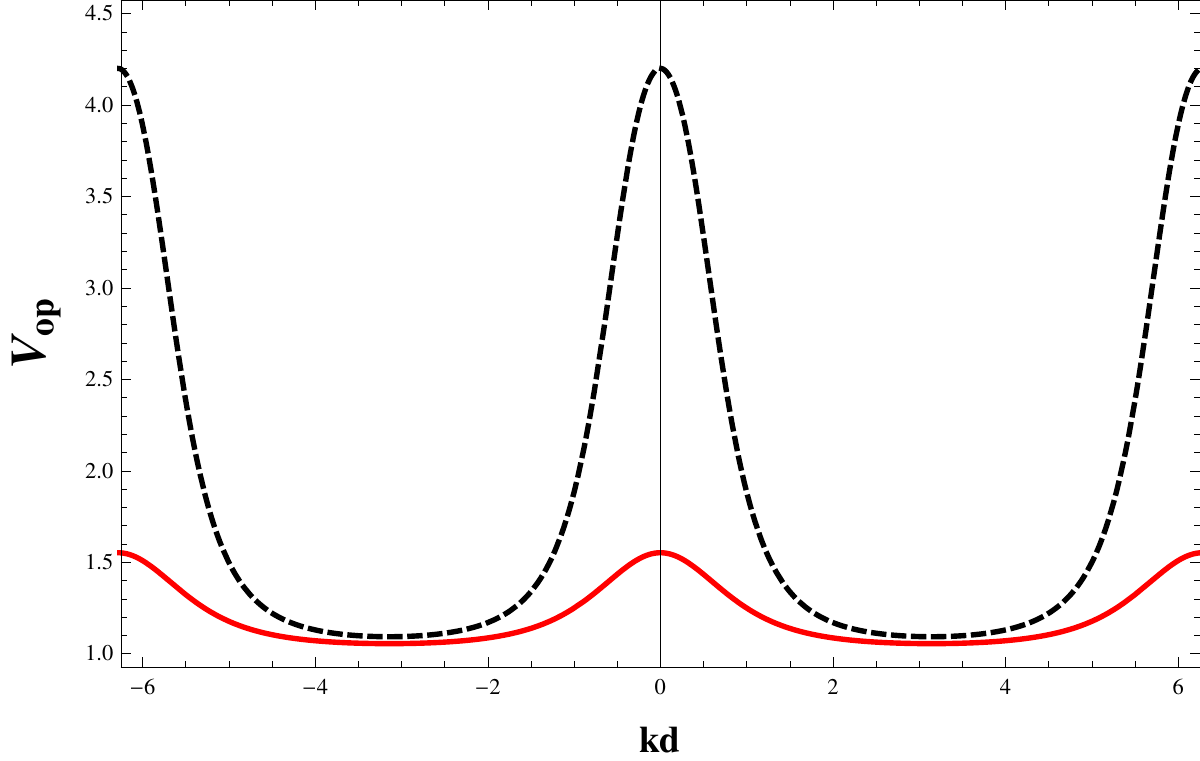}
\caption{The plot of the optical lattice potential as a function of $kd$ for $\frac{\Delta_C}{\kappa}$= 5.5 (bold, red) and 3.5 (dashed, black).}
\centering
\end{figure}

\subsubsection{Bloch energy and effective mass} 

Assuming mean- field solutions and substituting the expression for $\phi_j$ in the Heisenberg equation for the bosonic field operator ($b_j$), the chemical potential of the system can be expressed as: -

\begin{equation}
\mu= U n_0- 2J_{eff}(k) cos(kd),
\end{equation}

where $J_{eff}(k) \equiv -E- \hbar U_0 J \Big( 1+ \frac{\eta^2}{\kappa^2+ (\Delta_c- 2J n_0 IU_0 cos(kd))^2} \Big)$, which must be positive since $E$ is determined by the kinetic energy term which is negative and is larger than the second term. The expression brings out the fact that as the pump amplitude increases, the tunnelling between neighbouring wells decreses since the height of the barrier increases. Tunnelling is minimized when $\Delta_c= 2Jn_0 I U_0 cos(kd)$. The energy per particle is defined as: -

\begin{equation}
\epsilon(k)= \frac{1}{n_0} \int \mu_0 dn_0.
\end{equation}

This yields

\begin{equation}
\epsilon(k)= \frac{Un_0 I}{2}+ 2 (E+ \hbar U_0 J)cos(kd)- \frac{\hbar\eta^2}{\kappa n_0^2} tan^{-1} \Big[ \frac{\Delta_c- 2U_0 n_0 I J cos(kd)}{\kappa}\Big],
\end{equation}

The third term in the above expression is the influence of the cavity photons on the Bloch energy due to the external pump and the second term reveals the tight binding expression for the energy of the Bloch state for a single particle in an optical lattice. An important role is played by the pump in manipulating the superfluid properties of the system. The energy per particle of stationary Bloch configuration consists of the motion of the whole condensate and carries current, which is constant in time and uniform in space (Bloch bands). The Bloch spectrum is found to be supressed at greater value of the optical lattice depth per photon. A detailed study of the Bloch energy can be found in \citep{35}.\\

	The expression for Bloch energy in the absence of pump reduces to the usual expression for BEC in an optical lattice in the absence of the cavity. The fact that the cavity and the pump parameters can be used as a tool to manipulate properties of BEC is an important result of the above analysis. The cavity modifies the expression for Bloch energy as seen in the previous discussion and we now proceed to discuss the effective mass by studying the low- $k$ behaviour of the lowest band $\epsilon(k)$. The expression for effective mass of the atoms in the optical lattice takes the form: -

\begin{equation}
m^{*}= \frac{-\hbar^2}{2d^2 (E+ \hbar U_0 J)} \Bigg[   \frac{n_0 \kappa^2 \Big[ 1+ ( \frac{\Delta_c}{\kappa}-\frac{2n_0 IU_0 J}{\kappa})^2\Big]}{n_0 \kappa^2 \Big[ 1+ (\frac{\Delta_c}{\kappa}- \frac{2n_0 I U_0 J}{\kappa})^2 \Big]+ \frac{\hbar U_0 J \eta^2}{(E+ \hbar U_0J)}}\Bigg].
\end{equation}

We reproduce here the plot of the effective mass as a function of the pump- cavity detuning $\frac{\Delta_c}{\kappa}$ for different values of pump amplitude (fig.7, left panel). Due to the increase in the effective optical potential with the pump, the tunneling decreases which is accompanied by an increase in the effective mass. Evidently, for positive detuning ($\omega_p > \omega_c$), there is a sharp increase in the effective mass and it shows a maxima at $\Delta_c= 2 n_0 IU_0 J$. The lattice depth is maximum here and the superfluid fraction is expected to be minimum at this condition. The same result will be verified in the next section where we shall aim to probe superfluidity of trapped cold atoms through transmission spectroscopy. Hence the detuning acts as a tool to alter the superfluid properties and nonlinear excitations such as solitons in optical cavities. The role of such interaction is to modify the effective mass as a result of the broadening of the wavefunction caused by the repulsion, which causes the tunneling.

\begin{figure}[h!]
\includegraphics[width=0.45\textwidth]{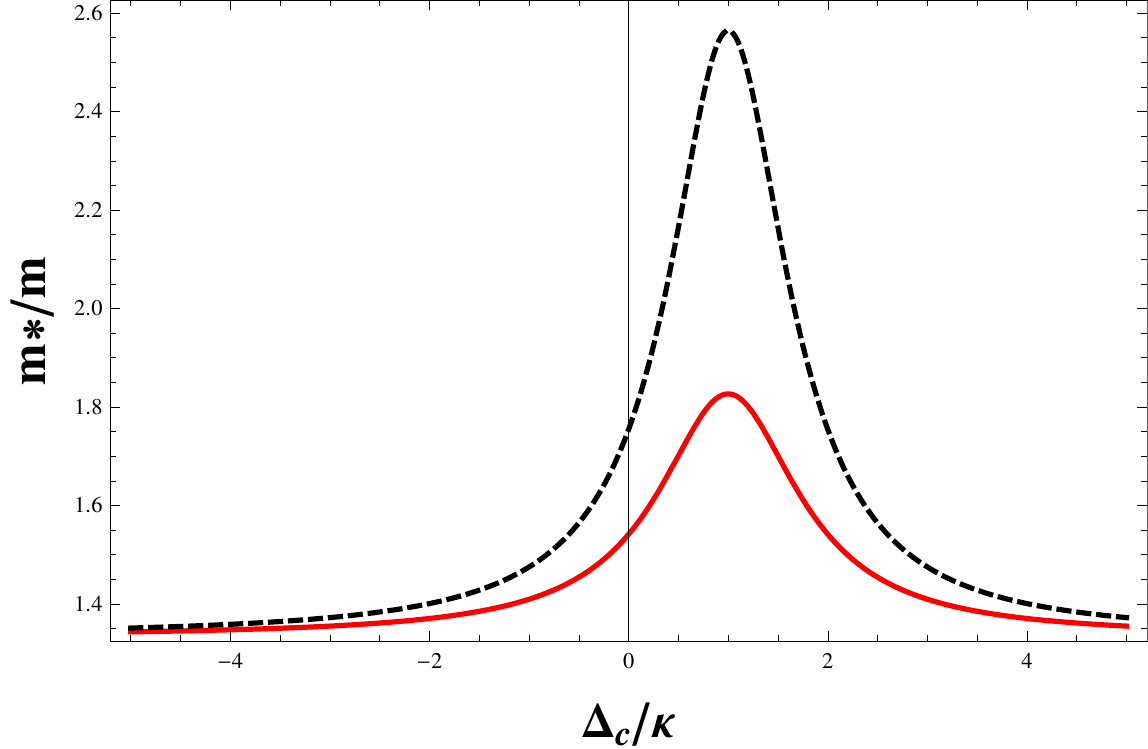}
\includegraphics[width=0.45\textwidth]{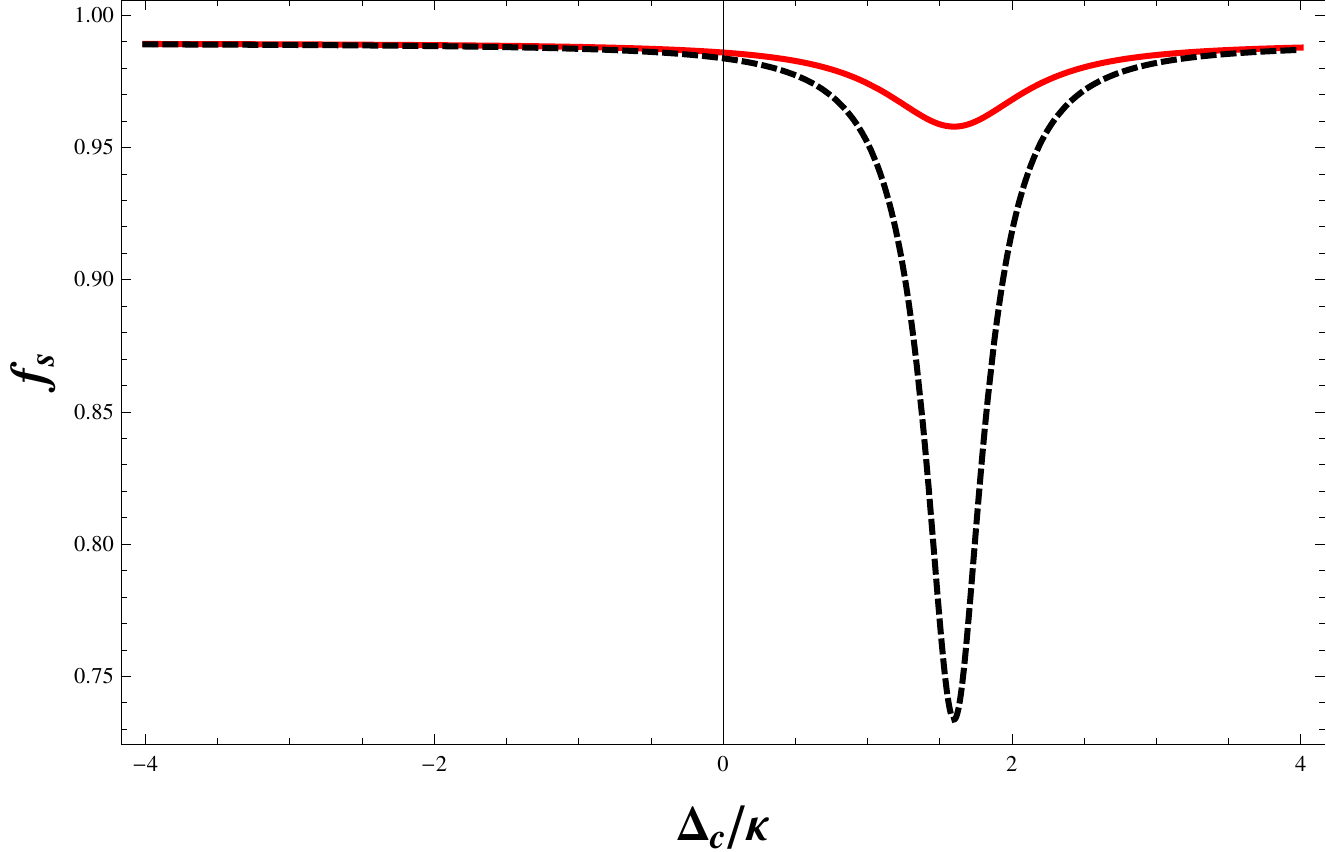}
\caption{Left- The variation of the effective mass as a function of the pump- cavity detuning for $\frac{\eta}{\kappa}$= 1.2 (black, dashed) and 9 (red, bold). The effective mass shows a maxima at $\Delta_c= 2n_0 I U_0 J$.\\
Right- The superfluid fraction as a function of pump- cavity detuning for $\frac{\eta}{\kappa}$= 1.0 (red, bold) and 1.2 (black, dashed).}
\centering
\end{figure}

\subsubsection{Superfluid fraction}

The existence of condensate in the interacting many body system brings out the concept of superfluidity. The one body density matrix requires to have atleast one macroscopic eigenvalue which determines the number of particles in the condensate and the corresponding eigenvector defines the condensate wave function $\phi_0= e^{i \Theta(\vec{r})} \mid \phi_0(\vec{r}) \mid$. A spatially varying condensate phase, $\Theta (\vec{r})$, which is associated with the superfluid velocity field for the condensate is given by: -

\begin{equation}
\vec{v_s}= \frac{\hbar}{m^{*}} \vec{ \bigtriangledown} \Theta (\vec{r}),
\end{equation}

which enables us to derive the expression for the superfluid fraction $f_s$. Let us consider a system with a finite linear dimension $L$ and ground state energy $E_0$, calculated with periodic boundary conditions. Imposing a linear phase variation with a very small twist angle $\theta << \pi$, the energy difference can be attributed to the kinetic energy, $T_s$, of the superflow generated by the phase gradient given as: -

\begin{equation}
E_{\theta}- E_{0}= T_s= \frac{1}{2} m^{*} N f_s \vec{v}_s^2,
\end{equation}

where $m^{*}$ is the effective mass of a single particle in the optical lattice and $N$ is the total number of particles with $mNf_s$  representing the total mass of the superfluid component. Using the expression for superfluid velocity, the equation for superfluid fraction takes the form: -

\begin{equation}
f_s= \frac{2m^{*}}{\hbar^2} \frac{L^2}{N} \frac{E_{\theta}- E_{0}}{\theta^2}= \frac{1}{N} \frac{E_{\theta}- E_{0}}{J^0_{eff} (\Delta\theta)^2}, 
\end{equation}

where $d$ is the distance between the sites and $\Delta\theta$ representing the phase variation over $d$. For the present case, $J^0_{eff} \equiv \hbar^2/ (2m^*d^2)$. We calculate the energy difference through second order perturbation theory assuming that the phase change $\Delta\theta$ is small. the expression for superfluid fraction is given by both the contributions of the first and second order terms of the perturbation expression which under Bogoliubov approximation takes the form: -

\begin{equation}
f_s= \frac{I}{N} \Big[ \phi^2+ \frac{1}{I} \sum\limits_{q} \mid v^q\mid^2 cos (qd) \Big],
\end{equation}

where $I$ represents the number of sites in the lattice and the summation runs over all quasi- momenta $q= \frac{2\pi}{Id} j$ with $j= 1, 2, ....(I-1)$. The normalization condition in the limit of zero lattice spacing is given by :-

\begin{equation}
I\phi^2+ \sum\limits_{q} \mid v^q \mid^2= N.
\end{equation}

These expressions give a complete insight of the superfluid fraction as a function of various cavity parameters. Right panel of fig. 7 shows the plot of superfluid fraction as a function of pump- cavity detuning. The plot of effective mass (fig. 7, left) showed maxima at $\Delta_c= 2n_0 IU_0J$, which is accompanied by a decrease in the superfluid fraction as shown in the right panel of fig. 7. These results are also consistent with earlier works of Christopher Maschler $\textit{et al.}$ \citep{34}, where the fluctuations in atom number was found to be enhanced (increase in superfluid fraction) as $\Delta_c$ was increased (decrease in lattice depth). Any change in atom- number fluctuations will bring out a change in photon- number fluctuations, which will be reflected in transmission spectra. We shall verify the same results in the next part where transmission spectroscopy will be employed for probing into the superfluid properties of trapped ultracold ensembles.

\subsection{Probing superfluidity of peridically trapped ultracold atoms by transmission spectroscopy}

For analyzing the system through transmission spectroscopy, we consider the same model as in previous cases ($\zeta$= 0) with minor modifications. The cavity field  is coupled to the external fields incident from the two side mirrors which are now partially transparent with  $\kappa_1$ and $\kappa_2$ denoting their corresponding loss coefficient. The partially transparent mirrors allow the input and output modes of the external and internal field to be analyzed through spectroscopy. The Hamiltonian of the system remains the same, and the quantum- Langevin equation for the single mode cavity takes the form \citep{36}: -

\begin{equation}
\dot{a}= -iU_0\Big[ J_0\sum\limits_{j} b_j^{\dagger}b_j+ J\sum\limits_{j} \Big( b_{j+1}^{\dagger} b_{j}+ b_{j+1}b_j^{\dagger} \Big) \Big] a+ \eta+ i\Delta_c a- \frac{\kappa_1}{2} a- \frac{\kappa_2}{2} a+ \sqrt{\kappa_1} a_{in}+ \sqrt{\kappa_2} b_{in},
\end{equation}

where all the symbols have the same meaning as in previous sections. The Heisenberg equation of motion for the bosonic field $b_j$ remains the same. $a_{in}$ and $b_{in}$ represents the internal and external input fields incident from the two side mirrors. We shall concentrate on the bad cavity limit where both $\kappa_1$ and $\kappa_2$ are the fastest timescales \textit{i.e.} the cavity decay rates are much larger than the oscillation frequency of the bound atoms in the optical lattice of the cavity. In such a limit, the intracavity field adiabatically follows the condensate wavefunction, and hence we can put $\dot{a}$= 0. In the frequency space, using the expression for the bosonic field operator, $b_{j}$ and cavity field operator $a$, we get: -

\begin{equation}
\tilde{a}(\omega)= \frac{\eta+ \sqrt{\kappa_1}\tilde{a}_{in}(\omega)+ \sqrt{\kappa_2}\tilde{b}_{in} (\omega)}{\frac{\kappa_1}{2}+ {\kappa_2}{2}- i(\Delta_c+ \omega- 2JNU_0 cos(kd))}.
\end{equation}

where $N$= $\sum\limits_{j} n_0$ is the total number of atoms and $U_0$ represents the backaction of the atoms on the field. The optical lattice potential in this case is related to $\tilde{a}^{\dagger}(\omega)\tilde{a}(\omega)$ and an inspection of the above expression reveals that at $\omega+ \Delta_c- 2JNU_0 cos(kd)$= 0, the lattice height has the maximum value. A greater lattice height implies the tendency of the condensate to be more localized within the wells, and hence a loss of superfluidity of the atoms as discussed in previous section. Using the boundary conditions at each mirror, the relation between the input and output modes can be found as: -

\begin{equation}
\tilde{a}_{out} (\omega)+ \tilde{a}_{in} (\omega)= \sqrt{\kappa_1} \tilde{a} (\omega), 
\end{equation}

\begin{equation}
\tilde{b}_{out} (\omega)+ \tilde{b}_{in} (\omega)= \sqrt{\kappa_2} \tilde{a} (\omega).
\end{equation}

We find,

\begin{equation}
\tilde{a}_{out} (\omega)= \frac{\eta \sqrt{\kappa_1}+ \Big( \frac{\kappa_1}{2}- \frac{\kappa_2}{2}+ i\Delta^{'} \Big)\tilde{a}_{in} (\omega)+ \sqrt{\kappa_1 \kappa_2} \tilde{b}_{in} (\omega)}{\frac{\kappa_1}{2}+ \frac{\kappa_2}{2}- i\Delta^{'}},
\end{equation}

where $\Delta^{'}$= $\Delta_c+ \omega- 2JNU_0 cos(kd)$. For identical mirrors, $\kappa_1$= $\kappa_2$= $\kappa$ and near resonance $\Delta^{'} \approx$ 0. The resonance point is one where the superfluid fraction is minimum. The above equation gets modified as: -

\begin{equation}
\tilde{a}_{out} (\omega)= \frac{\sqrt{\kappa}\eta+ \kappa \tilde{b}_{in}(\omega)}{(\kappa- i\Delta^{'})}.
\end{equation}

This shows that the cavity now behaves like a shifted through- pass Lorentzian filter. The input field will be completely reflected if $\Delta^{'} >> \kappa$ $\textit{i.e.}$ the atoms are in deep superfluid regime, $\tilde{a}_{out} (\omega) \approx -\tilde{a}_{in} (\omega)$. Clearly the BEC and the parameter $\Delta_c= 2JNU_0cos(kd)$ controls the superfluid fraction, and by monitoring the reflected and transmitted field, one can easily estimate the superfluid fraction. Thus the cavity will act either as a through- pass Lorentzian filter when the superfluid fraction is minimum, or completely reflect the input light when the superfluid fraction is maximum. Photon loss through the mirror can be minimized by using high- Q cavities, thereby ensuring that the light remains quantum mechanical for the duration of the experiment.

\subsection{Dynamics of small fluctuations}

The previous sections have well described the interesting scenario of the light- atom coupling through optical cavities with ultracold ensembles. In this section, we shall consider the fluctuations of the atom field and the condensate and study the dynamics that leads to the splitting of the normal mode into two modes (NMS). The optical force changes the frequency and damping constant of the Bogoliubov modes (collective density excitations) of the BEC and the normal mode splits due to the mixing of the fluctuations of the cavity and condensate, which vanishes for small value of the two- body interaction. The density excitations of the condensate can be used to squeeze the output quantum fluctuations of the light beam. We consider the same system described in previous sections with $V_{cl}$ as the classical potential and the Heisenberg- Langevin equations for the bosonic field operator ($b_j$) and internal cavity mode operator ($a$) takes the form \citep{38}: -

\begin{equation}
\dot{b_j}= -i \Big(U_0 a^{\dagger} a+ \frac{V_{cl}}{\hbar} \Big) \Big ( J_0 b_j+ J(b_{j+1}+ b_{j-1})\Big)- \frac{iE}{\hbar} (b_{j+1}+ b_{j-1})- \frac{iU}{\hbar}b_j^{\dagger}b_j b_j- \frac{iE}{\hbar}b_j- \frac{\Gamma_b}{2}b_j + \sqrt{\Gamma_b}\xi_b (t)
\end{equation}

and

\begin{equation}
\dot{a}= -iU_0\Big(J_0 \sum\limits_{j} b_j^{\dagger} b_j+ J\sum\limits_{j}(B_{j+1}^{\dagger}b_j+ b_{j+1}b_j^{\dagger})\Big) a+ \eta+ i\Big(\Delta_c - \frac{\kappa}{2} \Big)a+ \sqrt{\kappa}\xi_p (t),
\end{equation}

where $\kappa$ and $\Gamma_b$ characterizes the dissipation of the cavity field and Bogoliubov excitations of the BEC respectively. $\xi_p$ and $\xi_b$ are the noise operators of the input field for photon and boson. We linearize the above equations by adding fluctuations and transforming to the following quadratures:- $X_p= (a+ a^{\dagger})$, $P_p= i(a^{\dagger}- a)$, $X_b= (b+ b^{\dagger})$ and $P_b= i(b^{\dagger}- b)$, while neglecting the higher order terms. The displacement spectra of the condensate in the Fourier space which is defined as: -

\begin{equation}
S_x (\omega)= \frac{1}{4\pi} \int d\omega' e^{-i(\omega+ \omega')t} <X_b(\omega) X_b(\omega')+ X_b(\omega') X_b (\omega)>,
\end{equation}

takes the form: -

\begin{equation}
S_x(\omega)= \frac{\beta_1^2}{\mid d(\omega) \mid^2} \Big[4\Gamma_b n_b+ \frac{8 g_c^2 \kappa (\Delta_d^2+ \omega^2+ \kappa^2/4)}{(\Delta_d^2- \omega^2+ \kappa^2/4)+ \omega^2\kappa^2} \Big],
\end{equation}

where

\begin{equation}
\mid d(\omega) \mid ^2= (\Omega_{eff}^2- \omega^2)^2+ \omega^2 \Gamma_{eff}^2,
\end{equation}

and the effective Bogoliubov mechanical frequency ($\Omega_{eff}$) and the effective Bogoliubov mechanical damping ($\Gamma_{eff}$) are given as: -

\begin{equation}
\Omega_{eff}^2= \beta_1 \beta_2+  \frac{4 \Delta_d g_c^2\beta_1 (\Delta_d^2- \omega^2+ \kappa^2/4)}{(\Delta_d^2- \omega^2+ \kappa^2/4)^2+ \omega^2\kappa^2}
\end{equation}

and

\begin{equation}
\Gamma_{eff}= \Gamma_b- \frac{4\Delta_d g_c^2 \beta_1 \kappa}{(\Delta_d^2- \omega^2+ \kappa^2/4)^2+ \omega^2\kappa^2}.
\end{equation}

Here, $\beta_1$= $\nu+ U_{eff}$ and $\beta_2= \nu+ 3U_{eff}$ and other expressions can be referred from Appendix B. The spectrum equation derived above is characterized by a mechanical susceptibility of the condensate, $\chi (\omega)= 1/d(\omega)$ that is driven by thermal noise and other quantum fluctuations of the radiation pressure. The expressions for $\Omega_{eff}$ shows the effect of the radiation pressure on the Bogoliubov excitations which is equivalent to the 'optical spring effect' in cavity optomechanical systems. The plot of the normalized effective Bogoliubov mechanical frequency (fig. 8, left) of the BEC versus normalized frequency shows that the former increases as the strength of the interaction with the cavity field increases. A higher two body interaction makes the condensate more robust and the Bogoliubov frequency of the condensate doesn't deviate from $\omega_m( =\sqrt{\beta_1 \beta_2})$. The plot for mechanical damping (fig. 8, right) versus normalized frequency reveals that a high atom loss is induced because of stronger coupling with the cavity photons. Such loss of atoms was also reported by Murch K W $\textit{et al.}$ \citep{37} and was found to be getting enhanced near resonance. The radiation pressure of the light couples the condensate to the optical mode, which behaves as an additional reservoir for the oscillator and thereby induces higher damping of the Bogoliubov mode for higher interaction cases. Hence the effective temperature of the Bogoliubov mode will be intermediate between the initial thermal reservoir temperature and optical reservoir. Therefore, one can easily approach the mechanical ground state of the condensate when the atom photon coupling is higher than the cavity damping rate. This explains the mechanical cooling of the Bogoliubov mode for strong radiation pressure coupling.

\begin{figure}[h!]
\includegraphics[width=0.45\textwidth]{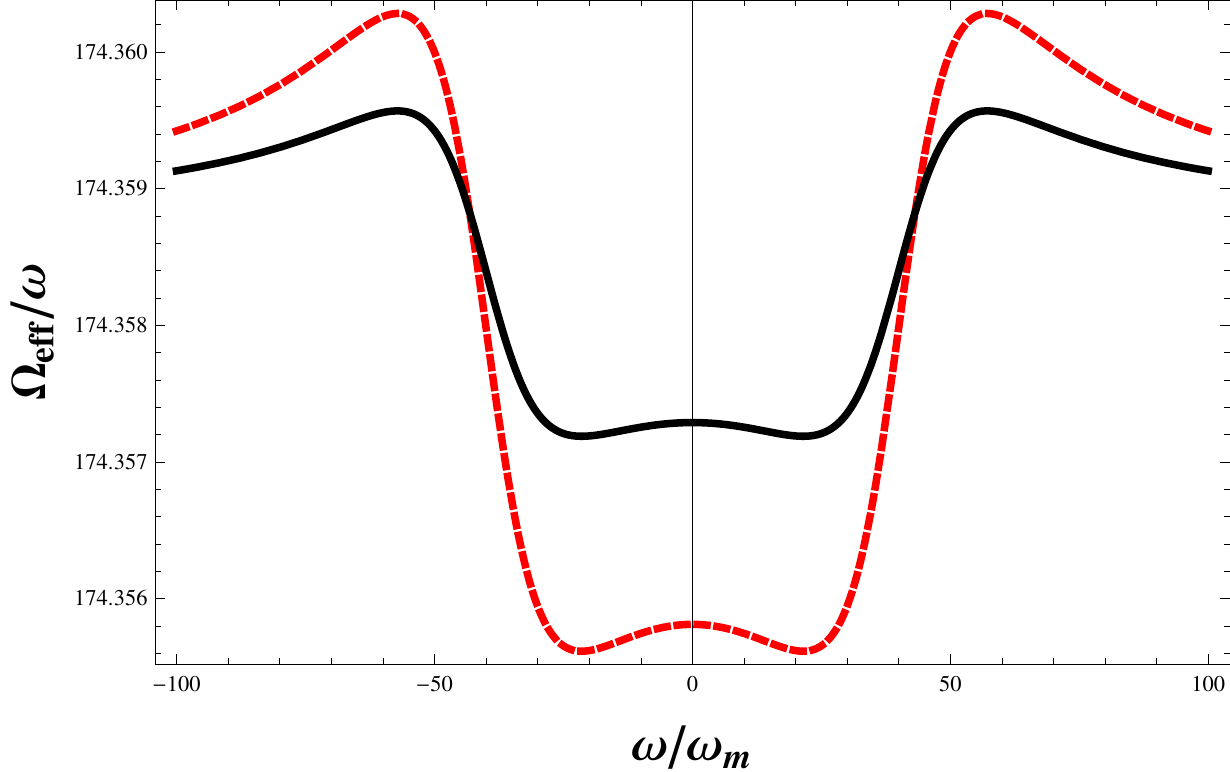}
\includegraphics[width=0.45\textwidth]{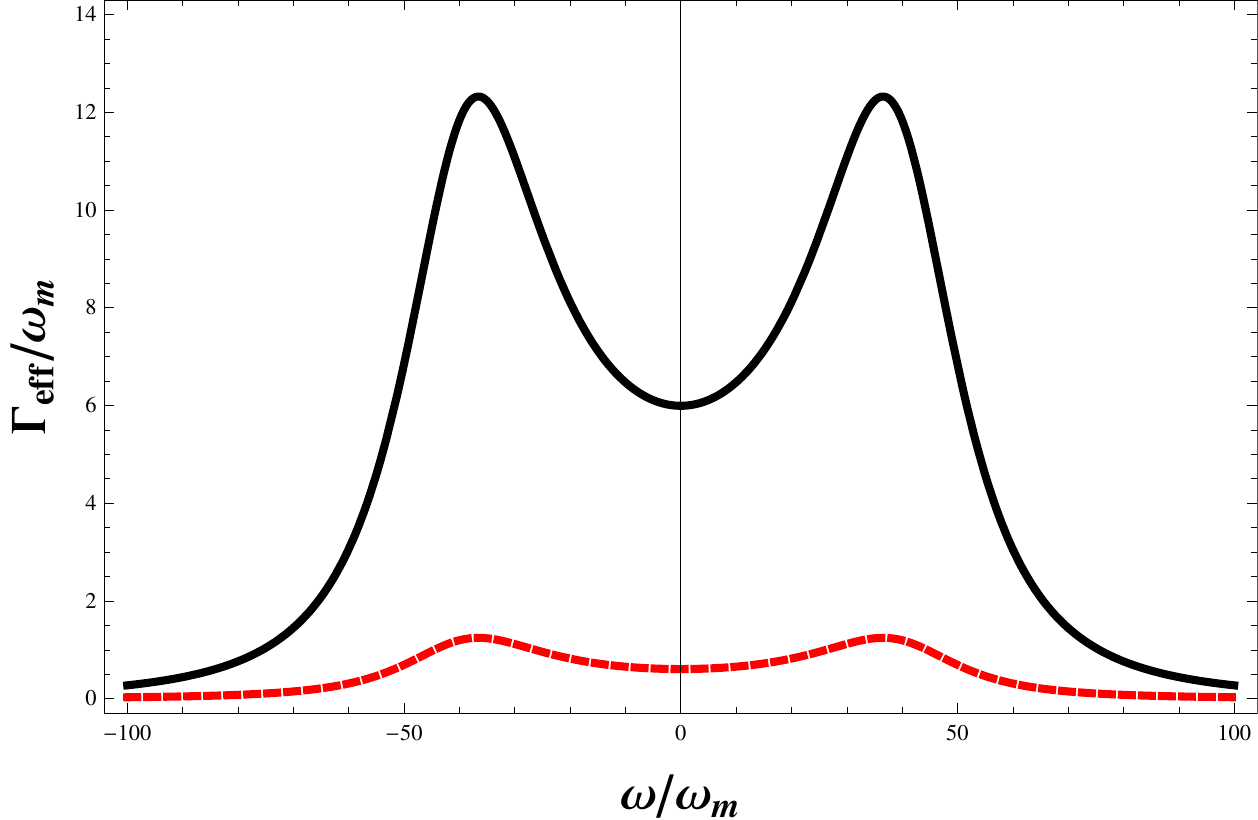}
\caption{Left- Plot of normalized effective Bogoliubov mechanical frequency versus normalized frequency. Parameters chosen were $\Gamma_b$= 0.025$\omega_m$, $\kappa$= 32.5$\omega_m$, $\Delta_d$= -40$\omega_m$, $U_{eff}$= 100$\omega_m$, $\nu= \omega_m$ for two values of atom- photon interaction parameter, $g_c$= 2.5$\omega_m$ (black, bold) and 3.5$\omega_m$ (red, dashed). \\
Right- Plot of normalized effective Bogoliubov mechanical damping as a function of normalized frequency. Parameters chosen were $\Gamma_b$= 0.025$\omega_m$, $\kappa$= 32.5$\omega_m$, $\Delta_d$= -40$\omega_m$, $U_{eff}$= 100$\omega_m$, $\nu= \omega_m$ for two values of atom- photon interaction parameter, $g_c$=10$\omega_m$ (black, bold) and 5$\omega_m$ (red, dashed).}
\centering
\end{figure}

The plot of the displacement spectrum (fig. 9) of the BEC within the cavity versus normalized frequency and normalized effective detuning shows the splitting of normal mode into two modes when the interactions are large, which ofcourse vanishes for weak interaction. This normal mode splitting is associated with the mixing of the fluctuations of the cavity field and condensate around their steady state and mean field respectively. The frequency of the Bogoliubov mode in low momentum limit is approximated as $\sqrt{U_{eff}}$, and hence vanishes for weak interaction. Such normal mode splitting has been observed in many experiments for large number of atoms coupled to the cavity field, however, the energy exchange between the two modes should take in a time scale faster than the decoherence of each mode.

\begin{figure}[h!]
\includegraphics[width=0.90\textwidth]{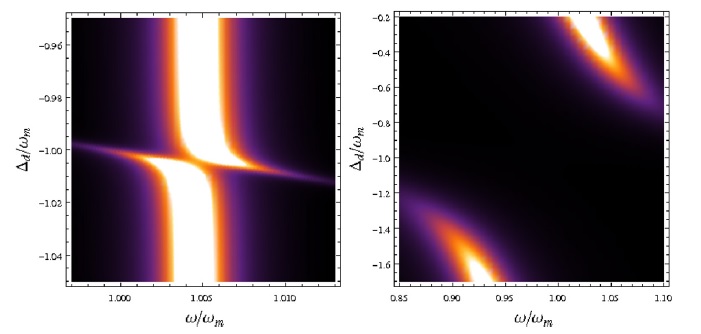}
\caption{The plot of the displacement spectrum of the BEC versus normalized frequency and normalized effective detuning for (a) U$_{eff}$= 150 $\times$ 10$^7$ Hz and (b)  U$_{eff}$= 150 $\times$ 10$^5$ Hz. Other parameters chosen were $\nu$= 4 $\times$ 10$^4$ Hz, $\Gamma_b$= 735 Hz, $n_b$= 10$^4$ Hz and g$_c, \kappa$= 7.35 $\times 10^6$ Hz. The normal mode splitting vanishes (right) as the effective interaction decreases. Reprinted with permission from Aranya B Bhattacherjee, J. Phys. B: At. Mol. Opt. Phys. 43 (2010) 205301 \citep{38}.} 
\centering
\end{figure}

\subsection{Experimental advancements in optical cavities}
\subsubsection{Bistability measurement}
Successful experiments has been realized both in microwave and optical domain, with the cavity field coupled to neutral atoms, Rydberg atoms and also artificial atoms. In this section we reproduce the experimental results and procedures demostrated by S.Ritter $\textit{et al.}$ \citep{39}. In their setup, a $^{87}$Rb BEC was coupled to the optical cavity with approximately 10$^{5}$ atoms trapped within a crossed beam dipole trap created by two far detuned laser beams operating perpendicular to the cavity axis. With trapping frequencies of 2$\pi \times$ (220, 48, 202) Hz for three directions, the atoms were prepared in the sub- level $|F, m_F>$= $|1, -1>$ of the 5S$_{1/2}$ ground state manifold, where m$_F$ is the magnetic quantum number and F the total angular momentum. The pump laser was blue detuned from the atomic transition frequency $\omega_a$ and the pump photons couple only to the $\sigma^+$ transitions due to the weak magnetic field along the cavity axis. The equations of motion for the system, with $\psi$ denoting the condensate wave function takes the form: -

\begin{equation} 
\iota\hbar\dot{\psi} (x,t)= \Big (\frac{-\hbar^2}{2m} \frac{\partial^2}{\partial x^2}+ \mid\alpha(t)\mid^2\hbar U_0 cos^2(kx)+ V_{ext}(x)+ g_{1D}\mid\psi\mid^2 \Big) \psi (x,t)
\end{equation}

\begin{equation} 
\iota \dot{\alpha}(x,t)= -(\Delta_c- U_0 N \beta+ \iota\kappa)\alpha(t)+ \iota\eta.
\end{equation}

where $\beta$= $<\psi|cos^2(kx)|\psi>$ and V$_{ext}$ denotes the external trapping potential, N representing the total number of atoms  and the atom- atom interaction strength denoted by g$_{1D}$. The steady state solution of the above equation with $\Delta_c= \omega_p- \omega_c$, denoting the cavity pump detuning can be represented as: -

\begin{equation} 
\mid\alpha\mid^2= \frac{\eta^2}{\kappa^2+ (\Delta_c- U_0 N \beta)^2}.
\end{equation}

The system exhibits bistable behaviour for sufficient pump strength, commonly known as the Kerr nonlinearity. The pump laser frequency was scanned slowly accross the resonance and from the measured photon count rate, the mean intracavity photon number is deduced. The  typical resonance curves for different pump strengths can be referred from fig.(3) of \citep{29}. The results showed good agreement between the experimental and theoretical curves with critical photon number n$_{cr}$= 0.21. However, as the pump strength is increased, the system deviates more and more from the steady state curves. Such dynamics of deviation for larger pump strengths goes beyond the physics of pure Kerr medium and is governed by the inertia of the refractive index medium.

\subsubsection{Energy spectrum of the system}

An important characteristic of the BEC- cavity coupled system is the energy spectrum which was first performed by Esslinger and his group in 2007. His system consisted of 2.2$\times 10^5$ $^{87}$Rb atoms trapped with a laser power of frequency 2$\pi \times $(290, 43, 277) Hz for three components. A transmission spectroscopy was performed to find the eigenenergies of the BEC coupled system, with a weak, linearly polarized probe laser of frequency $\omega_p$. The transmitted cavity light was monitored as a function of its detuning. From individual recordings of the transmitted light, the low excitation spectrum was mapped as a function of $\Delta_c= \omega_c- \omega_a$. Fig.(3) of \citep{25} depicts the energy spectrum of the system which also reveals a second avoided crossing at probe frequencies resonant with the bare atomic transition, located at the cavity detuning where the eigenenergy branch of the system with no atoms would intersect the atomic lines of the transition. The solid lines in the plot represents the results of the theoretical model with electric dipole and rotating wave approximation. Considering additional mode with the same coupling but detuned from the TEM$_{00}$ mode by $\Delta_t$ would shift the resonant frequency, which results in a clearly visible change of the energy spectrum with respect to a system with single cavity mode. The experimental results shows good agreement with the theoretical predictions for N= 154,000 atoms in the $|1, -1>$ state and for 2,700 atoms distributed over the Zeeman sublevels of the $|F= 2>$ state.

\section{Optomechanical Cavities}

Successful fabrication of nanomechanical oscillators with optical cavities has helped us reveal the dynamics of the interaction between the photons and phonons to a large extend. Replacing one of the high finesse mirrors with a cantilever allows the radiation pressure to displace the mirror and produce a phase shift in the reflected light which on interaction with the ultracold ensembles between the cavity produce interesting physics like back action cooling, normal mode splitting and many more. However the tricky part that connects the macroscopic and microscopic regime is the cooling of the nanomechanical mirror to its ground state, thereby allowing the mirror to generate quantized phonons rather than oscillations. The ability of radiation pressure to cool the mirror to ground state has been a subject of investigation from decades in the context of interferometers by Braginsky and his group. The study revealed that the retarding nature of the force provided either damping or anti damping of the mechanical motion. Later, starting in 1990, many other aspects of optomechanical systems has been observed and theorized including squeezing of light, quantum non demolition detection of light intensity, effects of Kerr medium and many more. Most recent advancements in this field, including the success of LIGO and VIRGO projects have added motivation to the rapidly growing interest in cavity optomechanics. Light sensitive optical detection of small forces, selective entanglement \citep{40, 41}, manipulation of mechanical motion through light, non classical states of light are among the recent works in the field. In this section we shall initially highlight the notion of ground state cooling of mirrors and then shed light on the recent theoretical and experimental developments in cavity optomechanics and discuss the implications of such in future experiments.\\

	We consider the same model described in previous sections with a minor modification of replacing one of the fixed mirror with a movable one. The two level cigar shaped BEC between the cavity strongly interacts with the one dimensional quantized cavity mode of frequency $\omega_c$ and an input laser of frequency $\omega_p$ drives the system through one of the mirror. The movable mirror moves freely with a mechanical frequency $\omega_m$ due to the radiation pressure. The schematic representation of the system has been shown below (fig.10).

\begin{figure}[h!]
\includegraphics[width=0.75\textwidth]{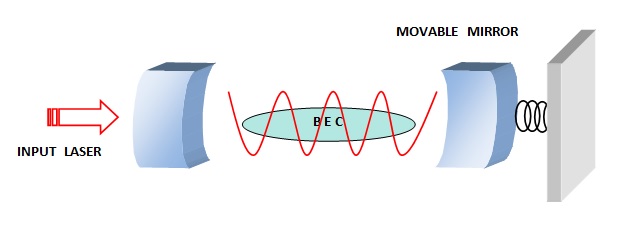}
\caption{The schematic representation of the optomechanical system with BEC trapped between the high finesse mirrors.}
\centering
\end{figure}

The effective Hamiltonian of such system, neglecting the tunnelling of the atoms into neighbouring wells can be written as \citep{42} :- 

\begin{eqnarray}
H&=&= K_0\sum\limits_{j} a_j^{\dagger}a_j- \hbar\Delta_c b^{\dagger}b +\hbar\omega_mc^{\dagger}c+ P_0(\hbar v_0b^{\dagger}b+ V_{cl}) \times \sum\limits_{j} a_j^{\dagger} a_j- \hbar\epsilon \omega_m b^{\dagger} b (c+ c^{\dagger}) \nonumber\\
&-& i\hbar\eta (b- b^{\dagger})+ \frac{v}{2}\sum\limits_{j} a_j^{\dagger}a_j^{\dagger}a_j a_j,
\end{eqnarray}

where $K_0= \int d\vec{r}$ $ w(\vec{r}- \vec{r}_j)${$(-\frac{\hbar^2\nabla^2}{2m})$}$ w(\vec{r}- \vec{r}_j)$ is the onsite kinetic energy of the atoms and the onsite potential energy of the atoms has been denoted by $P_0= \int d\vec{r}$ $ w(\vec{r}- \vec{r}_j) cos ^2 (k'x) w(\vec{r}- \vec{r}_j)$. The last term in the above Hamiltonian defines the two body atom- atom coupling where $v$ represents the effective on- site atom- atom interaction energy and is given by $\frac{4\pi a_s \hbar^2}{m} \int d\vec{r}$ $ \mid w(\vec{r}) \mid ^4$. The lowering (raising) operator of light mode, vibrational mode and bosonic field has been represented by $ b (b^{\dagger}), c (c^{\dagger})$ and $a_j (a_j^{\dagger})$ respectively. The mirror- photon coupling has been represented by $\epsilon$.

\subsection{Achieving the quantum ground state of the mechanical mirror}

The radiation pressure exerted by the cavity light on the mechanical mirror forms a system which acts as another reservoir connected to the mechanical oscillator when the cavity is properly detuned. As a result, the effective temperature of the vibrational mode is the temperature between the initial bath and the effective optical reservoir temperature. This effective temperature is extremely small in practice and hence the quantum ground state is achieved when the coupling to the initial reservoir is much smaller than the coupling to the effective optical reservoir. This explains the fact that the radiation pressure coupling should be strong for significant cooling of the mechanical oscillator. In this section, we shall discuss two main techniques for achieving ground state cooling, namely the back- action cooling and cold damping feedback scheme. Back action cooling dynamics has been realized in many experimental techniques which utilizes the randomness present in the unavoidable stochastic back action forces in the cavity due to photon shot noise to generate the radiation pressure, which in turn cools the mechanical mirror. The cold damping feedback scheme is an advanced and improved technique of cooling the mirror by overdamping it without increasing the thermal noise of the system. It involves a negative derivative feedback technique and the displacement of the cantilever is measured through homodyne detection of the cavity output which is fed back to the oscillator with a force proportional to the oscillator velocity.\\

	We start with the quantum Langevin equations (QLEs) of the system, which for the boson field operator $a_j$, cavity photon operator $b$ and movable mirror mode operator $c$ takes the form: -

\begin{equation}
\dot{a}_ j(t)= -i \frac{K_0}{\hbar} a_j (t)- i\frac{P_0}{\hbar}(\hbar v_0 b^{\dagger} (t) b(t)+ V_{cl})a_j (t)- i\frac{v}{\hbar} a_j^{\dagger} (t) a_j (t) a_j(t), 
\end{equation}

\begin{equation}
\dot{b} (t)= -i P_0 v_0 b(t) \sum\limits_{j} a_j ^{\dagger} (t) a_j (t)+ i \Delta_c b(t)+ i\epsilon\omega_m b(t) \times [c(t)+ c^{\dagger} (t)]+ \eta- \frac{\kappa}{2} b(t)+ \sqrt{\kappa} b_{in} (t), 
\end{equation}

\begin{equation}
\dot{c} (t)= -i \omega_mc(t)+ i\epsilon \omega_m b^{\dagger}(t) b(t)- \Gamma_m c(t)+ \sqrt{\Gamma_m} \xi_{in} (t).
\end{equation}

where $\kappa$ and $\eta$ have the same usual meaning as in previous sections and $\Gamma_m$ denotes the damping rate of the mechanical mode. $b_{in}(t)$ represents the vacuum radiation input noise. Linearizing the QLEs around their steady states and by introducing the amplitude and phase quadratures as $\delta q_a  (t)=  [\delta a (t)+ \delta a^{\dagger} (t)], \delta p_a (t)= i[\delta a^{\dagger} (t)- \delta a (t)], \delta q_b (t)= [\delta b(t)+ \delta b^{\dagger} (t)], \delta p_b (t)= i[\delta b^{\dagger} (t)- \delta b (t)], \delta q (t)= [\delta c (t)+ \delta c ^{\dagger} (t)], \delta p (t)= i[\delta c^{\dagger} (t)- \delta c (t)], q_{in} (t)= [b_{in} (t)+ b_{in}^{\dagger} (t)]$ and $p_{in} (t)= i[b_{in}^{\dagger} (t)- b_{in} (t)]$, the equations of the system takes the form: -

\begin{equation}
\delta \dot{q}_a (t)= \beta_1 \delta p_a (t),
\end{equation}

\begin{equation}
\delta \dot{p}_a (t)= -\beta_2 \delta q_a (t)- 2 g_c \delta q_b (t), 
\end{equation}

\begin{equation}
\delta \dot{q}_b (t)= -\frac{\kappa}{2} \delta q_b (t)+ \sqrt{\kappa} q_{in} (t) - \Delta_d \delta p_b (t),
\end{equation}

\begin{equation}
\delta \dot{p}_b (t)= - \frac{\kappa}{2} \delta p_b (t)- 2 g_c \delta q_a (t)+ 2G\beta\delta q(t)+ \sqrt{\kappa} p_{in} (t)+ \Delta_d \delta q_b (t),
\end{equation}

\begin{equation}
\delta \dot{q} (t)= \omega_m \delta p (t),
\end{equation}

\begin{equation}
\delta \dot{p} (t)= -\omega_m \delta q (t)+ 2G\beta \delta q_b (t)- \Gamma_m \delta p (t) + W (t),
\end{equation}

where the constants and correlation functions can be referred from Appendix C.

\subsubsection{Back action cooling}

As discussed earlier, this technique emphasizes on the stochastic back action force that creates the radiation pressure to cool the mirror to its ground state. In this section, we shall calculate the displacement spectrum and effective damping rate of the mirror for back- action cooling scheme for experimentally feasible parameters and show how ground state of the oscillator is achieved. The displacement spectrum in Fourier space is evaluated from: -

\begin{equation}
S_q (\omega)= \frac{1}{4 \pi} \int d\omega' e^{-i(\omega+ \omega')t } <\delta q(\omega) \delta q(\omega')+ \delta q(\omega') \delta q(\omega) >.
\end{equation}

The correlations used in Fourier space are highlighted in Appendix C. Thus, the displacement spectrum in the Fourier space for the movable mirror is given as: -

\begin{equation}
S_q (\omega)= \mid \chi_{eff} (\omega) \mid ^2 [ S_{th} (\omega)+ S_{rp} (\omega, \Delta_d)],
\end{equation}

where $S_{th}$ and $S_{rp} (\omega, \Delta_d)$ are the thermal noise spectrum and radiation pressure spectrum arising from the Brownian motion of the mirror and quantum fluctuation of the condensate respectively. $\chi_{eff} (\omega)$ is the effective susceptibility of the oscillator. The parameters are defined as: -

\begin{equation}
S_{th}(\omega)= \frac{\Gamma_m}{\omega_m} \omega coth \Big( \frac{\hbar \omega}{2K_B T} \Big),
\end{equation}

\begin{equation}
S_{rp} (\omega, \Delta_d)= \frac{4 G^2\beta^2\kappa (\omega^2- \beta_1 \beta_2 )^2 (\Delta_d^2+ \omega^2+ \frac{\kappa^2}{4})}{X (\omega )},
\end{equation}

\begin{equation}
\chi_{eff} (\omega)=  \frac{\omega_m}{[(\omega_m^2- \omega^2+ i\omega \Gamma_m)+ \chi_1 (\omega)]},
\end{equation}

where $X (\omega)$ and $\chi_1 (\omega)$ are highlighted in Appendix C. $\chi_{eff} (\omega)$ is the effective susceptibility of the oscillator altered by the radiation pressure and condensate fluctuations with

\begin{equation}
\mid \chi_{eff} \mid ^2= \frac{\omega_m^2}{[(\omega_m^{eff} (\omega)^2- \omega^2)^2+ \omega^2\Gamma_m^{eff} (\omega)^2]}.
\end{equation}

The effective mechanical susceptibility of the oscillator gives us the effective resonance frequency and effective damping rate as: -

\begin{equation}
\omega_m^{eff} (\omega)= [\omega_m^2+ \omega_m^{op}]^{1/2},
\end{equation}

where

\begin{equation}
\omega_m^{op}= \frac{4G^2\beta^2\Delta_d\omega_m (\omega^2- \beta_1 \beta_2) \Big[(\omega^2- \beta_1 \beta_2) \Big( \Delta_d^2+ \frac{\kappa^2}{4}-\omega^2 \Big)- 4g_c^2\Delta_d\beta_1 \Big]}{X(\omega)},
\end{equation}

and

\begin{equation}
\Gamma_m^{eff} (\omega)= \Gamma_m- \frac{4G^2\beta^2 \Delta_d \omega_m \kappa (\omega^2- \beta_1 \beta_2)^2}{X(\omega)}.
\end{equation}

Evident from the above three equations that the mechanical frequency of the mirror gets modified by the quantum fluctuations of the mirror due to the condensate fluctuations and radiation pressure. This is so called optical spring effect, which was also discussed in previous section when we were dealing with the dynamics of small fluctuations in optical cavities. The loss of photons through the cavity minimizes the energy of the cavity mode which can be reduced by using high finesse optical cavities. We produce here the plot of normalized effective mechanical frequency and normalized effective damping rate as a function of dimensionless frequency in the presence and absence of BEC.

\begin{figure}[h!]
\includegraphics[width=0.45\textwidth]{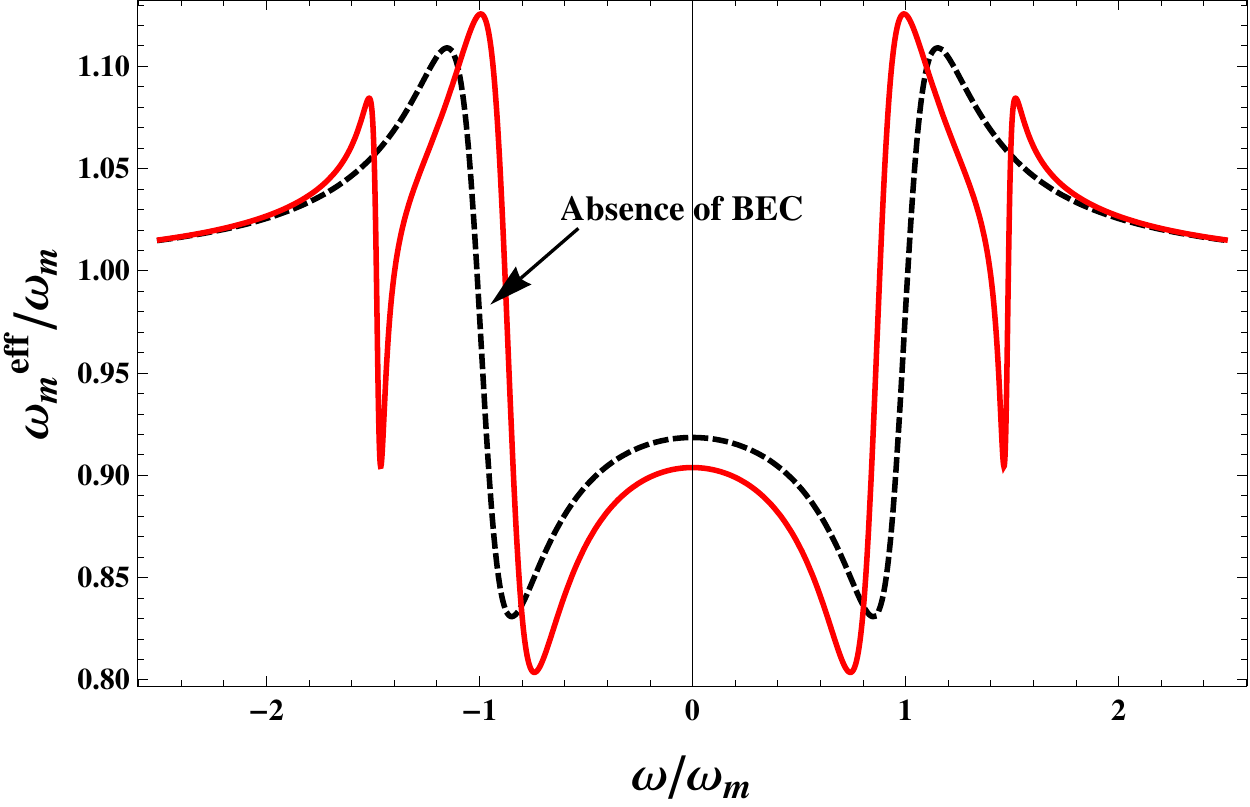}
\includegraphics[width=0.45\textwidth]{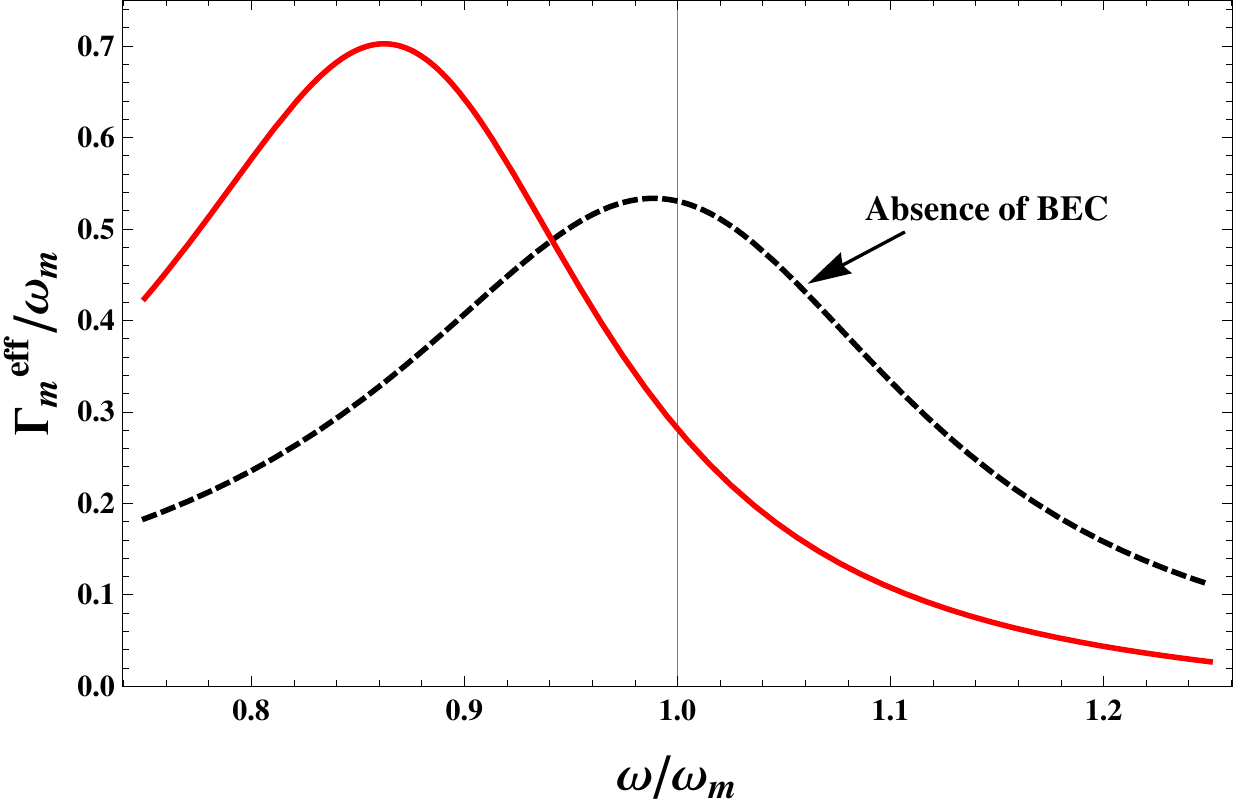}
\caption{Left- Plot of normalized effective mechanical frequency ($\omega_m^{eff}/\omega_m)$ as a function of dimensionless parameter.\\
Right- Plot of normalized effective damping rate ($\Gamma_m^{eff}/\omega_m)$ as a function of dimensionless parameter. Parameters chosen were $\Gamma_m= 10^{-5}\omega_m, \Delta_d= -\omega_m, \kappa= 0.5\omega_m, G= 4.5\omega_m, \beta= 0.06\omega_m, v= 0.02\omega_m, g_c= 0.3\omega_m$.}
\centering
\end{figure}

From the left panel of fig.11, the normalized effective mechanical frequency $(\omega_m^{eff}/\omega_m)$ of the oscillating mirror in the presence of BEC (red, bold) is compared with the case when BEC is not present (black, dashed). This shows an extra resonance dip in the presence of BEC. Significant deviation is observed from $\omega_m$ around $\omega= \pm \omega_m$ which gets enhanced in the presence of BEC. Similarly, from the other plot (fig.11, right), below resonance ($\omega$ $ <$ $ \omega_m$), the effective damping is more in the presence of the BEC and vice versa for cases above resonance ($\omega$ $ >$ $ \omega_m$). Thus it is clear from the plots that the presence of BEC enhances the ground state cooling of the mirror. Two body interaction parameter of the BEC ($U_{eff}$) also plays an important role in cooling the oscillator to the ground state. A detailed analysis reveals that a higher condensate two body interaction enhances the effective damping of the mirror below resonance, and vice versa for cases above resonance. Similarly the deviation of the mechanical frequency of the mirror from its resonance frequency $\omega_m$ decreases with the increase in the two body interaction parameter as the condensate becomes more robust with higher $U_{eff}$ \citep{42}. A detailed study of the back action cooling scheme can be found in Sonam Mahajan $\textit{et al.}$ \citep{42}.

\subsubsection{Cold damping feedback scheme}

An alternative advanced technique to improve the cooling of the mechanical oscillator by overdamping it witout increasing the thermal noise of the system is the cold damping feedback scheme. This technique has been realized experimentally and involves a negative derivative feedback technique. The displacement of the cantilever is homodyne detected which is fed back to the resonator with a force proportional to the oscillator velocity. The schematic representation of the configuration has been shown in fig. 12. 

\begin{figure}[h!]
\includegraphics[width=0.65\textwidth]{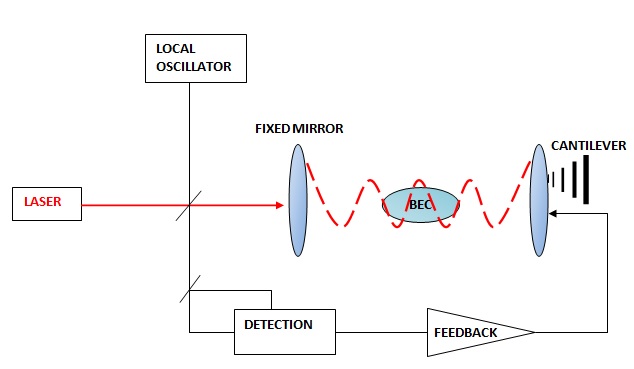}
\caption{The schematic representation of the setup involving an additional feedback loop with a force proportional to the oscillator velocity. The cavity output field is homodyne detected using the beam splitter as shown in the figure.}
\centering
\end{figure}

The QLEs remains same as in previous sub section except for: -

\begin{equation}
\delta \dot{p}= -\omega_m \delta q(t)+ 2G\beta\delta q_b (t)- \Gamma_m\delta p(t)+ W(t)- \int_{-\infty}^{\infty} ds g(t- s) \delta p_{est}(s), 
\end{equation}

where $\beta_1= v+ U_{eff}$ and $\beta_2= v+ 3U_{eff}$ and the filter function $g(t)$ is the causal kernal such that,

\begin{equation}
g(\omega)= \frac{-i\omega g_{cd} \sqrt{\lambda}}{1- i\omega/\omega_{fb}},
\end{equation}

which is the Fourier transform of $g(t)$ with $g_{cd}$ as the positive feedback gain and $\lambda$ quantifies the homodyne detection efficiency of the detector which is less than 1, if additional noise is considered. $\omega_{fb}^{-1}$  denotes the feedback loop delay time. The QLEs in the presence of the feedback term are solved in the frequency domain and a distinct displacement spectrum using the correlations of Appendix C can be written as: -

\begin{equation}
S_q^{cd} (\omega)= \mid \chi_{eff}^{cd} \mid ^2[ S_{th} (\omega)+ S_{rp} (\omega, \Delta_d)+ S_{fb} (\omega)],
\end{equation}

where $S_{th} (\omega)$ and $S_{rp} (\omega, \Delta_d)$ were defined in Eq. (55) and Eq. (56) respectively in the back- action cooling scheme. The position spectrum consists of an additional feedback induced term which arises since the cold damping loop feeds back the measurement noise into the dynamics of the cantilever. Similarly, mechanical susceptibility ($\chi_{eff}^{cd} (\omega)$) also gets modified by the filter function. Appendix D contains all the bulky equations for the effective mechanical susceptibility ($\chi_{eff}^{cd}(\omega)$), resonance frequency ($\omega_m^{eff, cd} (\omega)$) and damping rate ($\Gamma_m^{eff, cd} (\omega)$) and we discuss here only the results of the analysis. \\

	Cold damping feedback scheme uses additional viscous force to over damp the mechanical oscillations through feedback techniques which is possible only when the estimated intra- cavity phase quadrature $\delta p_{est}$ is proportional to the oscillator position $\delta q(t)$. The same is achieved in the bad cavity limit and we limit our discussion only to cases where $\kappa$ $>$ $\omega_m$. Fig. 13 shows the variation of the normalized effective mechanical frequency ($\omega_m^{eff}/ \omega_m)$ as a function of dimensionless frequency ($\omega/\omega_m$).

\begin{figure}[h!]
\includegraphics[width=0.45\textwidth]{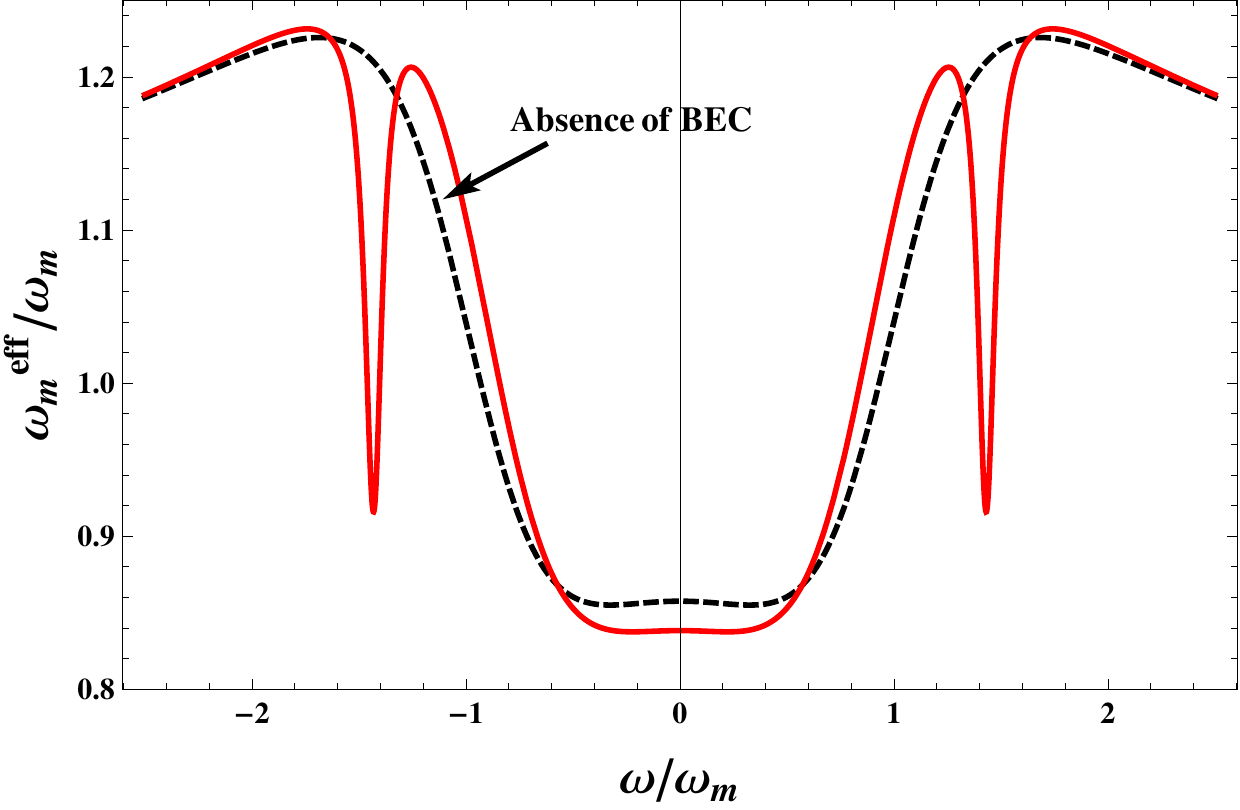}
\caption{The plot of the normalized effective mechanical frequency as a function of dimensionless frequency. General parameters chosen were same as in previous section with $g_{cd}= 0.8, \lambda= 0.8$.}
\centering
\end{figure}

It is evident from fig. 13 that there is no significant shift in the frequency due to the presence of BEC in the chosen parameter regime. This suggests a nominal optical spring effect as $\omega_m^2$ dominates $\omega_m^{op, cd}$ for higher resonance frequency. The black, dashed line shows the variation of normalized effective mechanical frequency in the absence of BEC and the red, bold line corresponds to the same in the presence of the cold atoms.

\begin{figure}[h!]
\includegraphics[width=0.45\textwidth]{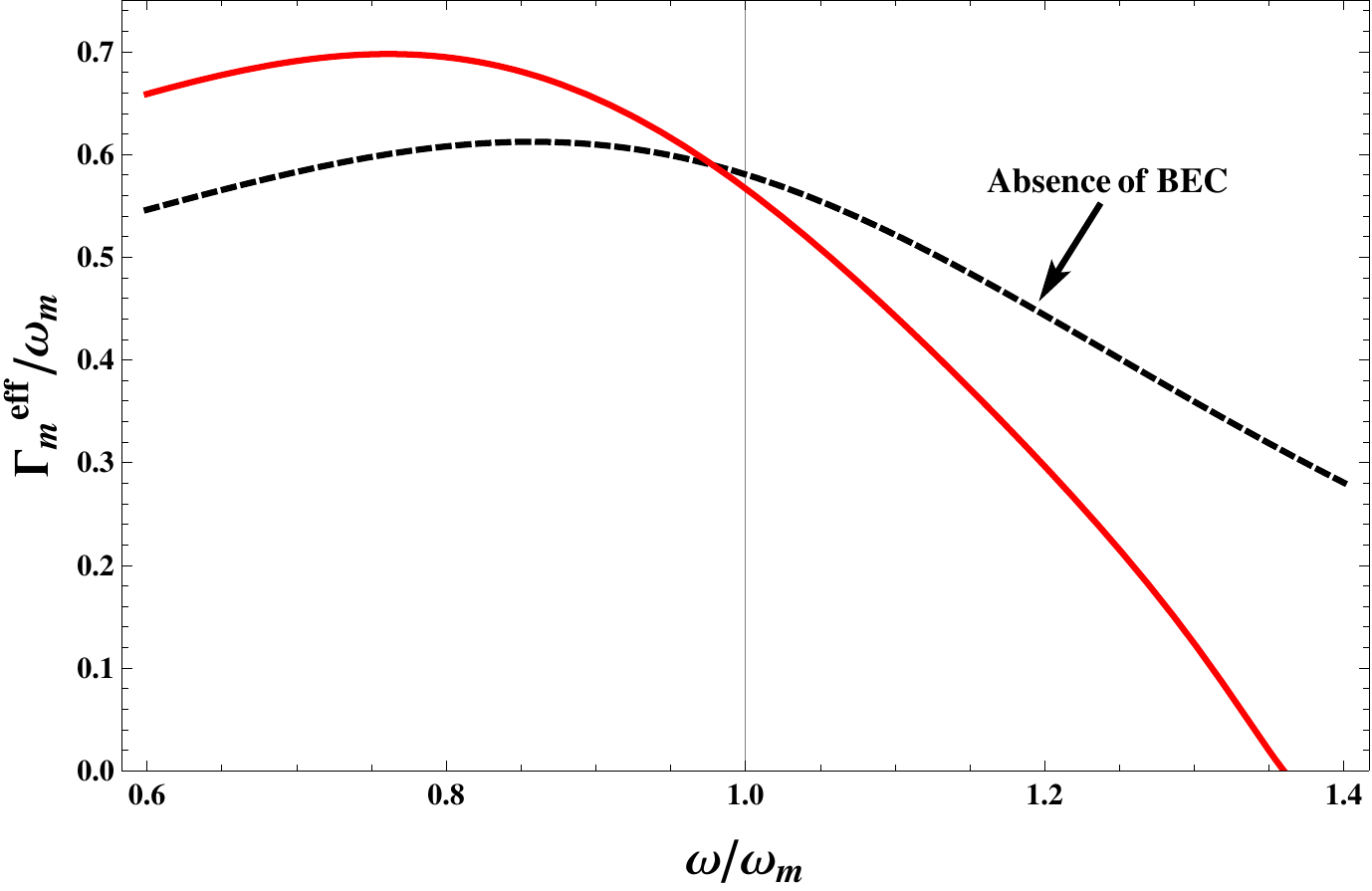}
\includegraphics[width=0.45\textwidth]{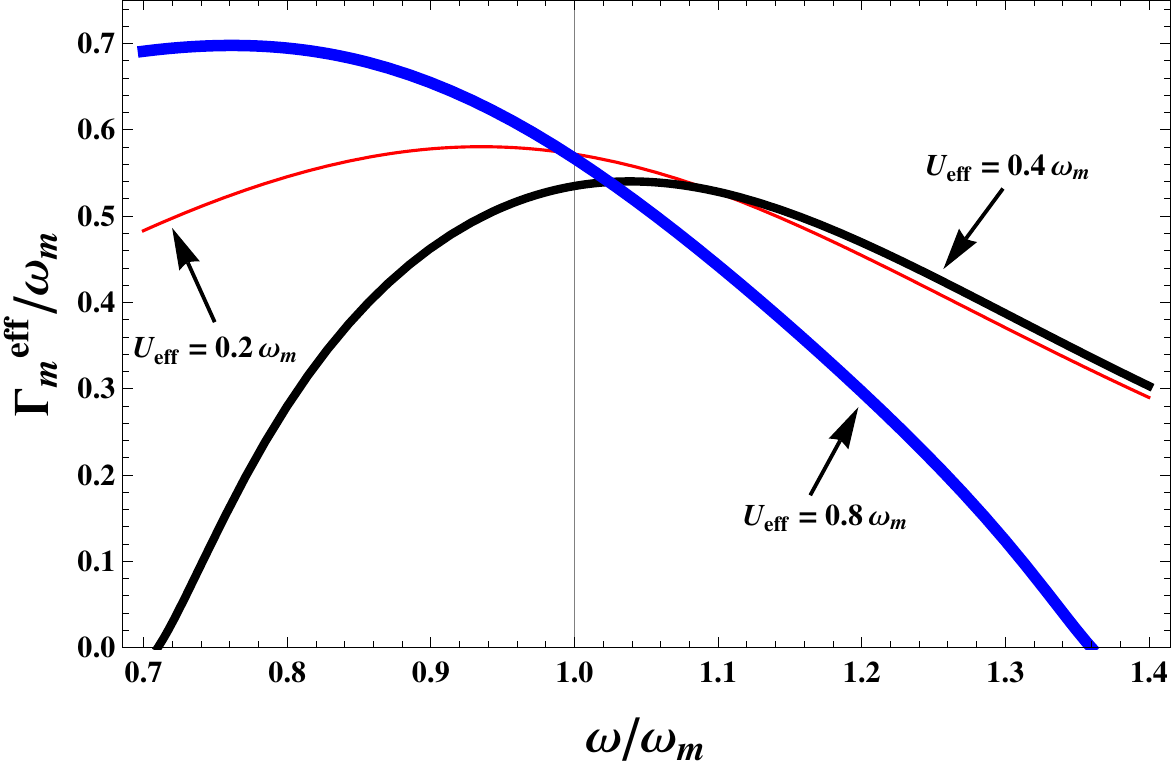}
\caption{Left- The plot of the normalized effective damping rate as a function of dimensionless frequency for cold damping feedback scheme.\\
Right- The effect of atomic two body interaction on the effective damping rate.  General parameters chosen were same as in previous section with $g_{cd}= 0.8, \lambda= 0.8$.}
\centering
\end{figure}

Fig. 14 shows the variation of ($\Gamma_m^{eff}/\omega_m$) as a function of $\omega/\omega_m$. It is clear that the mechanical damping rate shows significant variation with the change in frequency. Below resonance ($\omega<$ $\omega_m$), effective damping increases by adding BEC to the system (fig. 14, left plot) and also by increasing the atom- atom interaction (fig. 14, right plot) with an exception for $U_{eff}= 0.4 \omega_m$. The plot on the right panel shows the variation in effective damping rate for $U_{eff}= 0.8 \omega_m$ (thick line), $ 0.4 \omega_m$ (thin line) and $0.2\omega_m$  (thinner red line). The exception at $U_{eff}= 0.4 \omega_m$ can be explained in terms of effective phonon number ($n_{eff}$) at $\Delta_d= -\omega_m$, which shows a sudden increase in $n_{eff}$ for $0 \le U_{eff} \le 0.57 \omega_m$ \citep{42}. A detailed explanation of the phenomenon can be referrred from Sonam Mahajan $\textit{et al.}$ \citep{42}. Hence by varying the atom- atom interaction, cooling process can be optimized to achieve the quantum ground state of the oscillator.\\

	Thus the quantum mechanical state of the oscillator can be approached by the two techniques described in this section which gets enhanced due to the presence of BEC between the cavities. The cold damping feedback scheme seems more advantageous over the back action cooling scheme in approaching the ground state for a wide range of effective detuning. However, the feedback scheme involves an additional viscous force which helps in further cooling of the oscillator. Presence of BEC enhances the cooling and the atom- atom interaction serves as a tool to alter the cooling in both the schemes.

\subsection{Bistable behaviour and Normal mode splitting}

Having discussed the ground state cooling of the mechanical mirror in the previous section, we are now in a position to consider the dynamics of the cantilever of an optical cavity coupled through radiation pressure. Scattering of light from different atomic states of the two level BEC trapped between the high finesse mirrors creates different quantum states of the scattered light, which can be studied by analyzing the displacement spectrum of the movable mirror \citep{43}. We consider an optomechanical system consisting of an elongated cigar- shaped BEC of N two level $^{87} Rb $ atoms in the $|F=1>$ state with frequency $\omega_a$ of the $|F=1> \rightarrow |F=2>$ transition of the $D_{2}$ line, strongly interacting with the standing wave of the cavity with frequency $\omega_c$. The cavity mode and the mechanical oscillator is coupled with frequency $\Omega_m$ via a dimensionless parameter $\epsilon$. The harmonic confinement along the directions perpendicular to the optical lattice is taken to be so large that the system reduces to one dimensional. In the rotating wave and dipole approximation, the Hamiltonian of the system takes the form \citep{43}: -

\begin{equation}
H= \frac{p^2}{2m}- \hbar\Delta_a \sigma^{+} \sigma^{-}- \hbar\Delta_c a^{\dagger} a+ \hbar \Omega_m a_m^{\dagger} a_m- i\hbar g(x)[\sigma^{+} a- \sigma^{-}a^{\dagger}]- i\eta(a- a^{\dagger})+ \hbar\epsilon\Omega_ma^{\dagger}a(a_m+ a_m^{\dagger}),
\end{equation}

where $\Delta_a$ and $\Delta_c$ are the atom- pump and cavity- pump detuning respectively. $a (a^{\dagger}), a_m (a_m^{\dagger})$ are the annihilation (creation) operator of the cavity photon and mechanical mode respectively. Since the cavity field is damped with rate $\kappa$, we consider an intrinsically open system and neglect any direct coupling between the atoms and cantilever. Using adiabatic elimination and retaining only lowest band with nearest neighbour interaction, the corresponding Bose- Hubbard Hamiltonian with $b_j$ as the bosonic annihilation operator, can be written as: -

\begin{eqnarray}
H&=& E_0 \sum \limits_{j} b_j^{\dagger} b_j+ E\sum\limits_{j} (b_{j+1}^{\dagger} b_j+ b_{j+1} b_j^{\dagger})+ (\hbar U_0 a^{\dagger} a+ V_{cl}) \times \Big( J_0 \sum\limits_{j} b_j^{\dagger} b_j+ J\sum\limits_j (b_{j+1}^{\dagger} b_j+ b_{j+1}b_{j}^{\dagger}) \Big) \nonumber\\
&+& \frac{U}{2}\sum\limits_{j} b_j^{\dagger} b_j^{\dagger} b_j b_j- \hbar \Delta_c a^{\dagger} a- i\hbar\eta (a- a^{\dagger})+ \hbar \Omega_m a_m^{\dagger} a_m+ \hbar \epsilon \Omega_m a^{\dagger} a (a_m+ a_m^{\dagger}),
\end{eqnarray}

where the constants $U, E_0, E, J_0$ and $J$ can be referred from Appendix A and the other symbols have the same meaning as in previous sections. The Hamiltonian derieved is valid only for weak atom- field nonlinearity and it can be shown that the intracavity field intensity is bistable and leads to a bistable optical lattice potential.  The Heisenberg Langevin equation of motion for the bosonic field operator $b_j$, internal cavity mode $a$ and the mechanical phonon mode $a_m$ takes the form: -

\begin{equation}
\dot{b}_j= -i \Big( U_0 a^{\dagger} a+ \frac{V_{cl}}{\hbar} \Big) (J_0 b_j + J(b_{j+1}+ b_{j-1}))- \frac{iE}{\hbar} (b_{j+1}+ b_{j-1})- \frac{iU}{\hbar} b_j^{\dagger}b_j b_j- \frac{iE_0}{\hbar} b_j,
\end{equation}

\begin{equation}
\dot{a}= -iU_0 \Big( J_0 \sum\limits_j b_j^{\dagger} b_j+ J\sum\limits_j (b_{j+1}^{\dagger}b_j+ b_{j+1}b_j^{\dagger}) \Big)a+ \eta+ i(\Delta_c- \epsilon\Omega_m (a_m+ a_m^{\dagger}))a- \frac{\kappa}{2}a+ \sqrt{\kappa} \xi_p (t),
\end{equation}

\begin{equation}
\dot{a}_m= \Big( -i \Omega_m- \frac{\Gamma_m}{2} \Big) a_m- i\epsilon\Omega_m a^{\dagger} a+ \sqrt{\Gamma_m}\xi_m (t),
\end{equation}

where $\kappa$ and $\Gamma_m$ characterizes the dissipation of the optical and mechanical degrees of freedom and $\xi_p (\xi_m)$ denotes the noise operators for input field (mechanical oscillator). The steady state analysis of the Heisenberg equations of motion reveals: -

\begin{equation}
x_{m, s}= \frac{-8\epsilon\Omega_m^2 a_s^{\dagger} a_s}{4\Omega_m^2+ \Gamma_m^2},
\end{equation}

where $x_{m,s}$ is the stedy state position quadrature and the expression for $a_s^{\dagger}a_s$ comes out as:-

\begin{equation}
a_s^{\dagger}a_s= \frac{\eta^2}{(\Delta_c- U_0 J_0 \hat{N}- \epsilon \Omega_m x_{m,s})^2+ \kappa^2/4}.
\end{equation}

Here $\hat{N}= b_j^{\dagger}b_j$ and it is clear from the above expression that the coupling of the mirror and the atoms alters the cavity resonance frequency and changes the field inside the cavity to induce a new stationary intensity. Considering the case of large number of atoms and within mean field framework and tight- binding approximation, we deduce a cubic equation of $x_{m,s}$ from the above expressions without ignoring the tunnelling term $J$ as: -

\begin{equation}
x_{m,s}^3- \frac{2\Delta}{\epsilon\Omega_m} x_{m,s}^2+ \frac{(\Delta^2+ \kappa^2 /4)}{\epsilon^2 \Omega_m^2}+ \frac{8\eta^2}{\epsilon (4\Omega_m^2+ \Gamma_m^2)}= 0.
\end{equation}

Here, $\Delta= \Delta_c- U_0 N[J_0+ 2J cos(kd)]$. 

\begin{figure}[h!]
\includegraphics[width=0.85\textwidth]{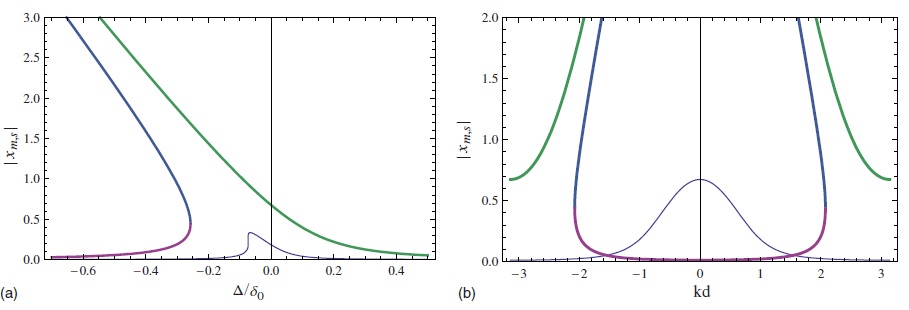}
\caption{Left- Steady state displacement spectrum as a function of $\Delta/\delta_0$. The thin (thick) line corresponds to $\eta/\delta_0= 0.2 (1.0)$.\\
Right- The displacement spectrum modified by the atomic back action through the cavity photons as a function of quasimomentum $(kd)$ for $\Delta_c /\delta_0= -0.5$ (thick) and $0.5$ (thin). Reprinted from Aranya B Bhattacherjee, Phy. Rev. A, 80, 043607 (2009).}
\centering
\end{figure}

The plots of $\mid x_{m,s} \mid$ versus $\Delta/\delta_0$ $(\delta_0= U_0J_0)$ (fig. 15) clearly shows a bistable behaviour for higher pump intensity. For pump rates, higher than the critical value, we find three steady state solutions for the mirror displacement, of which two of them are stable. The system prepared below resonance will follow the steady state curve until non steady dynamics is excited on reaching a lower turning point. The time scale of the mechanical motion dominates the dynamics of the system because the cavity damping is almost two orders of magnitude faster. Due to the strong coupling, the phonons develop a quasimomentum dependence with the condensate mediated through the cavity photons. The atom- field interaction changes as the condensate moves along the Brillouin zone and as a result the cantilever displacement spectra gets continuously modified. A bistable behaviour is seen when the condensate is at the edge of the Brillouin zone ($\Delta_c /\delta_0= -0.5$), which is otherwise absent when $\Delta_c /\delta_0= 0.5$.

In this part we discuss the splitting of the normal mode into three modes due to the coupling of the mechanical oscillator, cavity field fluctuations and the condensate fluctuations (Bogoliubov mode). The normal mode splitting does not appear in the steady state spectra but rather manifests itself in the mirror displacement spectra. Linearizing the Heisenberg- Langevin equations about their steady state values and transforming to the following quadratures: $X_m= (a_m+ a_m^{\dagger}), P_m= i(a_m^{\dagger}- a_m), X_p= (a+ a^{\dagger}), P_p= i(a^{\dagger}- a), X_b= (b+ b^{\dagger})$ and $P_b= i(b^{\dagger}- b)$, the displacement spectrum in the Fourier space is found as: -

\begin{equation}
S_x (\omega)= \frac{x_0^2}{2\pi} \Omega_m^2 \mid \chi (\omega) \mid ^2 \Big[ \Gamma_m n_m- \frac{\Delta_d^2+ \omega^2+ \kappa^2/4}{2\Delta_d \Omega_m}\Gamma_s (\omega)\Big],
\end{equation}

where $n_m$ represents the equilibrium occupation number for the mechanical oscillator and

\begin{equation}
\chi^{-1} (\omega)= \Omega_m^2+ 2\Omega_m\Omega_s (\omega)- \omega^2- i\omega[\Gamma_m+ \Gamma_s (\omega)].
\end{equation}

The expressions for $\Omega_s(\omega)$ and $\Gamma_s (\omega)$ can be referred from Appendix E. The displacement spectrum is characterized by a mechanical susceptibility $\chi (\omega)$ which is driven by thermal noise and quantum fluctuations of the radiation pressure and condensate. The plot below (fig. 16) shows the variation of displacement spectrum $S_x (\omega)$ for different values of atomic two body interaction. 

\begin{figure}[h!]
\includegraphics[width=0.85\textwidth]{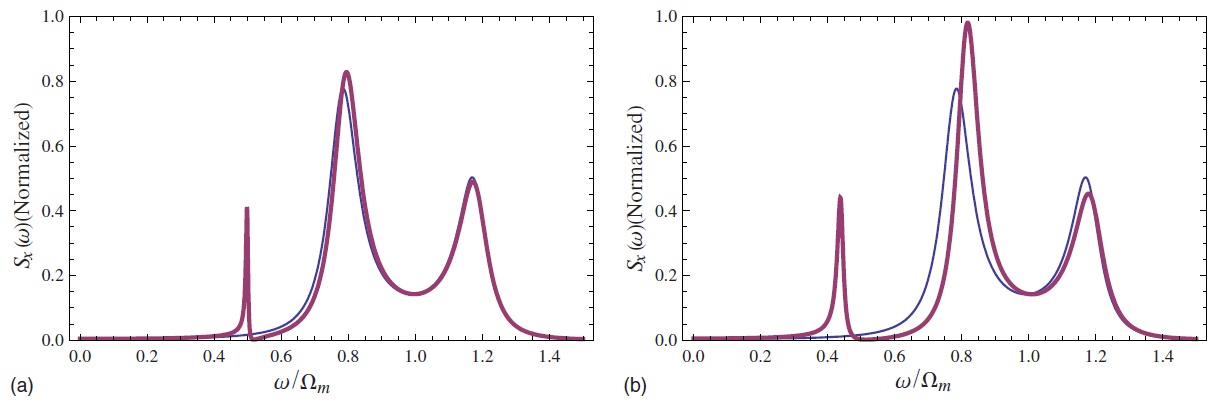}
\caption{Left- Normalized plot of the displacement spectrum $S_x(\omega)$ for different values of atomic two- body interaction. $E_{eff}/\Omega_m$= 0 (thin line) and 0.3 (thick line).\\
Right- Normalized plot of $S_x(\omega)$ with a stronger atom- photon coupling. The NMS gets prominent as compared to (a). Reprinted from Aranya B Bhattacherjee, Phy. Rev. A, 80, 043607 (2009) \citep{43}.}
\centering
\end{figure}

Evidently, the normal mode splits into two modes in the absence of interactions (thin line) and as the atom- atom interaction is increased $(U_{eff}/\Omega_m)= 0.3$, the normal mode splits into three modes which is associated with the mixing of mechanical mode and fluctuations of the cavity field and condensate around the mean field. The frequency of the Bogoliubov mode in the low momentum limit $\approx$ $\sqrt{U_{eff}}$, which clearly shows its absence when there in no atom- atom interactions. It is noteworthy to mention here that in order to observe Normal mode splitting, the exchange energy between the three modes must take place on a time scale faster than the decoherence of each mode.

\subsection{Experimental advancements with optomechanical cavities}

Optical dipole force and radiation pressure of laser light allows the construction of  hybrid optomechanical systems with mirror and micromechanical membrane. Vibrations of the mechanical membrane alters the standing wave, thereby coupling the centre of mass motion of the atoms. The atoms, conversely modulates the radiation pressure on the membrane and creating an optomechanical system where laser acts as a coupling agent between the two modes. Many successful experiments showing coupling between the optical and mechanical mode has been reported so far. In this section, we reproduce the experimental results of Theodor W. Hansch \citep{27} and his group in creating a hybrid atom- membrane optomechanical system.

\subsubsection{The experimental setup}

The system consists of a SiN membrane which retroreflects the light from a laser of power P and frequency $\omega$, which is red detuned with respect to the atomic transition. The reflected light overlaps with the incoming beam thereby creating an optical lattice potential. A displacement of the oscillator by $x_m$ results in force F= m$\omega_{at}^2x_m$ on each atom, where $\omega_{at}$ and m represents the trap frequency in harmonic approximation and atomic mass of each atom respectively. The membrane motion couples with the ultracold ensembles trapped in the lattice potential. The SiN membrane in a second room temperature vacuum chamber acts as a partially reflective end mirrorfor the 1 D optical lattice. A Michelson interferometer reads the membrane displacement and the two lasers are separated with a half wave plate (WP), a polarizing beam splitter (PBS) and a dichroic mirror (DM). The figure below shows the experimental setup used by the group.

\begin{figure}[h!]
\includegraphics[width=0.5\textwidth]{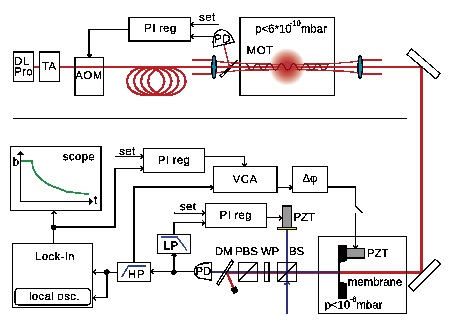}
\caption{Schematic representation of the experimental setup used by Theodor W. Hansch and his group. Reprinted with permission from Stephan Camerer $\textit{et al.}$, Phys. Rev. Lett. 107, 223001 (2011) \citep{27}.}
\centering
\end{figure}

\subsubsection{Backaction of the atoms and membrane dissipation}

Applying laser cooling at a rate $\gamma_c$ to the atoms manipulates the atomic damping rate $\gamma_{at}$= $\gamma_c$+ $\gamma_{\phi}$, which also includes the additional dephasing of the c.m motion at a rate $\gamma_{\phi}$. To maintain the steady state on much slower time of membrane dynamics, $\gamma_{at}$ (motional damping of the atoms) was much higher than the coupling constant, g and the motional damping rate of the membrane, $\gamma_m$. In such conditions, the dacay rate of the membrane vibrations can be approximated as: -

\begin{equation} 
\Gamma= \gamma_m+ \gamma_{at} \frac{g^2 \alpha}{\delta^2+ (\gamma_{at}/2)^2}.
\end{equation}

with $\delta$ denoting the detuning ($\omega_{atoms}- \omega_{membrane}$) and $\alpha$ is the product of the reflectivity (of SiN membrane) and transmittivity (between atoms and the membrane). The second term in the above expression represents additional dissipation due to the coupling between the membrane and ultracold atoms. The backaction of the atoms on the membrane was observed in membrane ringdown measurements and alternating experiments were performed by the group in the presence and absence of atoms to determine the respective decay rates $\Gamma$ and $\gamma_m$. The plot below shows the backaction of atoms on membrane as a function of laser power P. The top plot shows the variation of additional membrane dissipation rate $\Delta\gamma$= $\Gamma$- $\gamma_m$ and the lower plots correspond to the atom number in the experiments. The solid line shows the theoretical curve which agrees well with the experimental data found by the group.\\

\begin{figure}[h!]
\includegraphics[width=0.9\textwidth]{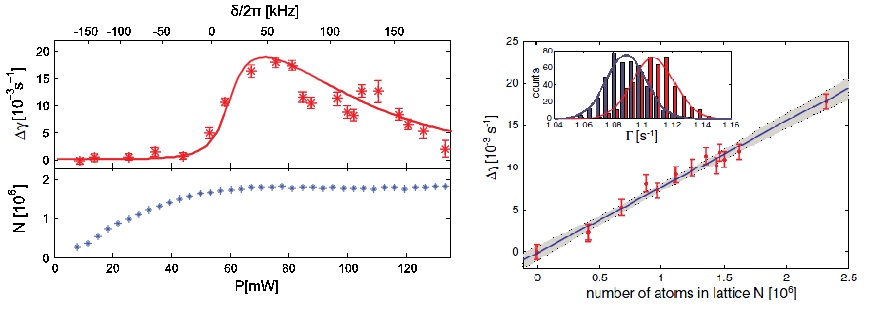}
\caption{Left- Plot of the backaction of the laser cooled atoms onto the membrane. Top: shows the measured additional membrane dissipation rate due to coupling to atoms as a function of P. Solid line refers to the theoretical fit. Bottom: Lattice atom number in the experiment.\\
Right- Measured additional membrane dissipation as a function of the atom number for resonant coupling. The line is a linear fit. The observed dependence agrees well with theory. Inset: Histogram of measurements of $\Gamma$ for N= 2.3 $\times 10^6$ (right) and N= 0 (left). Reprinted with permission from Stephan Camerer $\textit{et al.}$, Phys. Rev. Lett. 107, 223001 (2011) \citep{27}.}
\centering
\end{figure}

To study the dependence of the membrane dissipation on the atom number, the system was prepared on resonance and the atomic number N was varied by altering the power of the MOT repump laser. The group reported a linear dependency of $\Delta \gamma$ on N, as shown above. This agrees well with the theoretical proposal of K. Hammerer $\textit{et al.}$ \citep{57}. Since the model does not account for finite temperature, spatial variation and lattice trap anharmonicity, the experimental agreement with the theory was a remarkable success.

\section{Dicke model}

We reconsider the model described in the third section with N two level atoms trapped between the optical cavities with two pumping lasers of amplitude $\eta$ and $\xi$. When $\eta$= 0, we have only the transverse pumping and such a system undergoes a quantum phase transition from superfluid state to a self organized state above a critical atom- cavity coupling strength. This refers to the basic Dicke model with N particles interacting coherently with the radiation mode. Physical realizations of Dicke model has been possible for single radiation mode systems, however, theoretical proposals for multimode systems have predicted interesting physics in the field of quantum simulation and quantum information. Below the threshold pump frequency, only the pump mode is present in the cavity, however, as the power is increased above the threshold, the atoms self organizes into a checkerboard pattern trapped in the interference pattern of the pump and cavity beams. This marks the onset of the superradiance in an effective non equilibrium Dicke model. In this section we shall initially discuss the experimental realization of Dicke model by Tilman Esslinger and his group \citep{26} and then elaborate the theoretical proposals for optomechanical and two atom Dicke model. \\
	
	The Dicke phase transition was observed in an optical cavity consisting of $^{87}$Rb BEC with approximately 10$^5$ atoms prepared in the ground state with $\Ket{k_x, k_z}= \Ket{0,0}$. The atoms were trapped inside the optical cavity of length 178$\mu$m and subjected to a transverse pump of frequency $\omega_p$, which was far detuned from the atomic transition frequency $\omega_a$. However, the pump frequency was near detuned to the cavity frequency $\omega_c$, resulting in efficient scattering from the pump beam into the optical cavity, with $\kappa$ as its decay rate. Absorption and emission of photons yield an effective two level 'spin' system with level splitting $\omega_0= 2\omega_r$, coupled through a pair of Raman laser, where $\omega_r= \hbar^2 k^2/2m$ is the atomic recoil energy due to absorption or emission of photons. The effective Hamiltonian of such a system can be written as ($\hbar= 0$ throughout the paper): -

\begin{equation}
H= \omega_0 S_z+ \omega a^{\dagger} a+ US_z a^{\dagger} a+ g(a^{\dagger}S^{-}+ a S^{+})+ g' (a^{\dagger}S^{+}+ a S^{-}),
\end{equation}
	
where $U= -(1/4)g_0^2/(\omega_a- \omega_c)$ describes the back- reaction of the cavity light field on the BEC and $\omega= \omega_c- \omega_p- N(5/8)g_0^2/ (\omega_a- \omega_c)$ and $g_0$ representing the strength of the cavity coupling. The level scheme corresponding to the experimental setup has been shown below.

\subsection{Experimental observation of the self organized state}
The Dicke model highlights the quantum phase transition from a superfluid to a self organized state above a certain threshold pump frequency. The phase transition is driven by long range interaction between the cold atoms, induced by cavity mode and pump field. The transition frequency of the two level atomic system exceeds the available dipole coupling strength and a Raman transition brings out the energy difference between the atomic modes from optical to much lower scale. To highlight the experimental process of \citep{26} in observing Dicke phase transition, the laser pump power was gradually increased over time while constantly monitoring the light leaking out of the cavity. As long as the pump power was below the threshold, no light was detected at the cavity output and the expected momentum distribution of the condensate loaded into the shallow standing wave formed by the cavity field was observed. Once the pump reached the critical value, a sudden abrupt increase in the mean intracavity photon number was observed, which marked the onset of the superradiance or the self organization phase. Simultaneously, the momentum distribution of the condensate recorded a striking change showing additional momentum components as ($p_x, p_z)= (\pm hk, \pm hk$). This marked the direct evidence of the phase transition in a non equilibrium Dicke model. Fig. (3) of \citep{26}, represents the phase transition and formation of additional momentum components as the pump power was gradually increased while monitoring the mean intracavity photon number.

\subsection{Mapping the phase diagram}
The phase diagram showing the normal and superradiant phase can be mapped by tracing intracavity photon number for different values of pump- cavity detuning. Evidently, the critical pump power scales linearly with the effective cavity frequency for large negative values of the detuning parameters. For $\omega$ (cavity mode frequency) $<$ 0, no real solutions can be found for critical coupling strength. Indeed, for cases when pump- cavity detuning is larger than the dispersively shifted cavity resonance, no light scattering was reported. The intracavity photon number increases as the pump cavity detuning approaches the shifted cavity resonance. The corresponding mapped data can be referred from fig.(5) of \citep{26}, where the red dotted lines correspond to the theoretical calculations by mean field description, including the external confinement of the atoms, cavity mode profiles, transverse pump and collisional atom- atom interaction. The theoretical and experimental agreement of the phase diagram is just excellent. In the sections to come, we shall reproduce the same plots through theoretical analysis and also define some phase regions that have only theoretical existence. We shall exploit some characteristics of the system and also show that the critical value can be altered by simple and efficient way.

\subsection{Theoretical approach: Optomechanical Dicke model and phase transition}

In this section, we introduce the optomechanical Dicke model reviewed here which involves a Fabry- Perot optical cavity with one fixed and another movable mirror of mass $M$ and oscillating freely with frequency $\omega_m$. Within the optical cavity, a cigar shaped $N$ two level BEC of mass $m$ is trapped with transition frequency $\omega_0$. The cold condensate is coupled to a single standing wave cavity mode of frequency $\omega_c$ trapped within the high finesse optical cavity of length $L$ and decay rate $\kappa$. The schematic representation of the system is shown in fig. (19). Note that in the first case, we discuss the results when the external force pump is non functional and we deal with the same in the second case. An external laser of frequency $\omega_l$ is applied perpendicular to the cavity axis and a tight harmonic potential of frequency $\omega_r$ is formed. Assuming atom- laser detuning $\Delta_0 (= \omega_0- \omega_l)$ to be very large, the adiabatic elimination of the electronically excited states can be justified. As a result, two level atomic system is formed with zero momentum state $\Ket{p}$= $\Ket{0}$ and excited momentum state $\Ket{p}$= $\Ket{\pm \hbar k}$, where $p$ is the momenta along the cavity axis and $k$ denotes the wave vector of the pump laser field. These momentum states are coupled through a pair of Raman channels such that $\omega_a$ is twice the atomic recoil frequency. When all the BEC atoms with ground and excited momentum states are coupled identically with the single- mode cavity field, the effective Hamiltonian of the system can be written as \citep{46}: -

\begin{equation}
H= \hbar \omega_a \hat{J}_z+ \hbar\omega_c\hat{b}^{\dagger}\hat{b}+ \hbar\omega_c\delta_0\hat{a}^{\dagger}\hat{a} (\hat{b}+ \hat{b}^{\dagger}) + \hbar\frac{\lambda}{\sqrt{N}} (\hat{a}+ \hat{a}^{\dagger})(\hat{J}_{+}+ \hat{J}_{-}),
\end{equation}

where $\hat{J}_z$, $\hat{J}_{+}$ and $\hat{J}_{-}$ are the collective atomic operators satisfying the standard angular momentum commutation relations. $\hat{a} (\hat{a}^{\dagger})$ and $\hat{b} (\hat{b}^{\dagger})$ represents the annihilation (creation) oprators of the photons of the cavity field and phonons of the oscillating mirror respectively, satisfying the usual commuation relation $[\hat{a} (\hat{b}), \hat{a}^{\dagger} (\hat{b}^{\dagger})]$= 1. $\delta_0$ represents the nonlinear dispersive coupling between the cavity field intensity and position quadrature of the mirror and $\delta_0<<$1. Introducing the c- number variables, $\alpha \equiv <\hat{a}>$, $\beta \equiv <\hat{b}>$, $w \equiv <\hat{J}_z >$ and $\gamma \equiv <\hat{J}_{-}>$, where $\alpha$, $\beta$, $\gamma$ represents the complex cavity field, mirror mode and atomic polarization amplitudes respectively. $w$ is the population inversion and is always real. 

\begin{figure}[h!]
\includegraphics[width=0.68\textwidth]{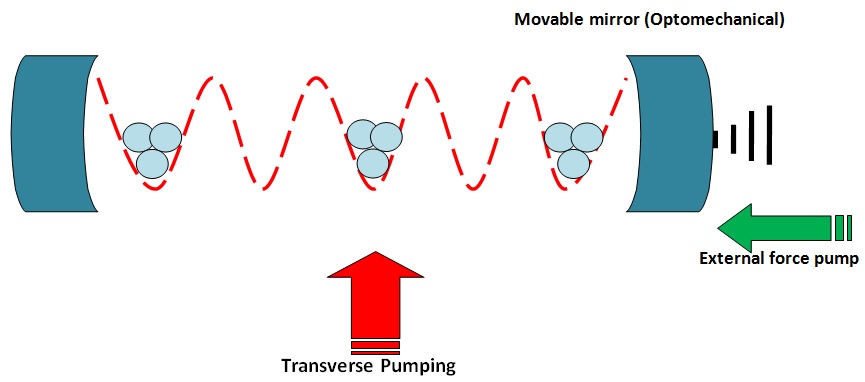}
\caption{The schematic representation of the optomechanical Dicke model considered in this paper. Above the critical atom-  cavity field coupling strength, the atoms self organizes in alternate (odd or even) potential wells to form a checkerboard pattern.}
\centering
\end{figure}

Using the mean field analysis of the system, the semi classical equations of motion for the system Hamiltonian (Eq.76) can be written as:-

\begin{equation}
\dot{\alpha}= -(\kappa+ i\omega_c)\alpha- i\omega_c\delta_0\alpha (\beta+ \beta^{*})- i\frac{\lambda}{\sqrt{N}} (\gamma+ \gamma^{*}),
\end{equation}

\begin{equation}
\dot{\beta}= -(\Gamma+ i\omega_m)\beta- i\omega_c\delta_0 \mid \alpha \mid^2,
\end{equation}

\begin{equation}
\dot{\gamma}= -i\omega_a\gamma+ 2i \frac{\lambda}{\sqrt{N}} (\alpha+ \alpha^{*})w,
\end{equation}

\begin{equation}
\dot{w}= i \frac{\lambda}{\sqrt{N}} (\alpha+ \alpha^{*}) (\gamma- \gamma^{*}),
\end{equation}

where $\Gamma$ represents the damping rate of the mirror due to the interaction of the mechanical mirror with the environment. Using the pseudo angular momentum constraint $w^2+ \mid \gamma \mid ^2= N^2/4$, we find the steady state values for different c- number variables by considering steady state condition and factorizing the obtained linear algebraic equations. The steady state analysis displays a bifurcation point at $\lambda= \lambda_c$, given as: -

\begin{equation}
\lambda_c= \frac{1}{2} \Big( \frac{\omega_a}{\omega_c}(\kappa^2+ \omega_c^2) \Big) ^{1/2},
\end{equation}

where $\lambda_c$ represents the critical value of the atom cavity coupling strength above which the phase transition to superradiant phase initiates. The steady state solutions for $\lambda < \lambda_c$ are given as: -

\begin{equation}
\alpha_s= \beta_s= \gamma_s= 0, w_s= \pm \frac{N}{2}.
\end{equation}

Above the critical value, $\lambda_c$, the above solutions become unstable and a new set of stable solutions appear, which can be obtained by solving the following cubic equation:-

\begin{equation}
w_s^3 \Big[ \frac{\lambda^2 \delta_0^2 \sigma (1- 2\bar{\epsilon})}{N\lambda_c^2} \Big]+ w_s \Big[1- \frac{N\lambda^2 \delta_0^2 \sigma (1- 2\bar{\epsilon})}{4\lambda_c^2}\Big]+ \frac{N\lambda_c^2}{2\lambda^2}= 0,
\end{equation}

where,

$\bar{\epsilon}= \frac{w_c^2}{\kappa^2+ \omega_c^2}$ and $\sigma= \frac{2\omega_m\omega_a}{\Gamma^2+ \omega_m^2}$. The above equations are solved numerically to obtain the steady state values above the critical value as: -

\begin{equation}
\gamma_s= \pm \Big(\frac{N^2}{4}- w_s^2\Big)^{1/2},
\end{equation}

\begin{equation}
\mid \alpha_s \mid= \pm \Big[\frac{N(\kappa^2+ \omega_c^2)}{4\lambda^2\gamma_s^2}- \frac{4\delta_0^2\omega_m \omega_c \bar{\epsilon}}{(\Gamma^2+ \omega_m^2)} \Big] ^{1/2},
\end{equation}

\begin{equation}
\beta_s= \frac{-\omega_c \delta_0 \mid \alpha \mid ^2 (\omega_m+ i\Gamma)}{\Gamma^2+ \omega_m^2}.
\end{equation}

We plot the above variables as a function of the normalized atom- cavity field coupling strength $\lambda / \omega_m$ for different values of mirror- photon coupling $\delta_0$. The bifurcation point in all the plots represents the critical value above which the superradiant phase exists. Clearly the system dynamics changes above the critical value which represents the self organized state.  The first plot (fig. 20, top- left panel) shows an abrupt in steady state state atomic population inversion above the critical point and $\gamma_s$ and $\mid \alpha_s \mid$ shows almost same symmetric behaviour above the bifurcation point for positive and negative value. The behaviour of the steady state semi classical solutions below and above the critical atom- cavity field coupling strength demonstrates the phase transition from normal (or inverted) state to a state of self organization. An increase in the steady state value of the cavity field amplitude due to the increase in the mirror- photon coupling (fig. (20), bottom- left panel, red- dotted line) naturally leads to the increase in the radiation pressure, which increases the steady state mechanical field amplitude of the vibrating mirror. Thus, by continuous monitoring of these solution, one can actually detect the Dicke phase transition in an optomechanical cavity. The discussion will become clear in the next part when we shall include the back action parameter to the same model and discuss its implications through a detailed study of the phase portraits.

\begin{figure}[h!]
\includegraphics[width=0.45\textwidth]{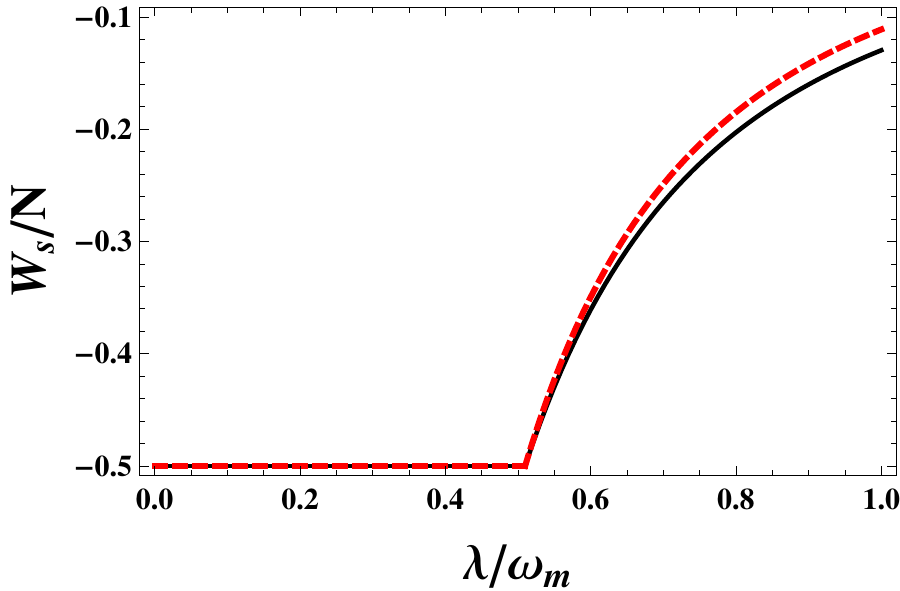}
\includegraphics[width=0.45\textwidth]{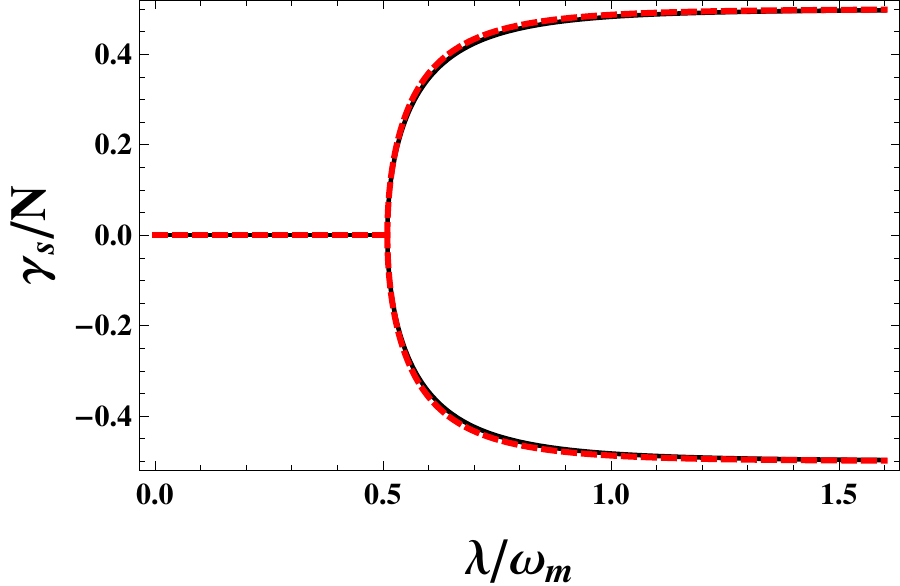}
\includegraphics[width=0.45\textwidth]{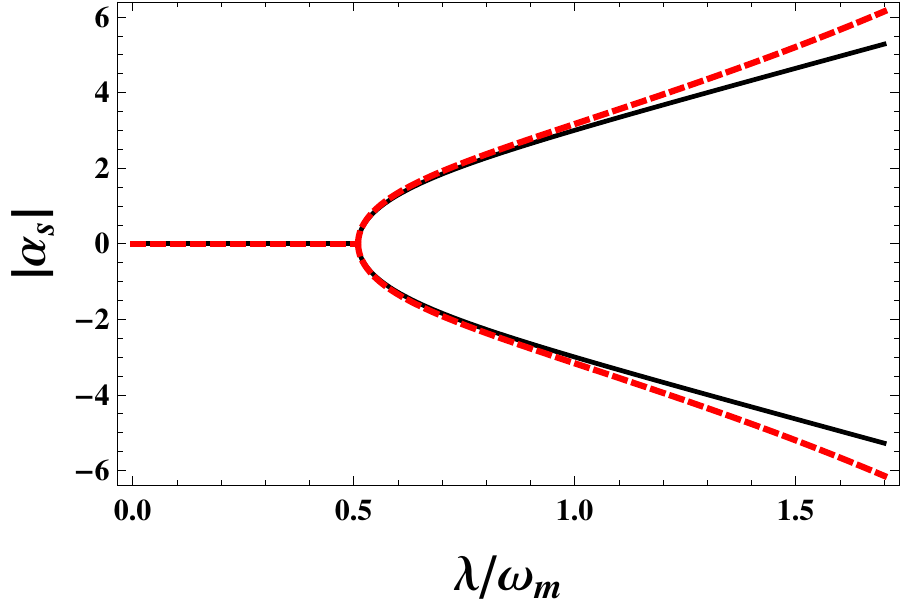}
\includegraphics[width=0.45\textwidth]{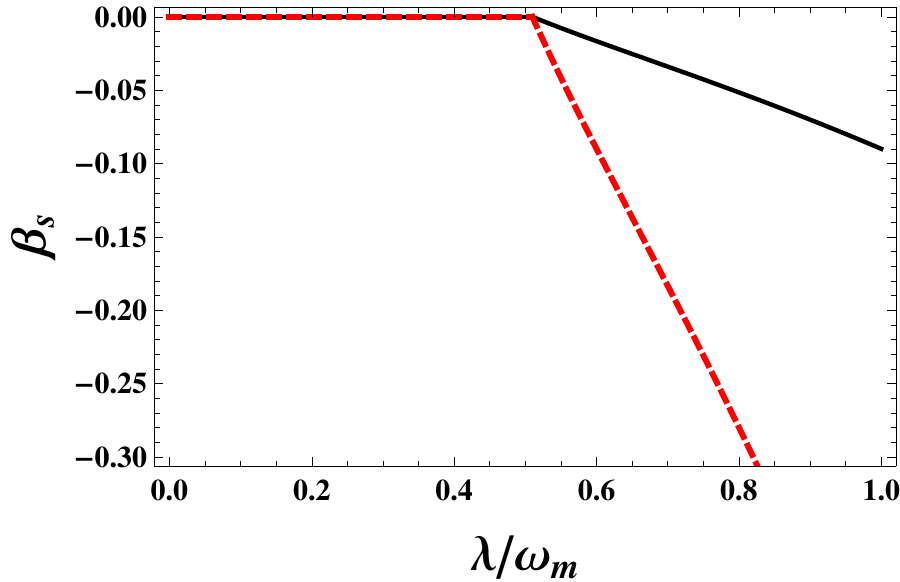}
\caption{ Plot of the steady state population inversion $w_s$ (top- left), polarization amplitude $\gamma_s$  (top- right), absolute value of cavity field amplitude $\mid \alpha \mid$ (bottom- left) and mirror mode amplitude $\beta_s$ (bottom right) as a function of dimensionless atom- photon coupling strength $\lambda/ \omega_m$ for different values of mirror- photon coupling. $\delta_0$= 0.01 (solid line) and 0.05 (dashed line) with $\omega_a= \omega_m$. Other prameters used are $\Gamma= 10^{-5}\omega_m$, $\kappa= 0.2\omega_m$ and $N$= 10.}
\centering
\end{figure}

	We now consider the optomechanical system in the presence of the external mechanical pump as shown in fig. 19, which was neglected in the pevious discussion. This external pump can be any mechanical object or any external laser that helps in oscillating the mirror via radiation pressure. In the presence of the mechanical object, the new Hamiltonian of the system can be written as:-

\begin{equation}
H= \hbar \omega_a\hat{J}_z+ \hbar\omega_c\hat{a}^{\dagger}\hat{a}+ \hbar\omega_m\hat{b}^{\dagger}\hat{b}+ \hbar\omega_c\delta_0\hat{a}(\hat{b}+ \hat{b}^{\dagger})+ \hbar\frac{\lambda}{N^{1/2}}(\hat{a}+ \hat{a}^{\dagger})(\hat{J}_{+}+ \hat{J}_{-})+ \hbar\eta_p(\hat{b}+ \hat{b}^{\dagger}),
\end{equation}

The last additional term in the above Hamiltonian represents the energy due to an external mechanical pump, where $\eta_p$ is the mechanical pump frequency which has been considered small throughout the paper, $\textit{i.e.}$ $\eta_p<<$1. The other symbols have the same usual meaning as in the previous discussion. The semiclassical equation of motion for the mirror in the presence of the external pump can be rewritten as: -

\begin{equation}
\dot{\beta}= -(\Gamma+ i\omega_m)\beta- i\omega_c\delta_0 \mid \alpha \mid^2- i\eta_p.
\end{equation}

In this case, the critical value of the atom- photon coupling strength gets modified as: -

\begin{equation}
\lambda_c '= \frac{\lambda}{\Big(1- \frac{\sigma \eta_p \delta_0(1- 2\bar{\epsilon})}{\omega_a}\Big)^{1/2}}.
\end{equation}

Clearly the bifurcation point has shifted from $\lambda_c$ to $\lambda_c'$ in the presence of the external mechanical pump, which leads to a change in the steady state expressions. For $\lambda<\lambda_c'$, the steady states are given as:- 

\begin{equation}
\alpha_s= \gamma_s= 0, \beta_s= \frac{-\eta_p(\omega_m+ i\Gamma)}{\Gamma^2+ \omega_m^2}, w_s= \pm\frac{N}{2}.
\end{equation}

Above the critical point, the new set of steady states expressions are given as: -

\begin{equation}
\gamma_s= \pm \Big( \frac{N^2}{4}- w_s^2\Big)^{1/2},
\end{equation}

\begin{equation}
\mid \alpha_s \mid= \pm \frac{X_2}{X_3},
\end{equation}

\begin{equation}
\beta_s= \frac{-(\omega_m+ i\Gamma)(\eta_p+ \omega_c\delta_0\mid a_s \mid^2)}{\Gamma^2+ \omega_m^2},
\end{equation}

where 

\begin{equation}
X_2= \Big[ 1+ \frac{4\delta_0\omega_m \bar{\epsilon} \eta_p}{(\Gamma^2+ \omega_m^2)} \Big]^{1/2},
\end{equation}

\begin{equation}
X_3= \Big[ \frac{N(\kappa^2+ \omega_c^2)}{4\lambda^2\gamma_s^2}- \frac{4\omega_m\omega_c\delta_0^2\bar{\epsilon}}{\Gamma^2+ \omega_m^2} \Big]^{1/2}.
\end{equation}

Fig. (21) represents the plot of the steady state atomic inversion $w_s$ and polarization amplitude $\gamma_s$ as a function of normalized atom- cavity field coupling strength for different values of external pump frequency ($\eta_p$= 0.3$\omega_m$ (dashed line) and $\eta_p$= 0 (bold line)). The plots clearly show the deviation of the system parameters in the presence of the external mechanical pump. It is evident that the additional mechanical pump shifts the bifurcation point to a lesser value of atom- photon coupling, which results the phase transition to occur at lesser critical value. All the plots of fig.(21) implies that the phase transition point can be controlled coherently by accessing the external mechanical pump. Thus by calibrating the device, one can aim to measure the weak force from the value of the critical transition point, as the external pump can be considered a weak force. In the next part, we remodel the system considering the detuning parameters and back action parameter and study the effect of the mechanical mirror through a detailed study of the phase portraits of the system. We shall encounter a very interesting feature of this model through analytical arguments, which was not visible when back action parameter was neglected in previous discussions.

\begin{figure}[h!]
\includegraphics[width=0.45\textwidth]{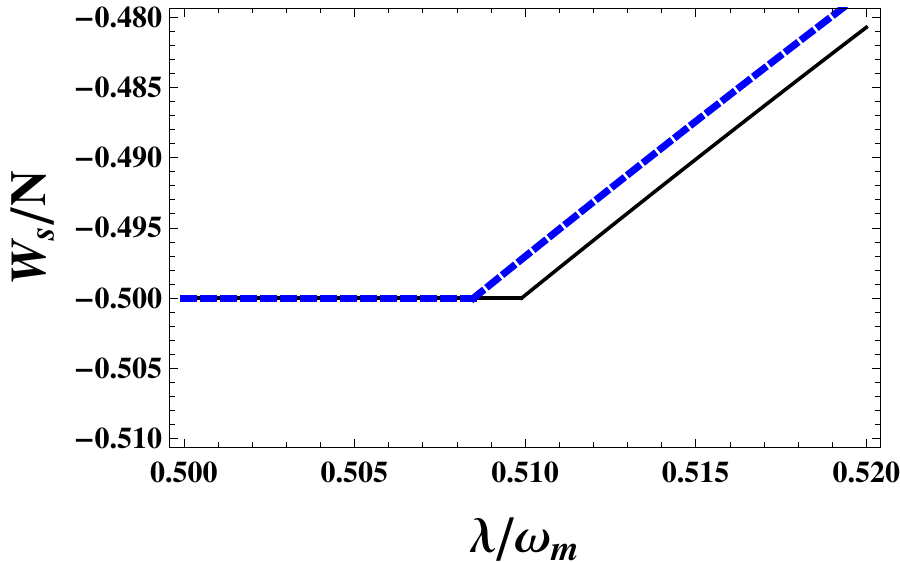}
\includegraphics[width=0.45\textwidth]{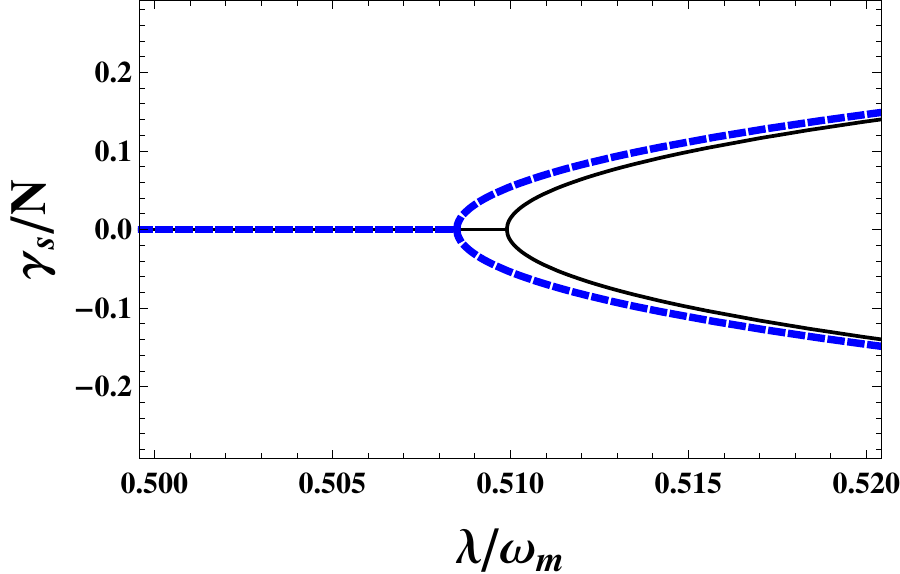}
\caption{ Plot of the steady state population inversion $w_s$ (left), polarization amplitude $\gamma_s$  (right) as a function of dimensionless atom- photon coupling strength $\lambda/ \omega_m$ for different values of mechanical pump frequency. $\eta_p$= 0 (solid line) and 0.3 $\omega_m$ (dashed line). Other prameters used are same as in previous plots.}
\centering
\end{figure}

\subsubsection{Dicke model with back action parameter $U$}

We consider the same optomechanical Dicke model with slight different parameters to retain the back action parameter terms in the Hamiltonian. Please refer \citep{58} for the derivation of hamiltonian (optomechanical considerations added separately) in terms of the cavity detuning and back action parameter, which originates due to a Rayleigh scheme, involving distinct momentum states rather than internal hyperfine states. This leads to the presence of the back action term in the Hamiltonian, given as \citep{56, 58}: -

\begin{equation}
H= \omega_a S_z- \omega_c a^{\dagger}a+ \omega_m b^{\dagger}b+ \omega_c' a^{\dagger}a (b+ b^{\dagger})+ g(a+ a^{\dagger})(S_{+}+ S_{-})+ US_za^{\dagger}a,
\end{equation}

where $\omega_a$ is the transition frequency, $\omega_c$ the cavity detuning and $\omega_c'$ represents the coupling constant. $U$ describes the back reaction of the cavity light on the BEC and may be interpreted as the a.c Stark shift due to the formation of the weak optical lattice. In the experiments described initially \citep{26}, $U$ was negative, however, both the signs are achievable experimentally and we shall deal with the both here. In the thermodynamic limit, the semi classical equations of motion for the open system described by the above Hamiltonian is given as: -

\begin{equation}
\dot{S_{-}}= -i(\omega_a+ U \mid a \mid^2)S_{-}+ 2ig(a+a^{\dagger})S_z,
\end{equation}

\begin{equation}
\dot{a}= -[\kappa+ i(\omega_c+ US_z+ \omega_c'(b+ b^{\dagger}))]a,
\end{equation}

\begin{equation}
\dot{b}= -i\omega_m b- i\omega_c'\mid a \mid ^2- \Gamma_m b,
\end{equation}

\begin{equation}
\dot{S}_z= -ig(a+ a^{\dagger})+ ig(a+ a^{\dagger})S_{-},
\end{equation}

where the operators have the same usual meaning as in the previous Hamiltonian. To study the dynamics of the system we consider the steady state condition of the above equations $(\dot{S}_{-}= \dot{a}= \dot{S}_z= \dot{b}= 0)$. For $U$= 0, the critical atom- coupling constant was calculated in the previous section by introducing c- number variables. In this section, we shall follow a different numerical technique. We define $a= a_1+ ia_2$, $b= b_1+ ib_2$ and $S_{\pm}= S_x \pm iS_y$ and substitute in the first three semi classical equations of motion and compare the real and imaginary parts, which yields 6 equations. From the first equation, we get: -

\begin{equation}
S_y (\omega_a+ U\mid a \mid ^2)= 0.
\end{equation}

Clearly, two possibilities exist, either $S_y$= 0 or $(\omega_a+ U\mid a \mid^2)$= 0. We define the first as the superradiant A (SRA) phase and the second as the superradiant B (SRB) phase, ofcourse the second type possible only for negative value of the back action parameter $U$. Note that the SRA phase is consistent with any value of $U$. SRA phase is defined as the transition from the normal (N) or inverted (I) state into regime of superradiance $\textit{i.e.}$ all the atoms pointing either upwards or downwards and no photons or phonons. In other words, $[S_x, S_y, S_z]= [0, 0, \pm N/2]$ defines the SRA phase. The semiclassical equations when separated into real and imaginary parts, can be represented in a matrix form as: -

\begin{equation}
\begin{vmatrix} 
\omega_{a} + U\mid a \mid^2 & 0 & -4gS_{z} & 0 \\ 0 &  \omega_{a} + U\mid a \mid^2 & 0 & 0 \\ 2g & 0 & \chi & \kappa  \\ 0 & 0 & -\kappa & -\chi \end{vmatrix}= 0, 
\end{equation}

where $\chi= (\omega_c+ US_z- (2\omega_c^2 \delta_0^2\mid a \mid^2)/\omega_m)$. Of the four semiclassical equations, the first equation brings out an important condition, $S_y(\omega_a+ U\mid a \mid ^2)$= 0. The two possibilities gives the two superradiant phases, namely $S_y$= 0, which defines Superradiant A (SRA) phase while $(\omega_a + U \mid a \mid^2)= 0$, which represents superradiant B (SRB) phase. The above determinant can be straightforwardly solved for $S_z$, which when equated to $\pm N/2$, yields the equation for $\omega_c$ that represents the dynamical phase diagrams of the system for the SRA phase. Similarly, ($\omega_a + U\mid a \mid^2)$= 0 determines the other phase. We plot here all the phase regions possible through analytical arguments and it is evident from fig. 22 that there are many regions which has not yet been observed experimentally. It is noteworthy to mention here that although all these phase regions can be investigated in various experimental conditions, however, not all will emerge in a single experiment. The designing of such a system to observe various phase regions discussed here is a matter of technological advancement in controlling the parameters of the system. Experiments reported by K. Baumann $\textit{et al.}$ \citep{26}, showed the system evolving from normal phase with all spins pointing downwards and no photons.

\begin{figure}[h!]
\includegraphics[width=0.48\textwidth]{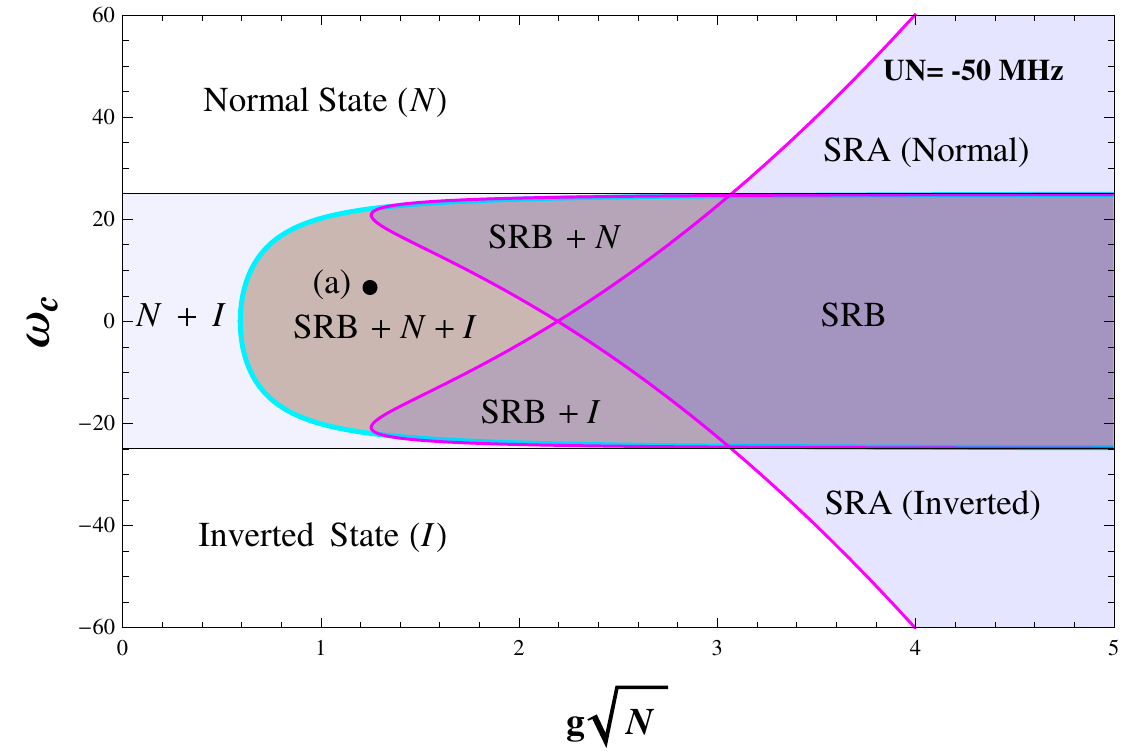}
\includegraphics[width=0.48\textwidth]{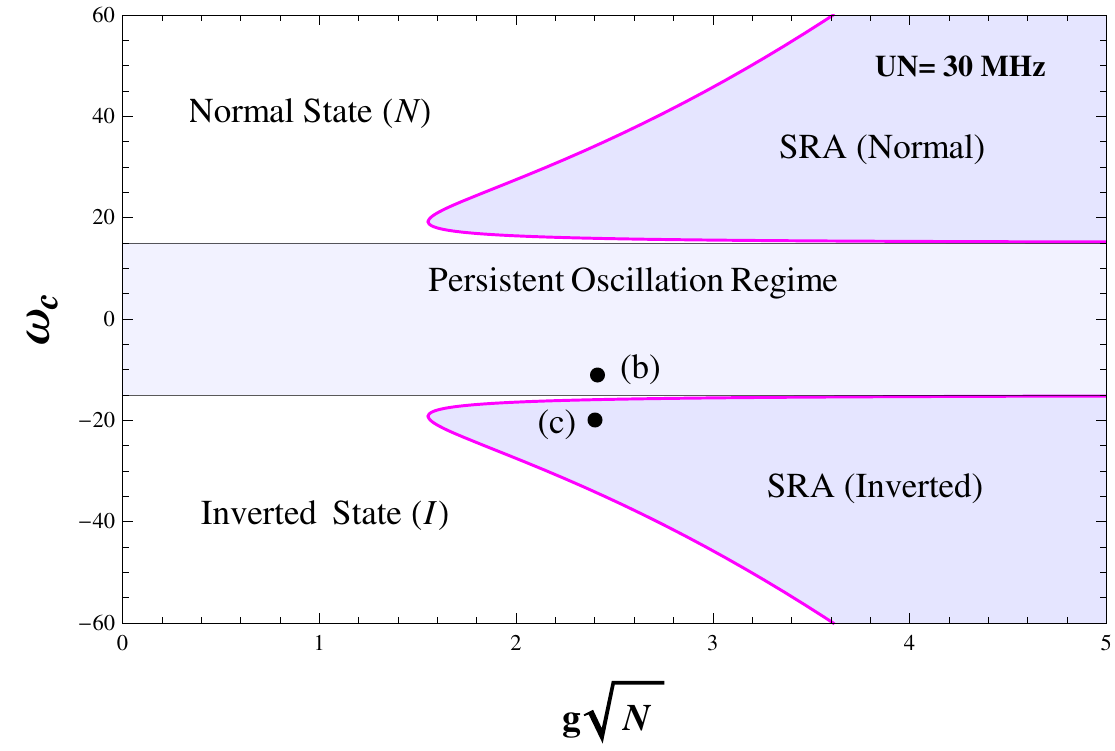}
\caption{The dynamical phase diagrams for UN= -50 MHz (left) and UN= 30 MHz (right). A complete detailed analysis can be found in \citep{56}.} 
\centering
\end{figure}

Determinant equation (Eq. 103) when solved for $S_z$, yields a quadratic equation which supports two roots of $S_z$. There are regions in the phase diagram where both the roots of $S_z$ are supported and such regions are known as 2SRA phase or more informatively as SRA (N)+ SRA (I) phase. Such regions remain prominent for optical cases but can be modulated in optomechanical systems. By altering the mirror frequency, $\omega_m$, these regions can be changed and the mirror therefore acts as a handle to modify the dynamical phase diagrams in an efficient and easy way. We don't produce the plots showing the coexisting region, however, interested readers can refer \citep{44, 45} for optical case and \citep{56} for optomechanical Dicke model. As the back action parameter is made positive, the SRB phase region vanishes for reasons determined by the boundary condition ($\omega_a+ U \mid a \mid^2)$= 0 and the phase boundaries shift in the opposite direction and tend to separate from each other towards higher value of $\omega_c$ (fig. 22, right panel). This leads to the formation of a persistent oscillation regime (fig. 24, right panel) which was otherwise a coexisting region (fig. 22, left panel) when the back action parameter was negative. As the name suggests, the persistent oscillation regime describes persistent oscillations and no steady state is reached even for long duration experiments, thereby predicting the presence of a limit cycle. The notion can be made clear from the plot below (fig.23) where we have plotted the relaxation time from point (c) which lies in the persistent oscillation regime, which shows all the system parameters oscillating periodically and no stable points can be reached even after long duration.

\begin{figure}[h!]
\includegraphics[width=0.48\textwidth]{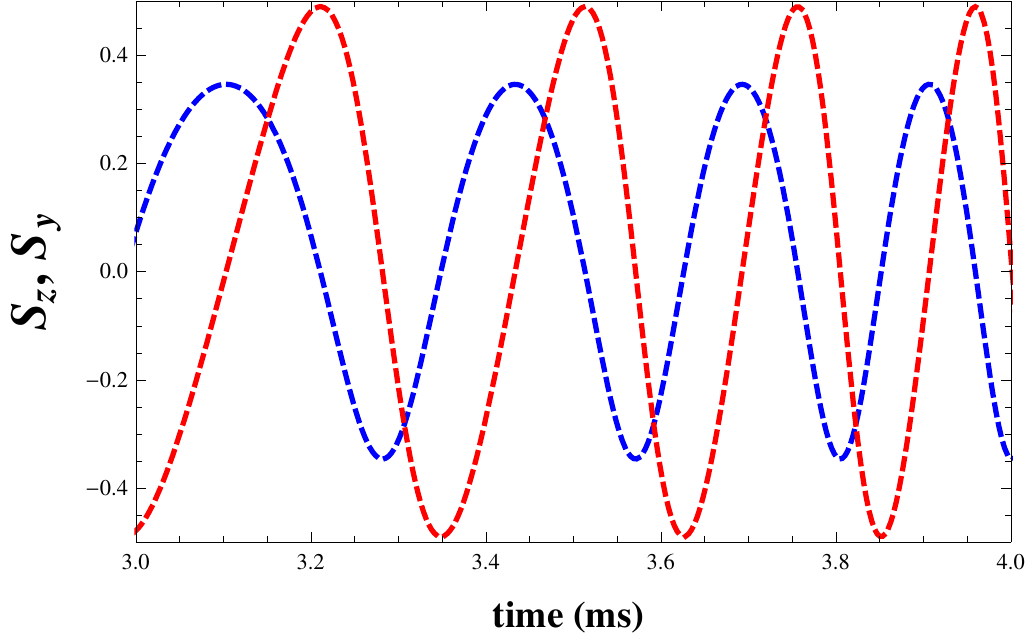}
\caption{The time evolution of point (c), marked in the plot of right panel of fig. (22). A complete detailed analysis can be found in \citep{56}.} 
\centering
\end{figure}

	Considering the external mechanical pump with frequency $\eta_p$ discussed in the previous model (the Hamiltonian remains identical as Eq.(97) with just the external pump energy term getting added), the determinant equation (Eq.103) modifies as: -

\begin{equation}
\begin{vmatrix} \omega_{a} + U \mid a \mid^2 & 0 & -4gS_{z} & 0 \\ 0 &  \omega_{a} + U \mid a \mid^2 & 0 & 0 \\ 2g & 0 & \chi' & \kappa  \\ 0 & 0 & -\kappa & -\chi' \end{vmatrix} = 0,
\end{equation}

where $\chi'$ = $\frac{\alpha}{\omega_{m}}$ and $\alpha =\Big(\omega_{c}\omega_{m} + US_{z}\omega_{m} - 2\omega_{c}^2\delta_{0}^2\mid a \mid^2 - 2\omega_{c}\delta_{0}\eta_{p}\Big)$. The modified determinant represented above when solved for S$_z$ and equated to $\pm$ N/2, describes the superradiant phase of first type. The equations are solved numerically and are too cumbersome to reproduce here. We plot the dynamical phase portraits separately for SRA and SRB phase for $\eta_p$= 1. The shift in the discussed bifurcation point reveals the change in the phase portrait and the addition of the external mechanical pump has actually created a new regime in both SRA and SRB phase denoted in the plots as $\eta$- SRA and $\eta$- SRB respectively (green shaded region). 

\begin{figure}[h!]
\includegraphics[width=0.4\textwidth]{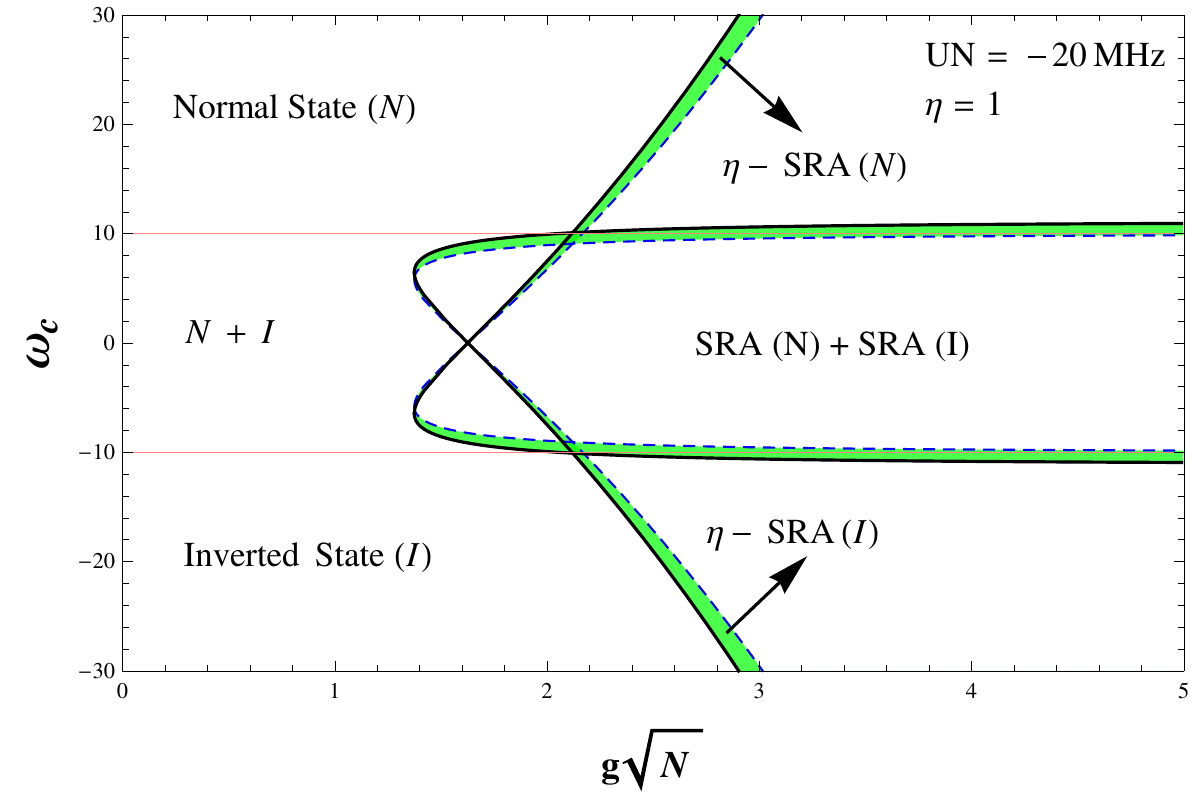}
\includegraphics[width=0.4\textwidth]{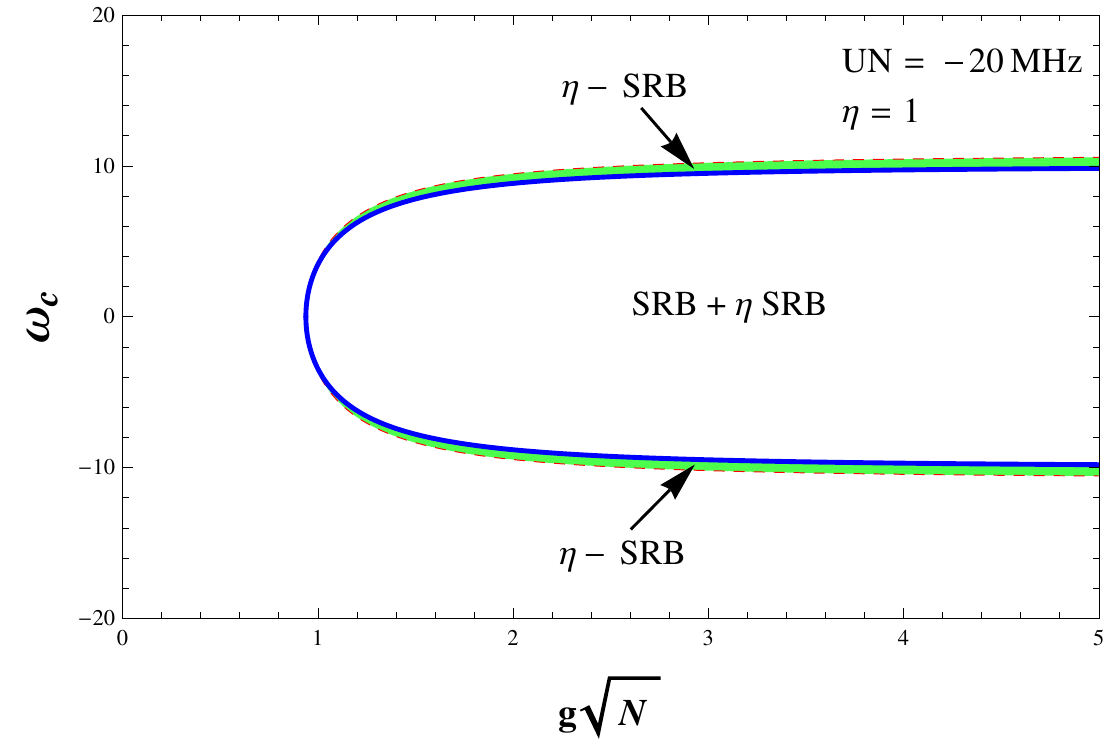}
\caption{$\eta$- SRA and $\eta$- SRB phase for $\eta_p$= 1 in lower two plots for UN= -20 MHz.\\
For detailed analysis, refer \citep{56}.} 
\centering
\end{figure}

The two plots (fig.24) showing the implication of mirror frequency in previous section well concludes the fact that the mirror and the external pump can be used to alter the phase portraits efficiently and can be used as a tool to enhance the phase regions. The phase portraits (fig.24, left panel) also  confirm the finding of \citep{46} that the external mirror pump shifts the bifurcation point to lesser of atom- photon coupling ($g\sqrt{N}$).  The time evolution of the system parameters in the presence of external pump gets modified slightly with no significant change in the nature of physics. It was also observed that by altering the detuning of the cavity light with respect to the atomic transitions and changing the mechanical pump frequency, the condensate energy can be changed. Such systems can also be used for the detection of weak forces as the external pump behaves as a ponder motive detector and provides us with efficient and better control of the phase transitions.

\subsection{Two atom Dicke model}

We consider here two species of atoms A and B inside an optical cavity coupled to the quantized field with transition frequencies $\omega_1$ and $\omega_2$ respectively. The frequency of the cavity mode is denoted by $\omega_c$, driven by an external pump laser of frequency $\omega_p$. $\lambda_1$ and $\lambda_2$ represents the light matter coupling of the two atoms. The detuning ($\omega_p - \omega_i$) is considered to be large so as to avoid spontaneous emission. The effective Hamiltonian of such system, shown schematically as below, takes the form:-

\begin{figure}[h!]
\includegraphics[width=0.65\textwidth]{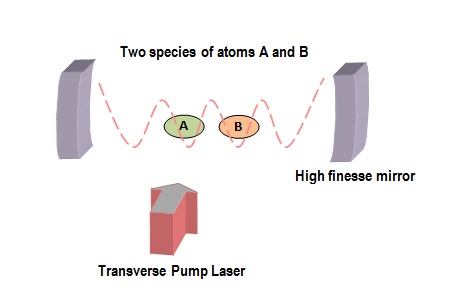}
\caption{The schematic representation of the optomechanical Dicke model with two atomic species.} 
\centering
\end{figure}

\begin{equation}
H= \hbar\omega_1 J_{1z}+ \hbar\omega_2 J_{2z}+ \hbar\omega_c\hat{a}^\dagger \hat{a}+ \frac{\hbar\lambda_1}{\sqrt{N_1}}(J_{1+}+ J_{1-})(\hat{a}+ \hat{a}^\dagger)+ \frac{\hbar\lambda_2}{\sqrt{N_2}}(J_{2+}+ J_{2-})(\hat{a}+ \hat{a}^\dagger),
\end{equation}

where J$_i$= (J$_{ix}$, J$_{iy}$, J$_{iz}$) is the effective collective spin of length of N$_i$/2 for two species of atoms and J$_{i\pm}$= J$_{ix} \pm\iota J_{iy}$. To discuss the non equilibrium dynamics of the above mode, we calculate the semi classical equations of motion: -

\begin{equation}
\dot{J}_{1z}= \frac{\iota\lambda_1}{\sqrt{N_1}} (\hat{a}^\dagger+ \hat{a})(J_{1-}- J_{1+}),
\end{equation}

\begin{equation}
\dot{J}_{2z}= \frac{\iota \lambda_2}{\sqrt{N_2}}(\hat{a}^\dagger+ \hat{a})(J_{2-}- J_{2+}),
\end{equation}

\begin{equation}
\dot{J}_{1-}= -\iota \omega_1 J_{1-}+ \frac{2\iota\lambda_1}{\sqrt{N_1}}(\hat{a}^\dagger+ \hat{a})J_{1z},
\end{equation}

\begin{equation}
\dot{J}_{2-}= -\iota\omega_2 J_{2-}+ \frac{2\iota\lambda_2}{\sqrt{N_2}}(\hat{a}^\dagger+ \hat{a})J_{2z},
\end{equation}

\begin{equation}
\dot{\hat{a}}= -(\kappa+ \iota\omega_c)\hat{a}- \frac{\iota\lambda_1}{\sqrt{N_1}}(J_{1+}+ J_{1-})- \frac{\iota\lambda_2}{\sqrt{N_2}}(J_{2+}+ J_{2-}).
\end{equation}

where $\kappa$ is the photon decay rate. Using pseuso angular momentum conservation and steady state analysis, these equations of motion leads to four types of steady states, namely (a= 0, J$_{1z}= \pm N_1/2, J_{2z}= \pm N_2/2$). Normal and inverted states are defined as (a= 0, J$_{1z}= -N_1/2, J_{2z}= -N_2/2$) and (a= 0, J$_{1z}= N_1/2, J_{2z}= N_2/2$) respectively. The presence of two species of atoms introduces mixed phases defined by (a= 0, J$_{1z}= N_1/2, J_{2z}= -N_2/2$) and (a= 0, J$_{1z}= -N_1/2, J_{2z}= N_2/2$). The critical coupling strength corresponding to the onset of superradiance starting from normal, inverted and mixed state is obtained by putting J$_i$= (0, 0, $\pm N_i /2)$ (i= 1, 2). We explore the fluctuation dynamics above the steady fixed points and instability of the normal ($\downarrow\downarrow$), inverted ($\uparrow\uparrow$) and mixed phases ($\uparrow\downarrow$ or $\downarrow\uparrow$) by fluctuating the system variables about their steady points and equating the coefficients with same time dependency. This gives a quadratic equation for $\omega$, whose roots characterize the possible instabilities. The various boundaries between exponentially growing and decaying fluctuations are given as: -

\begin{figure}[h!]
\includegraphics[width=0.35\textwidth]{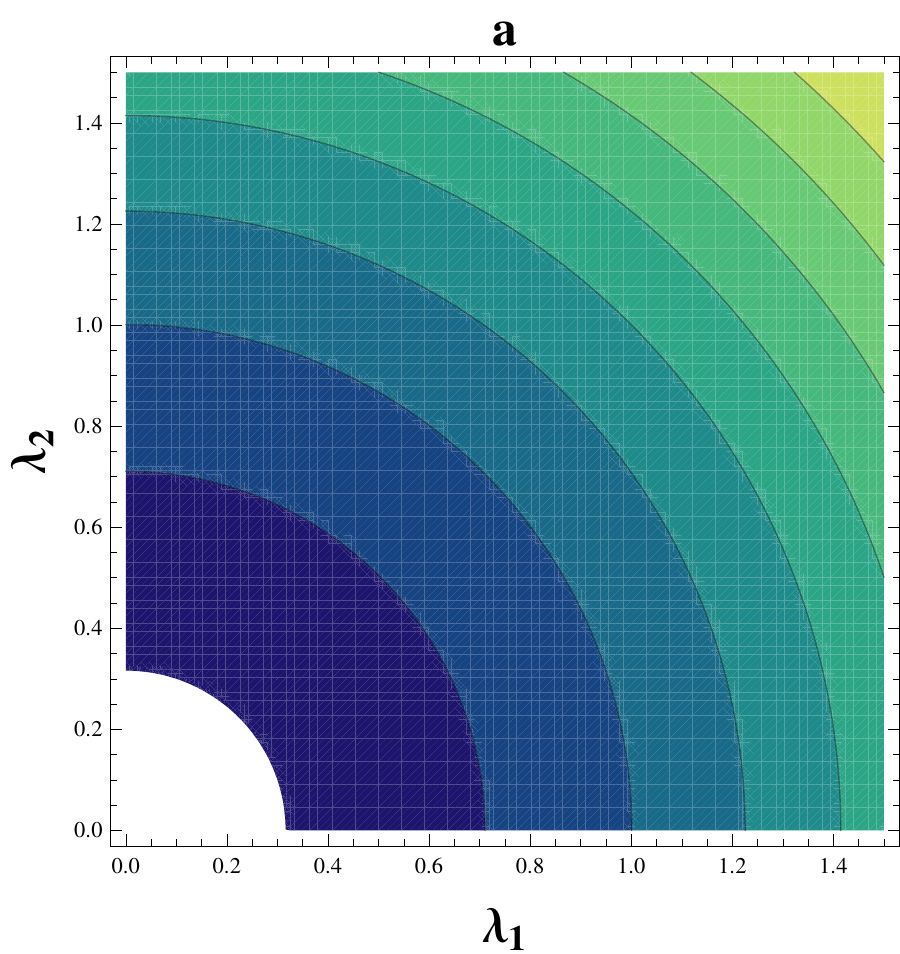}
\includegraphics[width=0.35\textwidth]{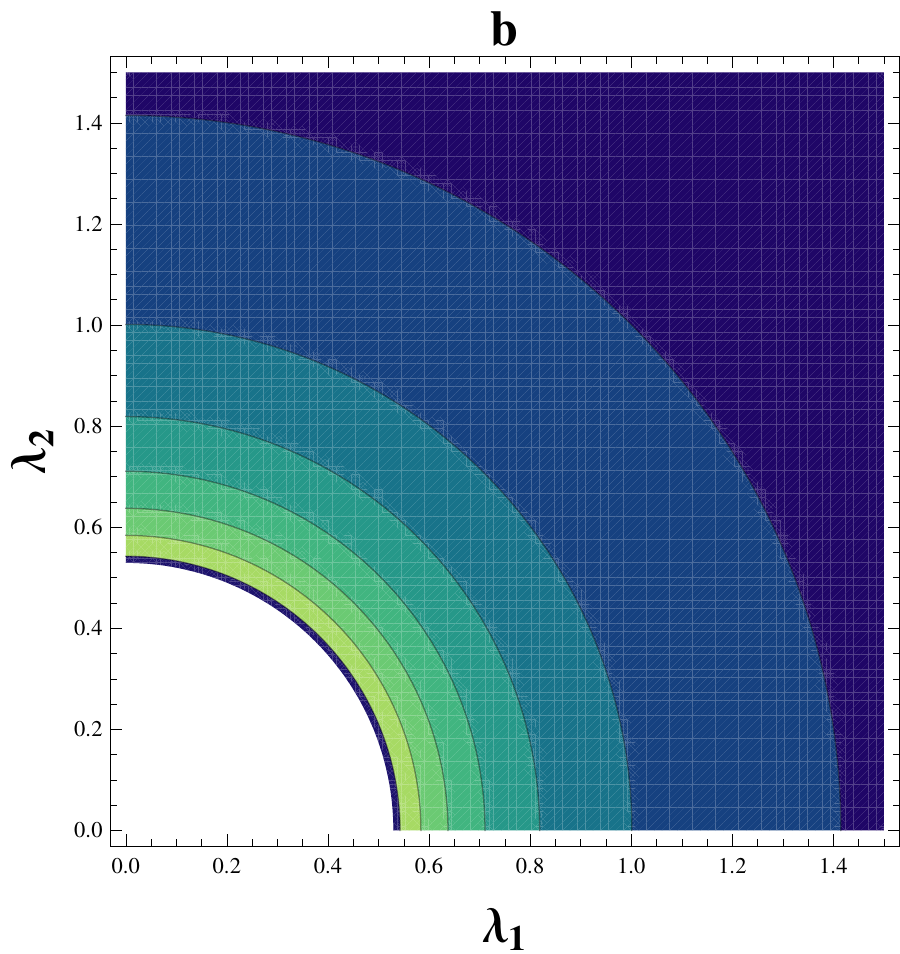}
\caption{The dynamical phase portraits corresponding to the normal phase. The white region is the non-superradiant normal phase while the superradiant phase is indicated by the contours. The darker region in the contour corresponds to low superradiance while light region corresponds to high superradiance. Left (Right) plot corresponds to $\omega_+ (\omega_-)$.} 
\centering
\end{figure}

Normal Phase:

\begin{equation}
\omega_{\pm}= 2\Big(\frac{\lambda_1^2}{\omega_1}+ \frac{\lambda_2^2}{\omega_2}\Big) \pm\sqrt{4\Big(\frac{\lambda_1^2}{\omega_1}+ \frac{\lambda_2^2}{\omega_2}\Big)^2- \kappa^2},
\end{equation}

Inverted Phase:

\begin{equation}
\omega_{\pm}= -2\Big(\frac{\lambda_1^2}{\omega_1}+ \frac{\lambda_2^2}{\omega_2}\Big) \pm\sqrt{4\Big(\frac{\lambda_1^2}{\omega_1}+ \frac{\lambda_2^2}{\omega_2}\Big)^2- \kappa^2},
\end{equation}

Mixed Phase 1:

\begin{equation}
\omega_{\pm}= 2\Big(\frac{\lambda_1^2}{\omega_1}- \frac{\lambda_2^2}{\omega_2}\Big) \pm\sqrt{4\Big(\frac{\lambda_1^2}{\omega_1}- \frac{\lambda_2^2}{\omega_2}\Big)^2- \kappa^2},
\end{equation} 

Mixed Phase 2:

\begin{equation}
\omega_{\pm}= -2\Big(\frac{\lambda_1^2}{\omega_1}- \frac{\lambda_2^2}{\omega_2}\Big) \pm\sqrt{4\Big(\frac{\lambda_1^2}{\omega_1}- \frac{\lambda_2^2}{\omega_2}\Big)^2- \kappa^2},
\end{equation}

\begin{figure}[h!]
\includegraphics[width=0.37\textwidth]{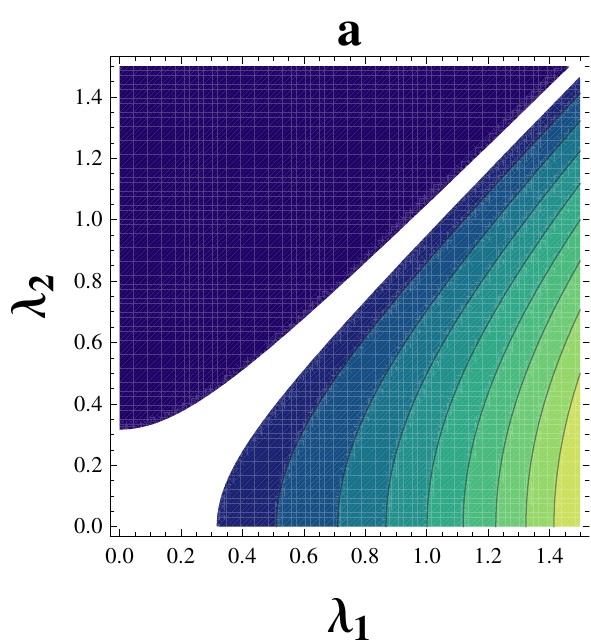}
\includegraphics[width=0.37\textwidth]{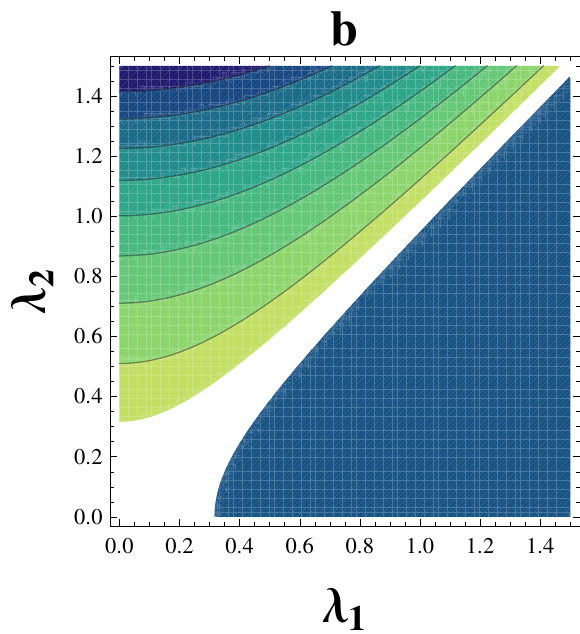}
\caption{The dynamical phase portraits corresponding to the mixed phases. The white region is the non-superradiant mixed phase 1 while the superradiant phase is indicated by the contours. Left (Right) plot corresponds to $\omega_+ (\omega_-)$.} 
\centering
\end{figure}

It is interesting to note here that the inverted phase is the inversion of the normal phase around $\omega$= 0, while mixed phase 2 is the mirror inversion of mixed phase 1. We reproduce here the dynamical phase diagrams (fig. 26) corresponding to the normal phase and mixed phase in the $\lambda_1, \lambda_2$ plane. The white region denotes the non superradiant normal phase (mixed phase in lower plots) while the contours denotes the superradiant phase. The left (right) plots denotes the $\omega_+ (\omega_-)$ roots. It is evident from the plots that as we move along the y axis ($\lambda_1= 0$), we reach the superradiant phase at $\lambda_{2c}= (\kappa^2+ \omega^2)\omega_2 /4\omega$. Thus along only x or y axis, the system behaves as if only one species of atoms are present. In any other direction, both the species contribute.

For the mixed phases, the phase diagram splits into two distinct superradiant regime separated by a non superradiant phase (fig. 27), keeping the two critical points along the axes same. There are regions in the plots where even when $\lambda_1> \lambda_{1c}$ and $\lambda_2> \lambda_{2c}$, the system stays in the superradiant phase. Interestingly, the superradiant phase can never be reached when $\lambda_1= \lambda_2$. The energy landscape in the ($\lambda_1, \lambda_2$) plane gives the impression of anti crossing of energy bands. It must be noted that for $\omega_1 \ne \omega_2$ case, the phase diagrams become asymmetrical. Within the framework of non equilibrium Dicke model, we reveal in this section a new and rich set of phase diagrams exhibiting interesting physics. For two atom Dicke models, there are even regions where the superradiant phase cannot exist even if the light- matter coupling is greater than the critical value for both set of atoms.

\section{Outlook}

	With almost two decades of research since the successful preparation of BECs, many of the theoretical proposals made initially have already been realized by many expertised experimental groups \citep{25, 26, 27, 48, 49} and the ultracold community observed experiments demonstrating better control over quantum system and many body physics through light matter interaction. Ultracold atoms prepared in magnetic and optical traps are now easily coupled to cavity and mechanical modes and the phenonemons like backaction force, Bogoliubov mode can be readily studied. Study of coupling and properties of recently observed photon BEC also remains a challenge for laser and ultracold physics. Recent experiment demonstrating subrecoil cavity cooling urged the replacement of evaporative cooling techniques by cavity cooling and direct preparation of quantum states. However lot many challenges such as trapping of molecular samples or suspended objects, multimode cavity systems, quantum entanglement in optical and optomechanical systems and efficient control of trapped ions for quantum information processing still persists in experimental regime \citep{55}.\\
	
Theoretical research also opened new paths for experimental verification such as superglass phases and physics of spin glasses \citep{53, 54}. More theoretical proposals and models need to be incorporated in the field of macroscopic quantum mechanics including preparation of quantum states \citep{50} involving superposition of mechanical states via optomechanical entanglement, and even probing the decoherence of such superposition states. The success of this field since the observation of radiation pressure in optomechanical systems,  prediction of BEC in 1926 and then preparing the same in 1995 and finally extending the optomechanical ideas even in diverse fields of gravitational wave detection \citep{51, 52}, LIGO projects has been tremendous.

\section{Acknowledgements}
	Aranya B Bhattacherjee acknowledge financial support from the Department of Science and Technology, New Delhi for financial assistance vide grant SR/S2/LOP-0034/2010.

\section{Appendix A}

The coupling matrix elements of Eq. 8 can be written as \citep{34}: -

\begin{equation}
E_{k, l}= \int d^3x w (\textbf{x}- \textbf{x}_{k}) \Big( -\frac{\hbar^2}{2m} \bigtriangledown ^2 \Big) w (\textbf{x}- \textbf{x}_l),
\end{equation}

\begin{equation}
J_{k, l}= \int d^3x w (\textbf{x}- \textbf{x}_{k}) cos ^2 (kx) w(\textbf{x}- \textbf{x}_l),
\end{equation}

\begin{equation}
\tilde{J}_{k, l}= \int d^3x w (\textbf{x}- \textbf{x}_{k}) cos (kx) w(\textbf{x}- \textbf{x}_l).
\end{equation}

Due to the presence of the cosine term in the above equations, which changes sign periodically, the two wells acquire different depths in the case of transverse pumping. This implies $\tilde{J}_{k, k}= -\tilde{J}_{k+1, k+1}$.\\

The constants used in Eq. (13), (66) can be written as \citep{35}: -

\begin{equation}
U= \frac{4 \pi a_s \hbar^2}{m} \int d^3x \mid w(\vec(r)) \mid ^4,
\end{equation}

\begin{equation}
E_0= \int d^3x w(\vec r- \vec r_j) \Big(- \frac{\hbar^2 \bigtriangledown ^2 }{2m} \Big) w(\vec{r}- \vec{r}_j),
\end{equation}

\begin{equation}
E= \int d^3x w(\vec r- \vec r_j) \Big(- \frac{\hbar^2 \bigtriangledown ^2 }{2m} \Big) w(\vec{r}- \vec{r}_{j\pm 1}),
\end{equation}

\begin{equation}
J_0= \int d^3x w(\vec r- \vec r_j) cos^2 (kx) w(\vec r- \vec r_j),
\end{equation}

\begin{equation}
J= \int d^3x w(\vec r- \vec r_j) cos^2 (kx) w(\vec r- \vec r_{j\pm 1}).
\end{equation}

The nearest neighbour non linear interaction terms are usually very small as compared to the onsite interaction and hence $J_0$ and $E_0$ are set to zero. Hence eq. (14) and (15) gives the corresponding Heisenberg equations of motion.

\section{Appendix B}

To keep consistency with the published literature \citep{ 38}, few symbols were defined separately to derive Eq. (38) and (39) which can be described as: -

\begin{equation}
U_{eff}= \frac{U n_0}{\hbar},
\end{equation}

\begin{equation}
g_c= U_0 J_0 \sqrt{N} \mid a_s \mid,
\end{equation}

where $U_0= \frac{g_0^2}{\Delta_z}$  is the optical lattice height per photon and can be interpreted as the atomic back action on the field. $n_0= N/M$ where there are total $N$ atoms in $M$ sites.

\begin{equation}
v= U_0 J_0 \mid a_s \mid^2+ \frac{V_{cl} J_0}{\hbar}+ \frac{E_0}{\hbar},
\end{equation}

\begin{equation}
\Delta_d= \Delta_c- U_0 N J_0,
\end{equation}

where $V_{cl}$ is the classical potential and $\Delta_d$ is the detuning with respect to the normalized resonance.

\section{Appendix C}

The correlations that are satisfied by the input noise operators and described in Eq. (53) are given as: -

\begin{equation}
<b_{in} (t) b_{in} (t')>= <b_{in}^{\dagger} (t) b_{in} (t')>= 0,
\end{equation}

\begin{equation}
<b_{in}(t) b_{in}^{\dagger} (t')>= \delta (t- t').
\end{equation}

$W(t)= i\sqrt{\Gamma_m}[\xi_m^{\dagger} (t)- \xi_m(t)]$ is the noise operator due to the Brownian motion of the mirror and satisfies the correlation given as: -

\begin{equation}
<W(t) W(t')>= \frac{\Gamma_m}{\omega_m}\int \frac{d\omega}{2\pi} e^{-i\omega(t- t')}\omega \Big[1+ coth \Big(\frac{\hbar \omega}{2k_B T} \Big)\Big],
\end{equation}

where $T$ is the finite temperature of the bath that is connected to the movable mirror and $k_B$ represents the Boltzmann constant. The correlations satisfied by the amplitude and phase quadrature of the input noise operator is given as: -

\begin{equation}
<q_{in}(\omega) q_{in} (\omega')>= 2\pi \delta(\omega+ \omega'),
<p_{in}(\omega) p_{in}(\omega')>= 2 \pi \delta (\omega+ \omega'),
\end{equation}

\begin{equation}
<q_{in}(\omega) p_{in}(\omega')>= 2\pi i \delta (\omega+ \omega'),
<p_{in} (\omega) q_{in}(\omega')>= -2 \pi i \delta (\omega+ \omega').
\end{equation}

The correlation function for the Brownian noise operator is given as: -

\begin{equation}
<W(\omega) W(\omega')>= 2\pi \frac{\Gamma_m}{\omega_m}\omega \Big[1+ coth \Big( \frac{\hbar \omega}{2k_B T} \Big) \Big] \delta (\omega+ \omega').
\end{equation} \\

The correlation function of the vacuum field quadrature in the Fourier space is $<p_v (\omega) p_v (\omega')>= 2 \pi \delta (\omega+ \omega')$.

$X (\omega)$ and $\chi_{1}(\omega)$ of Eq. (56) and (57) respectively can be written as: -

\begin{equation}
X (\omega)= 16g_c^4 \Delta_d^2\beta_1^2- 8g_c^2 \Delta_d \beta_1 (\omega^2- \beta_1 \beta_2)\Big( \Delta_d^2+ \frac{\kappa^2}{4}- \omega^2 \Big)+ (\omega^2- \beta_1 \beta_2^2)^2 \Big[ \omega^2 \kappa^2+ \Big(\Delta_d^2+ \frac{\kappa^2}{4}- \omega^2 \Big)^2 \Big],
\end{equation}

where 
\begin{equation}
\chi_1(\omega)= \frac{4G^2\beta^2\Delta_d \omega_m (\omega^2- \beta_1 \beta_2)}{\Big[(\omega^2- \beta_1\beta_2)\Big(\Delta_d^2+ \frac{\kappa^2}{4}- \omega^2+ i\omega\kappa \Big)- 4g_c^2\Delta_d\beta_1 \Big]}. 
\end{equation}

\section{Appendix D}

Various symbols undefined in the main text, in Eq. (64) are \citep{42}: -

\begin{eqnarray}
S_{fb} (\omega)&=& \frac{\Big[2G\beta\sqrt{\lambda}(\omega^2- \beta_1\beta_2) \Bigg( \Delta_d(\omega^2- \beta_1\beta_2) \Big(\omega^2- \frac{\kappa^2}{4}-\Delta_d^2 \Big)+ 4g_c^2\beta_1 \Big(\Delta_d^2+ \frac{\kappa^2}{2}\Big) \Bigg)(g(\omega)+ g(-\omega)) \Big]}{X(\omega)} \nonumber\\
&-& \frac{\Big[2iG\omega\kappa\sqrt{\lambda} (\omega^2- \beta_1\beta_2) \Bigg(\Delta_d (\omega^2- \beta_1\beta_2)- 4g_c^2\beta_1 \Bigg)\Big(g(\omega)- g(-\omega) \Big)\Big]}{X(\omega)} \nonumber \\
&+& \mid g(\omega) \mid^2 \Bigg[ \frac{1}{\kappa}- \frac{\Big[4g_c^2\beta_1\lambda\kappa \Big(\Delta_d(\omega^2- \beta_1\beta_2)- 4g_c^2\beta_1\Big)\Big]}{X(\omega)}\Bigg],
\end{eqnarray}

which arises due to the feedback of the measurement noiseinto the dynamics of the movable mirror. The mechanical susceptibility, $\chi_{eff}^{cd} (\omega)$, modified by the filter function is given as: -

\begin{equation}
\chi_{eff}^{cd} (\omega)= \frac{\omega_m}{[(\omega_m^2- \omega^2+ i\omega\Gamma_m)+ \chi_1^{cd}(\omega)]},
\end{equation}

where,

\begin{equation}
\chi_1^{cd} (\omega)= \frac{2G\beta\omega_m(\omega^2- \beta_1\beta_2){g(-\omega)\sqrt{\lambda} (i\omega+ \frac{\kappa}{2})+ 2G\beta\Delta_d}}{{(\omega^2- \beta_1\beta_2)(\Delta_d^2+ \frac{\kappa^2}{4}- \omega^2+ i\omega\kappa)- 4g_c^2\Delta_d\beta_1}}.
\end{equation}

From Eq. (63), it gives the effective resonance frequency and damping rate as: -

\begin{equation}
\omega_m^{eff, cd}(\omega)= [\omega_m^2+ \omega_m^{op, cd}]^{1/2},
\end{equation}

where,

\begin{eqnarray}
\omega_m^{op, cd}&=& X_1 (\omega) \Big[ (\omega^2- \beta_1\beta_2) \Big[ \Delta_d^2+ \frac{\kappa^2}{4}-\omega^2 \Big]- 4g_c^2\Delta_d\beta_1 \Big] \times \Bigg(4G\beta\Delta_d+ \frac{2\omega^2g_{cd}\omega_{fb}\lambda}{(\omega^2+ \omega_{fb}^2)} \Big( \frac{\kappa}{2}- \omega_{fb} \Big) \Bigg) \nonumber \\
&+& X_1 (\omega)(\omega^2- \beta_1\beta_2) \Big( \frac{2\omega^2g_{cd}\omega_{fb}\lambda\kappa}{(\omega^2+ \omega_{fb}^2)} (\omega^2+ \frac{\omega_{fb}\kappa}{2}) \Big).
\end{eqnarray}

\begin{eqnarray}
\Gamma_m^{eff, cd} (\omega)&=& \Gamma_m+ X_1 (\omega) \Big[ (\omega^2- \beta_1\beta_2) \Big[\Delta_d^2+ \frac{\kappa^2}{4}- \omega^2 \Big] -4 g_c^2\Delta_d\beta_1 \Big] \Bigg( \frac{2g_{cd}\omega_{fb}\lambda}{(\omega^2+ \omega_{fb}^2)} \Big( \omega^2+ \frac{\omega_{fb}\kappa}{2} \Big)\Bigg)\nonumber\\
&-& X_{1} (\omega) \kappa (\omega^2- \beta_1 \beta_2) \times  \Bigg(4G\beta \Delta_d+ \frac{2\omega^2g_{cd}\omega_{fb}\lambda}{(\omega^2+ \omega_{fb}^2)} (\frac{\kappa}{2}- \omega_{fb}) \Bigg),
\end{eqnarray}

where

\begin{equation}
X_1(\omega)= \frac{G\beta \omega_m (\omega^2- \beta_1 \beta_2)}{X(\omega)}.
\end{equation}

\section{Appendix E}

The expressions for $\Omega_s$ and $\Gamma_s$ used in Eq. (74) are \citep{43}: -

\begin{equation}
\Omega_s (\omega)= \frac{\Delta_d g_m^2 Y^4\Omega_{\nu}^2}{2(Y^8+ \kappa^2\omega^2\Omega_{\nu}^4)},
\end{equation}

\begin{equation}
\Gamma_s (\omega)= -\frac{\kappa\Delta_d g_m^2 \Omega_m \Omega_{\nu}^4}{(Y^8+\kappa^2\omega^2\Omega_{\nu}^4)},
\end{equation}

where

\begin{equation}
Y^4= \Bigg[ \Big( \frac{\kappa^2}{4}+ \Delta_d^2- \omega^2 \Big) (\omega^2- (\nu+ U_{eff})(\nu+ 3U_{eff}))- 4\Delta_dg_c^2(\nu+ U_{eff}) \Bigg],
\end{equation}

\begin{equation}
\Omega_{\nu}^2= ({\omega^2- (\nu+ U_{eff})(\nu+ 3U_{eff})}).
\end{equation}

\end{document}